\documentclass[%
twocolumn,
reprint,
superscriptaddress,
amsmath,amssymb,
prx,
longbibliography,
floatfix,
rc]{revtex4-2}
\usepackage{comment}

\usepackage{graphicx}
\usepackage{dcolumn}
\usepackage{bm}
\usepackage{xcolor}
\usepackage{siunitx}
\usepackage[hidelinks]{hyperref}

\begin{document}

\preprint{APS/123-QED}

\title{Acceptor-induced bulk dielectric loss in superconducting circuits on silicon}

%

\author{Zi-Huai Zhang}
\thanks{Z.Z. and K.G. contributed equally.}
\affiliation{
Department of Electrical Engineering and Computer Sciences, University of California,  Berkeley, Berkeley, California 94720, USA
}
\affiliation{
 Materials Sciences Division, Lawrence Berkeley National Laboratory, Berkeley, California 94720, USA
}
\affiliation{
Department of Physics, University of California, Berkeley, Berkeley, California 94720, USA
}%

\author{Kadircan Godeneli}
\thanks{Z.Z. and K.G. contributed equally.}
\affiliation{
Department of Electrical Engineering and Computer Sciences, University of California,  Berkeley, Berkeley, California 94720, USA
}
\affiliation{
 Materials Sciences Division, Lawrence Berkeley National Laboratory, Berkeley, California 94720, USA
}%

\author{Justin He}
\affiliation{
Department of Electrical Engineering and Computer Sciences, University of California,  Berkeley, Berkeley, California 94720, USA
}

\author{Mutasem~Odeh}
\affiliation{
Department of Electrical Engineering and Computer Sciences, University of California,  Berkeley, Berkeley, California 94720, USA
}
\affiliation{
 Materials Sciences Division, Lawrence Berkeley National Laboratory, Berkeley, California 94720, USA
}%

\author{Haoxin Zhou}
\affiliation{
Department of Electrical Engineering and Computer Sciences, University of California,  Berkeley, Berkeley, California 94720, USA
}
\affiliation{
 Materials Sciences Division, Lawrence Berkeley National Laboratory, Berkeley, California 94720, USA
}
\affiliation{
Department of Physics, University of California, Berkeley, Berkeley, California 94720, USA
}%

\author{Srujan Meesala}
\affiliation{Institute for Quantum Information and Matter, California Institute of Technology, Pasadena, California 91125, USA.}
\author{Alp Sipahigil}
\email{Corresponding author: alp@berkeley.edu}

\affiliation{
Department of Electrical Engineering and Computer Sciences, University of California,  Berkeley, Berkeley, California 94720, USA
}
\affiliation{
 Materials Sciences Division, Lawrence Berkeley National Laboratory, Berkeley, California 94720, USA
}
\affiliation{
Department of Physics, University of California, Berkeley, Berkeley, California 94720, USA
}%

\date{\today}

\begin{abstract}

%

The performance of superconducting quantum circuits is primarily limited by dielectric loss due to interactions with two-level systems (TLS). State-of-the-art circuits with engineered material interfaces are approaching a limit where dielectric loss from bulk substrates plays an important role. However, a microscopic understanding of dielectric loss in crystalline substrates is still lacking. In this work, we show that boron acceptors in silicon constitute a strongly coupled TLS bath for superconducting circuits. We discuss how the electronic structure of boron acceptors leads to an effective TLS response in silicon. We sweep the boron concentration in silicon and demonstrate the bulk dielectric loss limit from boron acceptors. We show that boron-induced dielectric loss can be reduced in a magnetic field due to the spin-orbit structure of boron. This work provides the first detailed microscopic description of a TLS bath for superconducting circuits, and demonstrates the need for ultrahigh purity substrates for next-generation superconducting quantum processors. 

\end{abstract}

\maketitle
Superconducting quantum processors are a leading platform for quantum computation~\cite{arute_quantum_2019,kim_2023} and  simulation~\cite{HartreeFock_2020}. The performance of superconducting quantum processors is currently limited by high error rates~\cite{googlequantumai_2023} from coupling of qubits to unwanted energy dissipation channels such as quasiparticles, vortices, radiation, parasitic modes, and two-level systems (TLS)~\cite{mcrae_2020a,siddiqi_2021a}. TLSs are atomic-scale defects that are described by the standard tunneling model~\cite{phillips_1972,Anderson_1972}. While the microscopic nature of TLSs remains elusive, they are primarily located inside amorphous materials at interfaces instead of the crystalline bulk substrate~\cite{gao_experimental_2008,wang_surface_2015}. TLSs can have strong electric and elastic dipoles~\cite{Müller_2019}. At the macroscopic level, this leads to dielectric loss, which is currently the dominant dissipation mechanism for superconducting qubits~\cite{martinis_2005a,wang_surface_2015}.  
Losses from interface TLSs can be reduced with improved surface treatments, material choices~\cite{place_2021}, and circuit designs with reduced surface participation~\cite{martinis_2022}. With advances on these fronts, state-of-the-art superconducting qubits now show lifetimes approaching 1~ms~\cite{ganjam_surpassing_2023}. For such low-loss devices, dielectric loss from the bulk substrate becomes non-negligible~\cite{read_2023b, crowley_2023a}. For instance, the best reported bulk loss tangent for sapphire~($\approx 2 \times 10^{-8}$) limits the quality factor to 50 million~\cite{read_2023b}, and the qubit lifetime to 1.5~ms. These results indicate the need for further advances in our understanding of bulk dielectric loss for substrates of superconducting circuits.

\begin{figure}[ht]
 	\centering
    \includegraphics[width=86mm]{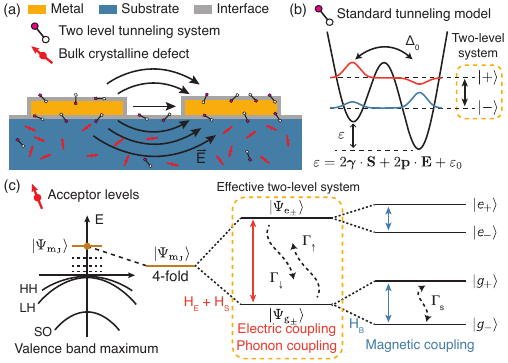}
 	\caption{\textbf{Comparison of amorphous two-level systems and boron acceptors in silicon.} 
  (a) Schematic of the cross-section of planar superconducting devices. TLS loss originates from defects or disorder in the amorphous interface layers and the bulk dielectric substrate. Bulk crystalline defects may also contribute to TLS loss. (b) Standard tunneling model of TLS. The asymmetry energy ($\varepsilon$) of the double-well potential couples strongly to electric field ($\textbf{E}$) and strain ($\textbf{S}$) through electric dipole ($\textbf{p}$) and deformation potential ($\bm{\gamma}$). $\Delta_0$ denotes the tunneling rate between the two wells. (c) Left: Valence band maximum of silicon showing the split-off (SO), heavy-hole (HH), and light-hole (LH) bands. Right: Electronic structure of boron acceptors in the hole picture showing coupling to electric, strain, and magnetic fields as described by the interaction Hamiltonians $H_E$, $H_S$, and $H_B$ in the main text. $\Gamma_\uparrow \,(\Gamma_\downarrow)$: orbital relaxation rate. $\Gamma_\text{s}$: relaxation rate between the generalized spin states.}
 	\label{fig1}
\end{figure}

Silicon is a widely used material for quantum devices based on superconducting, spintronic, mechanical, and photonic systems~\cite{bergeron_2020,redjem_2020,komza_2022,higginbottom_2022a,mi_2018, eichenfield_2009a}. High-resistivity (${>1000\,\Omega \cdot\text{cm}}$), float-zone~(FZ) grown silicon substrates show bulk dielectric loss tangents below $5\times 10^{-7}$~\cite{gambetta_2017,woods_2019,melville_2020}, and are a standard choice to realize high-performance superconducting devices. However, a microscopic understanding of the origin of bulk dielectric loss in silicon is currently missing. In this work, we identify crystalline point defects associated with boron acceptors in silicon as a bulk TLS bath with strong electric dipolar coupling to superconducting circuits. We begin with a theoretical description of how the microscopic electronic structure of boron in silicon can result in TLS loss for superconducting microwave circuits. We use superconducting resonator loss measurements under varying  doping concentration, microwave power, temperature, and device geometry to confirm the theoretical predictions. In addition, we experimentally show that the spin-orbit nature of boron acceptors results in a reduction of loss saturation power under a magnetic field. These observations are explained by the four-level fine structure of boron and support the identification of boron acceptors as the TLS bath. These results provide guidelines on silicon substrate purity requirements for low-loss qubits, and constitute the first microscopic identification of a TLS bath in a crystalline substrate.

For superconducting microwave circuits operating at low temperature and in the single-photon regime, defect-induced dielectric loss can arise from coupling to TLSs and bulk crystalline defects~(Fig.~\ref{fig1}(a)). TLSs can be described by the standard tunneling model (STM) (Fig.~\ref{fig1}(b)). They strongly couple to electric (\textbf{E}) and strain (\textbf{S}) fields through the asymmetry energy of the double-well potential: $\varepsilon = \varepsilon_0 + 2\bm{\gamma}\cdot \textbf{S} + 2\textbf{p}\cdot \textbf{E}$. Typical TLS has a deformation potential ($\bm{\gamma}$) of 1~eV~\cite{jackle_1972} and an electric dipole moment ($\textbf{p}$) of 3~D~\cite{sarabi_2016b}. The TLS can be described by the following Hamiltonian~\cite{phillips_1987a}:
\begin{equation}
\label{TLS_H}
H = \frac{1}{2}\Delta E \sigma_z + \left(\bm{\gamma} \cdot \textbf{S} + \textbf{p} \cdot \textbf{E}\right)\left(\frac{\varepsilon_0}{\Delta E}\sigma_z + \frac{\Delta_0}{\Delta E} \sigma_x\right)
\end{equation}
where $\Delta_0$ is the tunneling rate between the two potential wells, $\varepsilon_0$ is the static asymmetry energy, $\Delta E = \sqrt{\varepsilon_0^2 + \Delta_0^2}$ is the energy splitting between the two levels, and $\sigma$ denotes the Pauli operator. Due to their amorphous nature, TLS parameters are sampled from a broad distribution.

In contrast to highly disordered TLS defects in amorphous materials, silicon hosts high-quality spin (spin-orbit) qubits based on crystalline donor (acceptor) defects~\cite{muhonen_2014,wolfowicz_2013,kobayashi_2021a}. The electronic structure of donors and acceptors are well studied and can be used to predict their impact on superconducting circuits. Donor~(e.g., phosphorus, bismuth) defects weakly couple to their environment via magnetic dipole interactions and show long spin lifetimes~\cite{muhonen_2014,wolfowicz_2013,mansir_2018}. Magnetic dipole interactions are much weaker compared to electric dipole interactions, and the donor spin resonances are typically detuned from circuits at zero magnetic field. Donor defects therefore cannot result in the frequently observed saturable TLS loss in superconducting circuits. In contrast, other bulk crystalline defects can display Hamiltonians similar to TLSs and cause dielectric loss. In particular, acceptor (e.g., boron, aluminum) defects have strong electric and elastic dipoles, and therefore exhibit short lifetimes~\cite{song_2011a,kopf_1992a,neubrand_1978}. In this work, we investigate acceptors as a potential origin of a bulk TLS bath in silicon. Our study focuses on boron defects, one of the most common acceptor-type contaminants in silicon.

The electronic structure of boron acceptors inherits the properties of the valence band maximum of silicon (Fig.~\ref{fig1}(c)). The ground state is an effective spin-3/2 system with a two-fold orbital degeneracy and a two-fold spin degeneracy. The linear coupling of the ground state to magnetic ($H_B$), electric ($H_E$) and strain ($H_S$) fields can be described as~\cite{bir_pikus_1974}:
\begin{align}
H_B = &\ \mu_B \left[g_1(J_x B_x + \text{c.p.}) + g_2(J_x^3 B_x + \text{c.p.})\right] \\
H_E = &\ \frac{p_B}{\sqrt{3}}\left(E_x\{J_y,J_z\}_+ + \text{c.p.}\right)\\
H_S = &\ \gamma_B S_{xx}J_x^2 + \frac{\gamma_B^\prime}{\sqrt{3}}S_{xy}\{J_x,J_y\}_+ + \text{c.p.}
\end{align}
where $J$ is the spin-3/2 operator, $\{\cdot,\:\cdot\}_+$ denotes the anticommutator, c.p. denotes cyclic permutation, $g_1 = -1.07$ and $g_2 = -0.03$ are the g-factors~\cite{kopf_1992a}, $p_B = $ 0.26~D is the electric dipole moment~\cite{kopf_1992a}, and $\gamma_B = -1.42$~eV and $\gamma_B^\prime = -3.7$~eV are deformation potentials~\cite{neubrand_1978}. The two-fold orbital degeneracy can be lifted with static lattice strain and/or electric field ($H_S + H_E$), which results in a TLS-like level structure. Within each orbital branch, the spin degeneracy can be lifted with a static magnetic field ($H_B$). The magnitudes of the electric dipole ($p_B$) and deformation potentials ($\gamma_B, \gamma_B^\prime$) for boron acceptors  are similar to those of conventional TLSs ($\textbf{p}$ and $\bm{\gamma}$ in Eq.~\ref{TLS_H}) in amorphous materials. The strong correspondence in orbital structures and dipole strengths between borons and TLSs suggests that boron acceptors should lead to saturable dielectric loss like conventional TLSs.

\begin{figure}[t]
    \centering
    \includegraphics[width=\columnwidth]{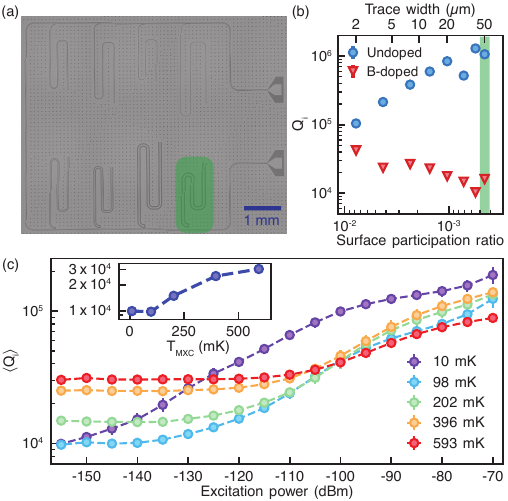}
    \caption{\textbf{Bulk two-level system loss in a boron doped silicon substrate.} (a) Optical image of the device with eight quarter-wave resonators capacitively coupled to a shared feedline. The resonator trace width is swept from 2~$\mu$m to 50~$\mu$m to access different surface participation ratios while maintaining 50~$\Omega$ impedance. The highlighted resonator (green) has the lowest surface participation. (b) Internal quality factor ($Q_i$) at single-photon level as a function of surface participation ratio. $Q_i$ of resonators on undoped (doped, $[\text{B}]= 7.4\times 10^{14}\,\text{cm}^{-3}$) silicon is measured at $\langle n\rangle\approx 0.1$ ($\langle n\rangle\approx0.01$). Data points shaded in green are measured from the lowest surface participation ratio resonator highlighted in~(a).  (c) Power dependent $Q_i$ at different temperatures for low surface participation ratio resonators on boron doped silicon. $\langle Q_{i}\rangle$ represents the average of power dependence from eight resonators. Inset: temperature dependence of low-power $Q_{i}$ (Excitation power -155 dBm, $\langle n \rangle < 0.1$).}
        \label{fig2}
\end{figure}

In the following, we test the hypothesis of boron acceptors constituting a bulk TLS bath and a strong dielectric loss channel. We use superconducting resonators to probe acceptor-induced loss under varying acceptor concentration, temperature, power, and magnetic fields. To access a bulk dielectric loss dominated regime, we first study the geometric dependence of loss under our fabrication procedure with quarter-wave resonators of varying dimensions (Fig.~\ref{fig2}(a)) and frequencies ranging from 4.3~GHz to 7.7~GHz. For resonators on both undoped ($[\text{B}]<8\times 10^{11}\, \text{cm}^{-3}$) and boron-doped ($[\text{B}]= 7.4\times 10^{14}\,\text{cm}^{-3}$) FZ silicon, we observe improvement of internal quality factors ($Q_i$) at high excitation powers~(see Supplemental Material Fig.~S2~\cite{Supp}). The power dependence shows the presence of TLS loss in both substrates. For undoped silicon, the single-photon $Q_i$ scales inversely with the surface participation ratio, indicating the loss is dominated by surface TLSs~(Fig.~\ref{fig2}(b))~\cite{wang_surface_2015}. On boron-doped silicon, the single-photon $Q_i$ does not show such an inverse scaling with surface participation (Fig.~\ref{fig2}(b)). The energy participation in the bulk silicon substrate is independent of geometry and is near unity. The absence of negative correlation between the surface participation ratio and $Q_i$ confirms that TLS loss is no longer surface limited on boron-doped silicon, and that we are probing TLSs in the bulk silicon substrate~\cite{crowley_2023a}.

We use the resonator design with the lowest surface participation to probe bulk TLS loss in boron-doped silicon ($[\text{B}]= 7.4\times 10^{14}\, \text{cm}^{-3}$). The sample contains eight quarter-wave resonators evenly spaced in a 1~GHz band centered around 6~GHz~(see Supplemental Material Fig.~S1(b)~\cite{Supp}). The average power-dependent $Q_i$ of the resonators at different mixing chamber temperatures ($\text{T}_{\text{MXC}}$) is plotted in Fig.~\ref{fig2}(c). For all resonators, $Q_i$ consistently shows strong reduction at low powers. At the same time, low-power $Q_i$ increases when thermal energy is comparable to the resonator frequency (Fig.~\ref{fig2}(c), inset). This thermal saturation feature is another indication that bulk loss tangent in boron doped silicon is dominated by \mbox{TLS-like} atomic defects~\cite{mcrae_2020a}. These observations confirm the TLS-acceptor correspondence discussed in Fig.~\ref{fig1}, and we conclude that boron acceptors act as a TLS loss channel in the bulk substrate. We note that the bulk TLS loss from boron defects appears broadband ($>1$~GHz) based on the consistent saturation behaviors of resonators at different frequencies. This broadband behavior is likely related to a broad inhomogeneous distribution of boron orbital splittings due to an inhomogeneous strain distribution near the metal-silicon interface (see Supplemental Material Fig.~S7 and Fig.~S8~\cite{pla_2018,Supp}) and the local disorder in the environment of individual defects. Therefore, performing direct loss tangent measurement on low or high strain samples may reveal a different loss behavior~\cite{checchin_2022c,read_2023b}.

We quantify the impact of acceptor-induced dielectric loss on superconducting qubits by studying superconducting resonator low-power $Q_i$ as a function of dopant type and doping concentration. Our first principles estimation~(see Supplemental Material Sec.~III~C \cite{Supp}) suggests acceptor-induced dielectric loss can be significant due to the near-unity energy participation of the bulk. However, the exact magnitude of acceptor-induced loss tangent strongly depends on the strain distribution and the electric dipole moment, and necessitates experimental characterization. 

\begin{figure}[!htbp]
 	\centering
 	\includegraphics[width=\columnwidth]{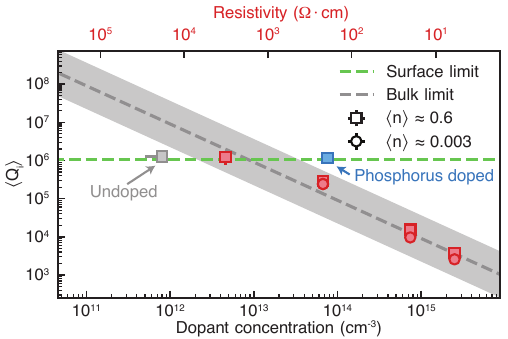}    
 	\caption{\textbf{Dielectric loss in silicon under different doping conditions.} For substrates with high boron concentration, we show the average $Q_i$ with $\langle n\rangle \approx 0.003$ (red circle) and  $\langle n\rangle \approx 0.6$ (red square). For substrates with low boron concentration, $Q_i$ is surface TLS limited, and we report the average $Q_i$ with $\langle n\rangle \approx 0.6$ due to the challenge of $Q_i$ extraction with an overcoupled ($Q_e \approx 4 \times 10^4$) resonator design (see Supplemental Material Fig.~S3~\cite{Supp}). $Q_i$ on phosphorus doped silicon is consistent with the surface TLS limit. The gray dashed line is a linear fit ($\log Q_i = - \log (a\times \rho$)) of loss at $\langle n\rangle \approx 0.003$ to boron concentration ($\rho$), and is extrapolated to a boron concentration of $5\times 10^{10}$ cm$^{-3}$. The gray band represents the 95\% confidence band. The boron concentration is measured using secondary ion mass spectrometry (SIMS) while phosphorus concentration is estimated using wafer resistivity. The boron concentration in undoped silicon is below the detection limit of SIMS ($8\times10^{11}$~cm$^{-3}$). The boron doped silicon with highest doping concentration is Czochralski grown whereas all other substrates are FZ grown. The top x-axis shows the expected room-temperature wafer resistivity from boron doping.
}
 	\label{fig3}
\end{figure}

We fabricate resonators identical to the ones studied in Fig.~\ref{fig2}(c) on undoped, phosphorus-doped, and boron-doped substrates~\cite{Supp}. The average low-power $Q_i$ at $\text{T}_{\text{MXC}} = 8$~mK as a function of doping concentration is summarized in Fig.~\ref{fig3}. For undoped silicon and lightly boron-doped ($[\text{B}]= 4.5\times 10^{12}\,\text{cm}^{-3}$) silicon, low-power $Q_i$ saturates around $10^6$, consistent with the surface TLS limit in Fig.~\ref{fig2}(b). For substrates with higher boron concentration, $Q_i$ shows strong anti-correlation with boron concentration. This anti-correlation is consistent with the assignment of boron acceptors as a bulk TLS bath. We extrapolate the bulk limit of $Q_i$ from boron doping for state-of-the-art devices using a linear fit of loss and boron doping. The extrapolation suggests bulk-limited $Q_i$ would be limited to $10^7$ and $10^8$ for boron concentrations of $10^{12}\, \text{cm}^{-3}$ and $10^{11}\,\text{cm}^{-3}$, respectively. We note that state-of-the-art quantum devices have lifetimes ($\text{T}_1$) around 1~ms, corresponding to a quality factor of $Q = 2\pi f \times \text{T}_1 \approx 3 \times 10^7$ at $f = 5$~GHz. Based on our extrapolation and considering bulk dielectric loss alone, realizing such high-performance devices on silicon requires the use of ultra high-purity silicon with boron concentration below $3 \times 10^{11}$~cm$^{-3}$ (resistivity $>50 \times 10^3$~$\Omega \cdot \text{cm}$). Further lifetime improvement will necessitate high-purity substrates free of boron defects and may require advancements in silicon wafer growth. We note that low-power $Q_i$ of resonators on phosphorus doped silicon does not show excess loss from the bulk. The lack of phosphorus induced loss is consistent with the prediction based on its electronic structure which does not contain any orbital degeneracy in its ground state (see Supplemental Material Fig.~S6~\cite{AGGARWAL1964163,Supp}).

\begin{figure*}[t!]
 	\centering
 	\includegraphics[width=2\columnwidth]{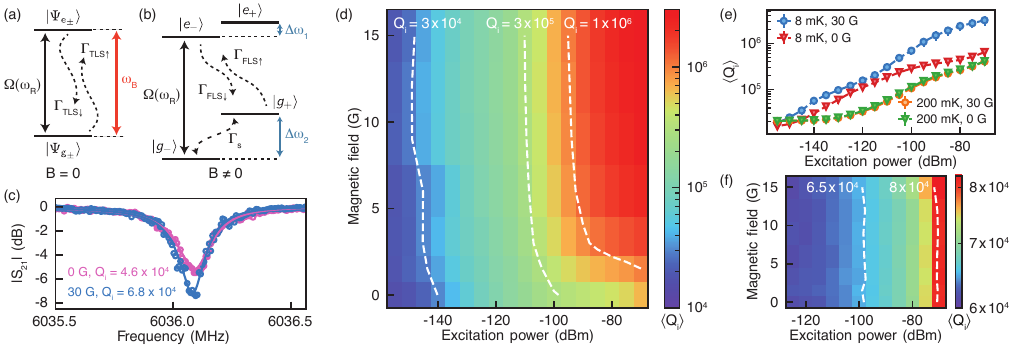}
 	\caption{\textbf{Magnetic response of four-level system dielectric loss in boron doped silicon.} 
  Simplified level structure of boron acceptors under (a) zero and (b) a small magnetic field. (a) $\Gamma_{\text{TLS}\uparrow} \,(\Gamma_{\text{TLS}\downarrow})$: orbital relaxation rate, $\omega_B$: boron resonance frequency. (b) The effective two-level structure of boron is modified to a four-level system in a magnetic field. $\Gamma_{\text{FLS}\uparrow}\, (\Gamma_{\text{FLS}\downarrow} )$: orbital relaxation rate between $\lvert e_{-} \rangle$ and $\lvert g_{+} \rangle$, orbital decay paths between other states are not shown for simplicity. $\Gamma_\text{s}$: relaxation rate between the generalized spin states $\lvert g_+\rangle$ and $\lvert g_-\rangle$. $\Omega(\omega_R)$: Rabi frequency at the resonator frequency ($\omega_R$). 
  (c) Low-power (-135 dBm) $\lvert \text{S}_{\text{21}} \rvert $ measurement of a resonator on boron doped silicon at $\text{T}_{\text{MXC}}=8$~mK.  $Q_i$ increases in a small magnetic field (30~G). 
  (d) Power-dependent $Q_i$ under different magnetic fields for resonators on boron doped silicon. The dashed lines are interpolated $Q_i$ isocontours. $Q_i$ shows an overall increase at higher magnetic fields. (e) Power-dependent $Q_i$ measured at $\text{T}_{\text{MXC}}=8$~mK and $200$~mK under zero magnetic field and 30~G. (f) Power-dependent $Q_i$ under different magnetic fields for resonators on undoped silicon with intentionally introduced amorphous TLS. No magnetic field response is observable. Data in (d,e) represent the average $Q_i$ of eight resonators while data in (f) represent the average $Q_i$ of three resonators.}
 	\label{fig4}
\end{figure*}

Having shown the bulk TLS behavior of boron acceptors, we turn to probing key distinctions between boron and conventional TLS~\cite{Müller_2019}. In the following, we demonstrate that the spin-orbit coupling and four-level nature of boron acceptors leads to a strong magnetic field dependence of boron-induced dielectric loss. Fig.~\ref{fig4}(a)  shows the  level structure of boron acceptors. Under zero magnetic field, the level structure resembles the structure of a conventional TLS, with stray strain and electric field determining the orbital splitting~($\omega_s$). Magnetic fields lift spin degeneracies and result in a four-level structure~(Fig.~\ref{fig4}(b)). The effective g-factors within the two orbital branches are determined by the exact strain environment, which in general leads to unequal Zeeman splittings~($\Delta \omega_1 \neq \Delta \omega_2$). When the orbital splitting is much greater than spin splitting, the lower orbital branch exhibits two long-lived generalized spin states~\cite{kobayashi_2021a}. 

We fabricate resonators with narrow traces (10~$\mu$m) and wide gaps (40~$\mu$m) on boron-doped silicon ($[\text{B}]= 7.4\times 10^{14} \,\text{cm}^{-3}$) to study how the loss is modified under an in-plane magnetic field due to the four-level structure of boron. The narrow-trace geometry is chosen to mitigate vortex formation~\cite{stan_2004}, while the wide-gap configuration helps maintain a low surface-participation ratio for detecting bulk loss. Under a small magnetic field (30~G), we observe an improvement in the low-power $Q_i$ compared to the zero-field case (Fig.~\ref{fig4}(c)). We further investigate how loss evolves under a magnetic field by measuring the power-dependent $Q_i$ as a function of magnetic field (Fig.~\ref{fig4}(d)). We observe three key features: (1)  the saturation power of boron-induced loss is reduced by an order of magnitude under a non-zero magnetic field ~(Fig.~\ref{fig4}(e) and Supplemental Material Fig.~S10~\cite{Supp}), (2) the loss reduction induced by magnetic field saturates at a magnetic field as low as 5~G, (3) high-power $Q_i$ shows a drastic improvement under a magnetic field. 

A simplified description of the change in saturation power in a magnetic field can be explained using the four-level system dynamics of boron acceptors in a  magnetic field. Under zero magnetic field, boron acceptors are equivalent to standard TLSs. Driving the resonator at $\omega_R$ excites the nearby boron defects ($\omega_s \approx \omega_R$) with a rate $\Omega$. Acceptor-induced dielectric loss is saturated when the excitation rate is comparable to the fast orbital relaxation rate $\Omega \sim \Gamma_{\text{TLS}\downarrow}/2$. For magnetic fields where the differential Zeeman splitting $\lvert \Delta \omega_2 - \Delta \omega_1 \rvert$ becomes greater than the resonator and boron linewidths ($\sim5$~G), transition associated with one of the ground states ($\lvert g_-\rangle$) is selectively driven. Therefore, the ground state population is pumped to the other long-lived dark state~($\lvert g_+\rangle$) with a rate of $\sim4|\Omega|^2/\Gamma_{\text{FLS}\downarrow}$. Such microwave pumped boron acceptors are trapped in  $\lvert g_+\rangle$ and decoupled from the resonator. This decrease of the effective number of TLSs results in reduced dielectric loss. In addition, the loss saturation occurs at a lower power where the  pumping rate equals the dark state decay rate ($4|\Omega|^2/\Gamma_{\text{FLS}\downarrow}\sim\Gamma_{\text{FLS}\uparrow}$), assuming  $\Gamma_{\text{s}} \ll \Gamma_{\text{FLS}\downarrow,\uparrow}$. We use the measured ratio of saturation powers with and without magnetic fields to estimate the thermal occupancy ($\bar{n}$) and the effective sample temperature. Our master equation model for four-level system saturation indicates an effective sample temperature of $\sim 75$\,mK based on the measured saturation power ratios (see Supplemental Material Fig.~S10 for a detailed analysis~\cite{Supp}). At elevated temperatures, the magnetic response becomes weaker (see Supplemental Material Fig.~S5~\cite{Supp}) due to exponential activation of thermal occupancy  and $\Gamma_{\text{FLS}\uparrow}/\Gamma_{\text{FLS}\downarrow}=\bar{n}/(1+\bar{n})$ where $\bar{n}<1$. We observe a complete disappearance of dielectric loss reduction in response to magnetic field at $\text{T}_{\text{MXC}}=200$~mK~(Fig.~\ref{fig4}(e)). Finally, we note that the loss reduction is higher at higher excitation powers, which may be accounted by the reduction of loss from a large ensemble of off-resonant boron acceptors. 

We also perform a magnetic field study of conventional TLSs at surfaces. To our knowledge, magnetic response of such amorphous TLSs has not been reported before. We intentionally introduce a high density of amorphous TLSs by drop casting Hydrogen Silsesquioxane (HSQ) on high-Q resonators fabricated on undoped silicon (see Supplemental Material Fig.~S1(d)~\cite{Supp}). In contrast to the acceptor-induced loss, the HSQ-induced loss remains constant in small magnetic fields (Fig.~\ref{fig4}(f)). We emphasize that the lack of magnetic field response in our experimental regime cannot rule out the existence of magnetic dipole moments in amorphous TLSs. Instead, the observation indicates that surface TLSs do not have a differential Zeeman splitting in different orbital states (Fig.~\ref{fig4}(b)).

In conclusion, we report the first microscopic identification of a TLS bath from bulk crystalline defects in silicon. Our study indicates that the acceptor-induced loss can act as a near-future limiting factor for state-of-the-art superconducting quantum devices on silicon. We show that the acceptor-induced dielectric loss can be reduced under a small magnetic field. These results also show the need for ultrahigh purity silicon wafers to enable next-generation superconducting qubits. We note that our observations apply to other acceptors in silicon which display similar electronic structures~\cite{birSpinCombinedResonance1963b,birSpinCombinedResonance1963c}. In addition to improvements to substrate purity, further loss suppression can be achieved using vacuum-gap capacitors~\cite{zemlicka_2023} or engineering the acceptor dynamics with phononic crystals~\cite{chen_2023a,odeh_2023}. Finally, further engineering of interactions with boron acceptors can enable their use as spin-orbit qubits strongly coupled to microwave and mechanical resonators ~\cite{ruskov_2013a,vanderheijden_2014,salfi_2016,salfi_2016b,kobayashi_2021a,zhang_2023a,samkharadze_2016b}.\\

\begin{acknowledgments} 
We thank Xueyue Zhang for fruitful discussions, as well as Mi Lei and Shimon Kolkowitz for feedback on the manuscript. This work was primarily funded by the U.S. Department of Energy, Office of Science, Basic Energy Sciences, Materials Sciences and Engineering Division under Contract No. DE-AC02-05CH11231 within the Nanomachine Program (device design and measurements). Material characterization (SRIM) and cryogenic RF instrumentation is supported by the U.S. Department of Energy, Office of Science, Office of Basic Energy Sciences, Materials Sciences and Engineering Division under Contract No. DE-AC02-05-CH11231 in the Phonon Control for Next-Generation Superconducting Systems and Sensors FWP (KCAS23). Additional support was provided for device fabrication by the ONR and AFOSR Quantum Phononics MURI program for lithography development. The devices used in this work were fabricated at UC Berkeley's NanoLab.
\end{acknowledgments}


\bibliography{main}

\providecommand{\noopsort}[1]{}\providecommand{\singleletter}[1]{#1}%
\begin{thebibliography}{53}%
\makeatletter
\providecommand \@ifxundefined [1]{%
 \@ifx{#1\undefined}
}%
\providecommand \@ifnum [1]{%
 \ifnum #1\expandafter \@firstoftwo
 \else \expandafter \@secondoftwo
 \fi
}%
\providecommand \@ifx [1]{%
 \ifx #1\expandafter \@firstoftwo
 \else \expandafter \@secondoftwo
 \fi
}%
\providecommand \natexlab [1]{#1}%
\providecommand \enquote  [1]{``#1''}%
\providecommand \bibnamefont  [1]{#1}%
\providecommand \bibfnamefont [1]{#1}%
\providecommand \citenamefont [1]{#1}%
\providecommand \href@noop [0]{\@secondoftwo}%
\providecommand \href [0]{\begingroup \@sanitize@url \@href}%
\providecommand \@href[1]{\@@startlink{#1}\@@href}%
\providecommand \@@href[1]{\endgroup#1\@@endlink}%
\providecommand \@sanitize@url [0]{\catcode `\\12\catcode `\$12\catcode `\&12\catcode `\#12\catcode `\^12\catcode `\_12\catcode `\%12\relax}%
\providecommand \@@startlink[1]{}%
\providecommand \@@endlink[0]{}%
\providecommand \url  [0]{\begingroup\@sanitize@url \@url }%
\providecommand \@url [1]{\endgroup\@href {#1}{\urlprefix }}%
\providecommand \urlprefix  [0]{URL }%
\providecommand \Eprint [0]{\href }%
\providecommand \doibase [0]{https://doi.org/}%
\providecommand \selectlanguage [0]{\@gobble}%
\providecommand \bibinfo  [0]{\@secondoftwo}%
\providecommand \bibfield  [0]{\@secondoftwo}%
\providecommand \translation [1]{[#1]}%
\providecommand \BibitemOpen [0]{}%
\providecommand \bibitemStop [0]{}%
\providecommand \bibitemNoStop [0]{.\EOS\space}%
\providecommand \EOS [0]{\spacefactor3000\relax}%
\providecommand \BibitemShut  [1]{\csname bibitem#1\endcsname}%
\let\auto@bib@innerbib\@empty
\bibitem [{\citenamefont {Arute}\ \emph {et~al.}(2019)\citenamefont {Arute}, \citenamefont {Arya}, \citenamefont {Babbush}, \citenamefont {Bacon}, \citenamefont {Bardin}, \citenamefont {Barends}, \citenamefont {Biswas}, \citenamefont {Boixo}, \citenamefont {Brandao}, \citenamefont {Buell} \emph {et~al.}}]{arute_quantum_2019}%
  \BibitemOpen
  \bibfield  {author} {\bibinfo {author} {\bibfnamefont {F.}~\bibnamefont {Arute}}, \bibinfo {author} {\bibfnamefont {K.}~\bibnamefont {Arya}}, \bibinfo {author} {\bibfnamefont {R.}~\bibnamefont {Babbush}}, \bibinfo {author} {\bibfnamefont {D.}~\bibnamefont {Bacon}}, \bibinfo {author} {\bibfnamefont {J.~C.}\ \bibnamefont {Bardin}}, \bibinfo {author} {\bibfnamefont {R.}~\bibnamefont {Barends}}, \bibinfo {author} {\bibfnamefont {R.}~\bibnamefont {Biswas}}, \bibinfo {author} {\bibfnamefont {S.}~\bibnamefont {Boixo}}, \bibinfo {author} {\bibfnamefont {F.~G. S.~L.}\ \bibnamefont {Brandao}}, \bibinfo {author} {\bibfnamefont {D.~A.}\ \bibnamefont {Buell}}, \emph {et~al.},\ }\bibfield  {title} {\bibinfo {title} {Quantum supremacy using a programmable superconducting processor},\ }\href {https://doi.org/10.1038/s41586-019-1666-5} {\bibfield  {journal} {\bibinfo  {journal} {Nature}\ }\textbf {\bibinfo {volume} {574}},\ \bibinfo {pages} {505} (\bibinfo {year} {2019})}\BibitemShut {NoStop}%
\bibitem [{\citenamefont {Kim}\ \emph {et~al.}(2023)\citenamefont {Kim}, \citenamefont {Eddins}, \citenamefont {Anand}, \citenamefont {Wei}, \citenamefont {Van Den~Berg}, \citenamefont {Rosenblatt}, \citenamefont {Nayfeh}, \citenamefont {Wu}, \citenamefont {Zaletel}, \citenamefont {Temme},\ and\ \citenamefont {Kandala}}]{kim_2023}%
  \BibitemOpen
  \bibfield  {author} {\bibinfo {author} {\bibfnamefont {Y.}~\bibnamefont {Kim}}, \bibinfo {author} {\bibfnamefont {A.}~\bibnamefont {Eddins}}, \bibinfo {author} {\bibfnamefont {S.}~\bibnamefont {Anand}}, \bibinfo {author} {\bibfnamefont {K.~X.}\ \bibnamefont {Wei}}, \bibinfo {author} {\bibfnamefont {E.}~\bibnamefont {Van Den~Berg}}, \bibinfo {author} {\bibfnamefont {S.}~\bibnamefont {Rosenblatt}}, \bibinfo {author} {\bibfnamefont {H.}~\bibnamefont {Nayfeh}}, \bibinfo {author} {\bibfnamefont {Y.}~\bibnamefont {Wu}}, \bibinfo {author} {\bibfnamefont {M.}~\bibnamefont {Zaletel}}, \bibinfo {author} {\bibfnamefont {K.}~\bibnamefont {Temme}},\ and\ \bibinfo {author} {\bibfnamefont {A.}~\bibnamefont {Kandala}},\ }\bibfield  {title} {\bibinfo {title} {Evidence for the utility of quantum computing before fault tolerance},\ }\href {https://doi.org/10.1038/s41586-023-06096-3} {\bibfield  {journal} {\bibinfo  {journal} {Nature}\ }\textbf {\bibinfo {volume} {618}},\ \bibinfo {pages} {500} (\bibinfo {year}
  {2023})}\BibitemShut {NoStop}%
\bibitem [{\citenamefont {Arute}\ \emph {et~al.}(2020)\citenamefont {Arute}, \citenamefont {Arya}, \citenamefont {Babbush}, \citenamefont {Bacon}, \citenamefont {Bardin}, \citenamefont {Barends}, \citenamefont {Boixo}, \citenamefont {Broughton}, \citenamefont {Buckley}, \citenamefont {Buell} \emph {et~al.}}]{HartreeFock_2020}%
  \BibitemOpen
  \bibfield  {author} {\bibinfo {author} {\bibfnamefont {F.}~\bibnamefont {Arute}}, \bibinfo {author} {\bibfnamefont {K.}~\bibnamefont {Arya}}, \bibinfo {author} {\bibfnamefont {R.}~\bibnamefont {Babbush}}, \bibinfo {author} {\bibfnamefont {D.}~\bibnamefont {Bacon}}, \bibinfo {author} {\bibfnamefont {J.~C.}\ \bibnamefont {Bardin}}, \bibinfo {author} {\bibfnamefont {R.}~\bibnamefont {Barends}}, \bibinfo {author} {\bibfnamefont {S.}~\bibnamefont {Boixo}}, \bibinfo {author} {\bibfnamefont {M.}~\bibnamefont {Broughton}}, \bibinfo {author} {\bibfnamefont {B.~B.}\ \bibnamefont {Buckley}}, \bibinfo {author} {\bibfnamefont {D.~A.}\ \bibnamefont {Buell}}, \emph {et~al.},\ }\bibfield  {title} {\bibinfo {title} {Hartree-fock on a superconducting qubit quantum computer},\ }\href {https://doi.org/10.1126/science.abb9811} {\bibfield  {journal} {\bibinfo  {journal} {Science}\ }\textbf {\bibinfo {volume} {369}},\ \bibinfo {pages} {1084} (\bibinfo {year} {2020})}\BibitemShut {NoStop}%
\bibitem [{\citenamefont {Acharya}\ \emph {et~al.}(2023)\citenamefont {Acharya}, \citenamefont {Aleiner}, \citenamefont {Allen}, \citenamefont {Andersen}, \citenamefont {Ansmann}, \citenamefont {Arute}, \citenamefont {Arya}, \citenamefont {Asfaw}, \citenamefont {Atalaya}, \citenamefont {Babbush} \emph {et~al.}}]{googlequantumai_2023}%
  \BibitemOpen
  \bibfield  {author} {\bibinfo {author} {\bibfnamefont {R.}~\bibnamefont {Acharya}}, \bibinfo {author} {\bibfnamefont {I.}~\bibnamefont {Aleiner}}, \bibinfo {author} {\bibfnamefont {R.}~\bibnamefont {Allen}}, \bibinfo {author} {\bibfnamefont {T.~I.}\ \bibnamefont {Andersen}}, \bibinfo {author} {\bibfnamefont {M.}~\bibnamefont {Ansmann}}, \bibinfo {author} {\bibfnamefont {F.}~\bibnamefont {Arute}}, \bibinfo {author} {\bibfnamefont {K.}~\bibnamefont {Arya}}, \bibinfo {author} {\bibfnamefont {A.}~\bibnamefont {Asfaw}}, \bibinfo {author} {\bibfnamefont {J.}~\bibnamefont {Atalaya}}, \bibinfo {author} {\bibfnamefont {R.}~\bibnamefont {Babbush}}, \emph {et~al.},\ }\bibfield  {title} {\bibinfo {title} {Suppressing quantum errors by scaling a surface code logical qubit},\ }\href {https://doi.org/10.1038/s41586-022-05434-1} {\bibfield  {journal} {\bibinfo  {journal} {Nature}\ }\textbf {\bibinfo {volume} {614}},\ \bibinfo {pages} {676} (\bibinfo {year} {2023})}\BibitemShut {NoStop}%
\bibitem [{\citenamefont {McRae}\ \emph {et~al.}(2020)\citenamefont {McRae}, \citenamefont {Wang}, \citenamefont {Gao}, \citenamefont {Vissers}, \citenamefont {Brecht}, \citenamefont {Dunsworth}, \citenamefont {Pappas},\ and\ \citenamefont {Mutus}}]{mcrae_2020a}%
  \BibitemOpen
  \bibfield  {author} {\bibinfo {author} {\bibfnamefont {C.~R.~H.}\ \bibnamefont {McRae}}, \bibinfo {author} {\bibfnamefont {H.}~\bibnamefont {Wang}}, \bibinfo {author} {\bibfnamefont {J.}~\bibnamefont {Gao}}, \bibinfo {author} {\bibfnamefont {M.~R.}\ \bibnamefont {Vissers}}, \bibinfo {author} {\bibfnamefont {T.}~\bibnamefont {Brecht}}, \bibinfo {author} {\bibfnamefont {A.}~\bibnamefont {Dunsworth}}, \bibinfo {author} {\bibfnamefont {D.~P.}\ \bibnamefont {Pappas}},\ and\ \bibinfo {author} {\bibfnamefont {J.}~\bibnamefont {Mutus}},\ }\bibfield  {title} {\bibinfo {title} {Materials loss measurements using superconducting microwave resonators},\ }\href {https://doi.org/10.1063/5.0017378} {\bibfield  {journal} {\bibinfo  {journal} {Review of Scientific Instruments}\ }\textbf {\bibinfo {volume} {91}},\ \bibinfo {pages} {091101} (\bibinfo {year} {2020})}\BibitemShut {NoStop}%
\bibitem [{\citenamefont {Siddiqi}(2021)}]{siddiqi_2021a}%
  \BibitemOpen
  \bibfield  {author} {\bibinfo {author} {\bibfnamefont {I.}~\bibnamefont {Siddiqi}},\ }\bibfield  {title} {\bibinfo {title} {Engineering high-coherence superconducting qubits},\ }\href {https://doi.org/10.1038/s41578-021-00370-4} {\bibfield  {journal} {\bibinfo  {journal} {Nature Reviews Materials}\ }\textbf {\bibinfo {volume} {6}},\ \bibinfo {pages} {875} (\bibinfo {year} {2021})}\BibitemShut {NoStop}%
\bibitem [{\citenamefont {Phillips}(1972)}]{phillips_1972}%
  \BibitemOpen
  \bibfield  {author} {\bibinfo {author} {\bibfnamefont {W.~A.}\ \bibnamefont {Phillips}},\ }\bibfield  {title} {\bibinfo {title} {Tunneling states in amorphous solids},\ }\href {https://doi.org/10.1007/BF00660072} {\bibfield  {journal} {\bibinfo  {journal} {Journal of Low Temperature Physics}\ }\textbf {\bibinfo {volume} {7}},\ \bibinfo {pages} {351} (\bibinfo {year} {1972})}\BibitemShut {NoStop}%
\bibitem [{\citenamefont {Anderson}\ \emph {et~al.}(1972)\citenamefont {Anderson}, \citenamefont {Halperin},\ and\ \citenamefont {Varma}}]{Anderson_1972}%
  \BibitemOpen
  \bibfield  {author} {\bibinfo {author} {\bibfnamefont {P.~W.}\ \bibnamefont {Anderson}}, \bibinfo {author} {\bibfnamefont {B.~I.}\ \bibnamefont {Halperin}},\ and\ \bibinfo {author} {\bibfnamefont {C.~M.}\ \bibnamefont {Varma}},\ }\bibfield  {title} {\bibinfo {title} {Anomalous low-temperature thermal properties of glasses and spin glasses},\ }\href {https://doi.org/10.1080/14786437208229210} {\bibfield  {journal} {\bibinfo  {journal} {The Philosophical Magazine: A Journal of Theoretical Experimental and Applied Physics}\ }\textbf {\bibinfo {volume} {25}},\ \bibinfo {pages} {1} (\bibinfo {year} {1972})}\BibitemShut {NoStop}%
\bibitem [{\citenamefont {Gao}\ \emph {et~al.}(2008)\citenamefont {Gao}, \citenamefont {Daal}, \citenamefont {Vayonakis}, \citenamefont {Kumar}, \citenamefont {Zmuidzinas}, \citenamefont {Sadoulet}, \citenamefont {Mazin}, \citenamefont {Day},\ and\ \citenamefont {Leduc}}]{gao_experimental_2008}%
  \BibitemOpen
  \bibfield  {author} {\bibinfo {author} {\bibfnamefont {J.}~\bibnamefont {Gao}}, \bibinfo {author} {\bibfnamefont {M.}~\bibnamefont {Daal}}, \bibinfo {author} {\bibfnamefont {A.}~\bibnamefont {Vayonakis}}, \bibinfo {author} {\bibfnamefont {S.}~\bibnamefont {Kumar}}, \bibinfo {author} {\bibfnamefont {J.}~\bibnamefont {Zmuidzinas}}, \bibinfo {author} {\bibfnamefont {B.}~\bibnamefont {Sadoulet}}, \bibinfo {author} {\bibfnamefont {B.~A.}\ \bibnamefont {Mazin}}, \bibinfo {author} {\bibfnamefont {P.~K.}\ \bibnamefont {Day}},\ and\ \bibinfo {author} {\bibfnamefont {H.~G.}\ \bibnamefont {Leduc}},\ }\bibfield  {title} {\bibinfo {title} {Experimental evidence for a surface distribution of two-level systems in superconducting lithographed microwave resonators},\ }\href {https://doi.org/10.1063/1.2906373} {\bibfield  {journal} {\bibinfo  {journal} {Applied Physics Letters}\ }\textbf {\bibinfo {volume} {92}},\ \bibinfo {pages} {152505} (\bibinfo {year} {2008})}\BibitemShut {NoStop}%
\bibitem [{\citenamefont {Wang}\ \emph {et~al.}(2015)\citenamefont {Wang}, \citenamefont {Axline}, \citenamefont {Gao}, \citenamefont {Brecht}, \citenamefont {Chu}, \citenamefont {Frunzio}, \citenamefont {Devoret},\ and\ \citenamefont {Schoelkopf}}]{wang_surface_2015}%
  \BibitemOpen
  \bibfield  {author} {\bibinfo {author} {\bibfnamefont {C.}~\bibnamefont {Wang}}, \bibinfo {author} {\bibfnamefont {C.}~\bibnamefont {Axline}}, \bibinfo {author} {\bibfnamefont {Y.~Y.}\ \bibnamefont {Gao}}, \bibinfo {author} {\bibfnamefont {T.}~\bibnamefont {Brecht}}, \bibinfo {author} {\bibfnamefont {Y.}~\bibnamefont {Chu}}, \bibinfo {author} {\bibfnamefont {L.}~\bibnamefont {Frunzio}}, \bibinfo {author} {\bibfnamefont {M.~H.}\ \bibnamefont {Devoret}},\ and\ \bibinfo {author} {\bibfnamefont {R.~J.}\ \bibnamefont {Schoelkopf}},\ }\bibfield  {title} {\bibinfo {title} {Surface participation and dielectric loss in superconducting qubits},\ }\href {https://doi.org/10.1063/1.4934486} {\bibfield  {journal} {\bibinfo  {journal} {Applied Physics Letters}\ }\textbf {\bibinfo {volume} {107}},\ \bibinfo {pages} {162601} (\bibinfo {year} {2015})}\BibitemShut {NoStop}%
\bibitem [{\citenamefont {Müller}\ \emph {et~al.}(2019)\citenamefont {Müller}, \citenamefont {Cole},\ and\ \citenamefont {Lisenfeld}}]{Müller_2019}%
  \BibitemOpen
  \bibfield  {author} {\bibinfo {author} {\bibfnamefont {C.}~\bibnamefont {Müller}}, \bibinfo {author} {\bibfnamefont {J.~H.}\ \bibnamefont {Cole}},\ and\ \bibinfo {author} {\bibfnamefont {J.}~\bibnamefont {Lisenfeld}},\ }\bibfield  {title} {\bibinfo {title} {Towards understanding two-level-systems in amorphous solids: insights from quantum circuits},\ }\href {https://doi.org/10.1088/1361-6633/ab3a7e} {\bibfield  {journal} {\bibinfo  {journal} {Reports on Progress in Physics}\ }\textbf {\bibinfo {volume} {82}},\ \bibinfo {pages} {124501} (\bibinfo {year} {2019})}\BibitemShut {NoStop}%
\bibitem [{\citenamefont {Martinis}\ \emph {et~al.}(2005)\citenamefont {Martinis}, \citenamefont {Cooper}, \citenamefont {McDermott}, \citenamefont {Steffen}, \citenamefont {Ansmann}, \citenamefont {Osborn}, \citenamefont {Cicak}, \citenamefont {Oh}, \citenamefont {Pappas}, \citenamefont {Simmonds},\ and\ \citenamefont {Yu}}]{martinis_2005a}%
  \BibitemOpen
  \bibfield  {author} {\bibinfo {author} {\bibfnamefont {J.~M.}\ \bibnamefont {Martinis}}, \bibinfo {author} {\bibfnamefont {K.~B.}\ \bibnamefont {Cooper}}, \bibinfo {author} {\bibfnamefont {R.}~\bibnamefont {McDermott}}, \bibinfo {author} {\bibfnamefont {M.}~\bibnamefont {Steffen}}, \bibinfo {author} {\bibfnamefont {M.}~\bibnamefont {Ansmann}}, \bibinfo {author} {\bibfnamefont {K.~D.}\ \bibnamefont {Osborn}}, \bibinfo {author} {\bibfnamefont {K.}~\bibnamefont {Cicak}}, \bibinfo {author} {\bibfnamefont {S.}~\bibnamefont {Oh}}, \bibinfo {author} {\bibfnamefont {D.~P.}\ \bibnamefont {Pappas}}, \bibinfo {author} {\bibfnamefont {R.~W.}\ \bibnamefont {Simmonds}},\ and\ \bibinfo {author} {\bibfnamefont {C.~C.}\ \bibnamefont {Yu}},\ }\bibfield  {title} {\bibinfo {title} {Decoherence in {{Josephson Qubits}} from {{Dielectric Loss}}},\ }\href {https://doi.org/10.1103/PhysRevLett.95.210503} {\bibfield  {journal} {\bibinfo  {journal} {Physical Review Letters}\ }\textbf {\bibinfo {volume} {95}},\ \bibinfo {pages}
  {210503} (\bibinfo {year} {2005})}\BibitemShut {NoStop}%
\bibitem [{\citenamefont {Place}\ \emph {et~al.}(2021)\citenamefont {Place}, \citenamefont {Rodgers}, \citenamefont {Mundada}, \citenamefont {Smitham}, \citenamefont {Fitzpatrick}, \citenamefont {Leng}, \citenamefont {Premkumar}, \citenamefont {Bryon}, \citenamefont {Vrajitoarea}, \citenamefont {Sussman}, \citenamefont {Cheng}, \citenamefont {Madhavan}, \citenamefont {Babla}, \citenamefont {Le}, \citenamefont {Gang}, \citenamefont {J{\"a}ck}, \citenamefont {Gyenis}, \citenamefont {Yao}, \citenamefont {Cava}, \citenamefont {De~Leon},\ and\ \citenamefont {Houck}}]{place_2021}%
  \BibitemOpen
  \bibfield  {author} {\bibinfo {author} {\bibfnamefont {A.~P.~M.}\ \bibnamefont {Place}}, \bibinfo {author} {\bibfnamefont {L.~V.~H.}\ \bibnamefont {Rodgers}}, \bibinfo {author} {\bibfnamefont {P.}~\bibnamefont {Mundada}}, \bibinfo {author} {\bibfnamefont {B.~M.}\ \bibnamefont {Smitham}}, \bibinfo {author} {\bibfnamefont {M.}~\bibnamefont {Fitzpatrick}}, \bibinfo {author} {\bibfnamefont {Z.}~\bibnamefont {Leng}}, \bibinfo {author} {\bibfnamefont {A.}~\bibnamefont {Premkumar}}, \bibinfo {author} {\bibfnamefont {J.}~\bibnamefont {Bryon}}, \bibinfo {author} {\bibfnamefont {A.}~\bibnamefont {Vrajitoarea}}, \bibinfo {author} {\bibfnamefont {S.}~\bibnamefont {Sussman}}, \bibinfo {author} {\bibfnamefont {G.}~\bibnamefont {Cheng}}, \bibinfo {author} {\bibfnamefont {T.}~\bibnamefont {Madhavan}}, \bibinfo {author} {\bibfnamefont {H.~K.}\ \bibnamefont {Babla}}, \bibinfo {author} {\bibfnamefont {X.~H.}\ \bibnamefont {Le}}, \bibinfo {author} {\bibfnamefont {Y.}~\bibnamefont {Gang}}, \bibinfo {author} {\bibfnamefont
  {B.}~\bibnamefont {J{\"a}ck}}, \bibinfo {author} {\bibfnamefont {A.}~\bibnamefont {Gyenis}}, \bibinfo {author} {\bibfnamefont {N.}~\bibnamefont {Yao}}, \bibinfo {author} {\bibfnamefont {R.~J.}\ \bibnamefont {Cava}}, \bibinfo {author} {\bibfnamefont {N.~P.}\ \bibnamefont {De~Leon}},\ and\ \bibinfo {author} {\bibfnamefont {A.~A.}\ \bibnamefont {Houck}},\ }\bibfield  {title} {\bibinfo {title} {New material platform for superconducting transmon qubits with coherence times exceeding 0.3 milliseconds},\ }\href {https://doi.org/10.1038/s41467-021-22030-5} {\bibfield  {journal} {\bibinfo  {journal} {Nature Communications}\ }\textbf {\bibinfo {volume} {12}},\ \bibinfo {pages} {1779} (\bibinfo {year} {2021})}\BibitemShut {NoStop}%
\bibitem [{\citenamefont {Martinis}(2022)}]{martinis_2022}%
  \BibitemOpen
  \bibfield  {author} {\bibinfo {author} {\bibfnamefont {J.~M.}\ \bibnamefont {Martinis}},\ }\bibfield  {title} {\bibinfo {title} {Surface loss calculations and design of a superconducting transmon qubit with tapered wiring},\ }\href {https://doi.org/10.1038/s41534-022-00530-6} {\bibfield  {journal} {\bibinfo  {journal} {npj Quantum Information}\ }\textbf {\bibinfo {volume} {8}},\ \bibinfo {pages} {26} (\bibinfo {year} {2022})}\BibitemShut {NoStop}%
\bibitem [{\citenamefont {Ganjam}\ \emph {et~al.}(2023)\citenamefont {Ganjam}, \citenamefont {Wang}, \citenamefont {Lu}, \citenamefont {Banerjee}, \citenamefont {Lei}, \citenamefont {Krayzman}, \citenamefont {Kisslinger}, \citenamefont {Zhou}, \citenamefont {Li}, \citenamefont {Jia}, \citenamefont {Liu}, \citenamefont {Frunzio},\ and\ \citenamefont {Schoelkopf}}]{ganjam_surpassing_2023}%
  \BibitemOpen
  \bibfield  {author} {\bibinfo {author} {\bibfnamefont {S.}~\bibnamefont {Ganjam}}, \bibinfo {author} {\bibfnamefont {Y.}~\bibnamefont {Wang}}, \bibinfo {author} {\bibfnamefont {Y.}~\bibnamefont {Lu}}, \bibinfo {author} {\bibfnamefont {A.}~\bibnamefont {Banerjee}}, \bibinfo {author} {\bibfnamefont {C.~U.}\ \bibnamefont {Lei}}, \bibinfo {author} {\bibfnamefont {L.}~\bibnamefont {Krayzman}}, \bibinfo {author} {\bibfnamefont {K.}~\bibnamefont {Kisslinger}}, \bibinfo {author} {\bibfnamefont {C.}~\bibnamefont {Zhou}}, \bibinfo {author} {\bibfnamefont {R.}~\bibnamefont {Li}}, \bibinfo {author} {\bibfnamefont {Y.}~\bibnamefont {Jia}}, \bibinfo {author} {\bibfnamefont {M.}~\bibnamefont {Liu}}, \bibinfo {author} {\bibfnamefont {L.}~\bibnamefont {Frunzio}},\ and\ \bibinfo {author} {\bibfnamefont {R.~J.}\ \bibnamefont {Schoelkopf}},\ }\href {http://arxiv.org/abs/2308.15539} {\bibinfo {title} {Surpassing millisecond coherence times in on-chip superconducting quantum memories by optimizing materials, processes, and
  circuit design}} (\bibinfo {year} {2023}),\ \bibinfo {note} {arXiv:2308.15539}\BibitemShut {NoStop}%
\bibitem [{\citenamefont {Read}\ \emph {et~al.}(2023)\citenamefont {Read}, \citenamefont {Chapman}, \citenamefont {Lei}, \citenamefont {Curtis}, \citenamefont {Ganjam}, \citenamefont {Krayzman}, \citenamefont {Frunzio},\ and\ \citenamefont {Schoelkopf}}]{read_2023b}%
  \BibitemOpen
  \bibfield  {author} {\bibinfo {author} {\bibfnamefont {A.~P.}\ \bibnamefont {Read}}, \bibinfo {author} {\bibfnamefont {B.~J.}\ \bibnamefont {Chapman}}, \bibinfo {author} {\bibfnamefont {C.~U.}\ \bibnamefont {Lei}}, \bibinfo {author} {\bibfnamefont {J.~C.}\ \bibnamefont {Curtis}}, \bibinfo {author} {\bibfnamefont {S.}~\bibnamefont {Ganjam}}, \bibinfo {author} {\bibfnamefont {L.}~\bibnamefont {Krayzman}}, \bibinfo {author} {\bibfnamefont {L.}~\bibnamefont {Frunzio}},\ and\ \bibinfo {author} {\bibfnamefont {R.~J.}\ \bibnamefont {Schoelkopf}},\ }\bibfield  {title} {\bibinfo {title} {Precision {{Measurement}} of the {{Microwave Dielectric Loss}} of {{Sapphire}} in the {{Quantum Regime}} with {{Parts-per-Billion Sensitivity}}},\ }\href {https://doi.org/10.1103/PhysRevApplied.19.034064} {\bibfield  {journal} {\bibinfo  {journal} {Physical Review Applied}\ }\textbf {\bibinfo {volume} {19}},\ \bibinfo {pages} {034064} (\bibinfo {year} {2023})}\BibitemShut {NoStop}%
\bibitem [{\citenamefont {Crowley}\ \emph {et~al.}(2023)\citenamefont {Crowley}, \citenamefont {McLellan}, \citenamefont {Dutta}, \citenamefont {Shumiya}, \citenamefont {Place}, \citenamefont {Le}, \citenamefont {Gang}, \citenamefont {Madhavan}, \citenamefont {Bland}, \citenamefont {Chang}, \citenamefont {Khedkar}, \citenamefont {Feng}, \citenamefont {Umbarkar}, \citenamefont {Gui}, \citenamefont {Rodgers}, \citenamefont {Jia}, \citenamefont {Feldman}, \citenamefont {Lyon}, \citenamefont {Liu}, \citenamefont {Cava}, \citenamefont {Houck},\ and\ \citenamefont {De~Leon}}]{crowley_2023a}%
  \BibitemOpen
  \bibfield  {author} {\bibinfo {author} {\bibfnamefont {K.~D.}\ \bibnamefont {Crowley}}, \bibinfo {author} {\bibfnamefont {R.~A.}\ \bibnamefont {McLellan}}, \bibinfo {author} {\bibfnamefont {A.}~\bibnamefont {Dutta}}, \bibinfo {author} {\bibfnamefont {N.}~\bibnamefont {Shumiya}}, \bibinfo {author} {\bibfnamefont {A.~P.~M.}\ \bibnamefont {Place}}, \bibinfo {author} {\bibfnamefont {X.~H.}\ \bibnamefont {Le}}, \bibinfo {author} {\bibfnamefont {Y.}~\bibnamefont {Gang}}, \bibinfo {author} {\bibfnamefont {T.}~\bibnamefont {Madhavan}}, \bibinfo {author} {\bibfnamefont {M.~P.}\ \bibnamefont {Bland}}, \bibinfo {author} {\bibfnamefont {R.}~\bibnamefont {Chang}}, \bibinfo {author} {\bibfnamefont {N.}~\bibnamefont {Khedkar}}, \bibinfo {author} {\bibfnamefont {Y.~C.}\ \bibnamefont {Feng}}, \bibinfo {author} {\bibfnamefont {E.~A.}\ \bibnamefont {Umbarkar}}, \bibinfo {author} {\bibfnamefont {X.}~\bibnamefont {Gui}}, \bibinfo {author} {\bibfnamefont {L.~V.~H.}\ \bibnamefont {Rodgers}}, \bibinfo {author} {\bibfnamefont
  {Y.}~\bibnamefont {Jia}}, \bibinfo {author} {\bibfnamefont {M.~M.}\ \bibnamefont {Feldman}}, \bibinfo {author} {\bibfnamefont {S.~A.}\ \bibnamefont {Lyon}}, \bibinfo {author} {\bibfnamefont {M.}~\bibnamefont {Liu}}, \bibinfo {author} {\bibfnamefont {R.~J.}\ \bibnamefont {Cava}}, \bibinfo {author} {\bibfnamefont {A.~A.}\ \bibnamefont {Houck}},\ and\ \bibinfo {author} {\bibfnamefont {N.~P.}\ \bibnamefont {De~Leon}},\ }\bibfield  {title} {\bibinfo {title} {Disentangling {{Losses}} in {{Tantalum Superconducting Circuits}}},\ }\href {https://doi.org/10.1103/PhysRevX.13.041005} {\bibfield  {journal} {\bibinfo  {journal} {Physical Review X}\ }\textbf {\bibinfo {volume} {13}},\ \bibinfo {pages} {041005} (\bibinfo {year} {2023})}\BibitemShut {NoStop}%
\bibitem [{\citenamefont {Bergeron}\ \emph {et~al.}(2020)\citenamefont {Bergeron}, \citenamefont {Chartrand}, \citenamefont {Kurkjian}, \citenamefont {Morse}, \citenamefont {Riemann}, \citenamefont {Abrosimov}, \citenamefont {Becker}, \citenamefont {Pohl}, \citenamefont {Thewalt},\ and\ \citenamefont {Simmons}}]{bergeron_2020}%
  \BibitemOpen
  \bibfield  {author} {\bibinfo {author} {\bibfnamefont {L.}~\bibnamefont {Bergeron}}, \bibinfo {author} {\bibfnamefont {C.}~\bibnamefont {Chartrand}}, \bibinfo {author} {\bibfnamefont {A.~T.~K.}\ \bibnamefont {Kurkjian}}, \bibinfo {author} {\bibfnamefont {K.~J.}\ \bibnamefont {Morse}}, \bibinfo {author} {\bibfnamefont {H.}~\bibnamefont {Riemann}}, \bibinfo {author} {\bibfnamefont {N.~V.}\ \bibnamefont {Abrosimov}}, \bibinfo {author} {\bibfnamefont {P.}~\bibnamefont {Becker}}, \bibinfo {author} {\bibfnamefont {H.-J.}\ \bibnamefont {Pohl}}, \bibinfo {author} {\bibfnamefont {M.~L.~W.}\ \bibnamefont {Thewalt}},\ and\ \bibinfo {author} {\bibfnamefont {S.}~\bibnamefont {Simmons}},\ }\bibfield  {title} {\bibinfo {title} {Silicon-{{Integrated Telecommunications Photon-Spin Interface}}},\ }\href {https://doi.org/10.1103/PRXQuantum.1.020301} {\bibfield  {journal} {\bibinfo  {journal} {PRX Quantum}\ }\textbf {\bibinfo {volume} {1}},\ \bibinfo {pages} {020301} (\bibinfo {year} {2020})}\BibitemShut {NoStop}%
\bibitem [{\citenamefont {Redjem}\ \emph {et~al.}(2020)\citenamefont {Redjem}, \citenamefont {Durand}, \citenamefont {Herzig}, \citenamefont {Benali}, \citenamefont {Pezzagna}, \citenamefont {Meijer}, \citenamefont {Kuznetsov}, \citenamefont {Nguyen}, \citenamefont {Cueff}, \citenamefont {G{\'e}rard}, \citenamefont {{Robert-Philip}}, \citenamefont {Gil}, \citenamefont {Caliste}, \citenamefont {Pochet}, \citenamefont {Abbarchi}, \citenamefont {Jacques}, \citenamefont {Dr{\'e}au},\ and\ \citenamefont {Cassabois}}]{redjem_2020}%
  \BibitemOpen
  \bibfield  {author} {\bibinfo {author} {\bibfnamefont {W.}~\bibnamefont {Redjem}}, \bibinfo {author} {\bibfnamefont {A.}~\bibnamefont {Durand}}, \bibinfo {author} {\bibfnamefont {T.}~\bibnamefont {Herzig}}, \bibinfo {author} {\bibfnamefont {A.}~\bibnamefont {Benali}}, \bibinfo {author} {\bibfnamefont {S.}~\bibnamefont {Pezzagna}}, \bibinfo {author} {\bibfnamefont {J.}~\bibnamefont {Meijer}}, \bibinfo {author} {\bibfnamefont {A.~{\relax Yu}.}\ \bibnamefont {Kuznetsov}}, \bibinfo {author} {\bibfnamefont {H.~S.}\ \bibnamefont {Nguyen}}, \bibinfo {author} {\bibfnamefont {S.}~\bibnamefont {Cueff}}, \bibinfo {author} {\bibfnamefont {J.-M.}\ \bibnamefont {G{\'e}rard}}, \bibinfo {author} {\bibfnamefont {I.}~\bibnamefont {{Robert-Philip}}}, \bibinfo {author} {\bibfnamefont {B.}~\bibnamefont {Gil}}, \bibinfo {author} {\bibfnamefont {D.}~\bibnamefont {Caliste}}, \bibinfo {author} {\bibfnamefont {P.}~\bibnamefont {Pochet}}, \bibinfo {author} {\bibfnamefont {M.}~\bibnamefont {Abbarchi}}, \bibinfo {author} {\bibfnamefont
  {V.}~\bibnamefont {Jacques}}, \bibinfo {author} {\bibfnamefont {A.}~\bibnamefont {Dr{\'e}au}},\ and\ \bibinfo {author} {\bibfnamefont {G.}~\bibnamefont {Cassabois}},\ }\bibfield  {title} {\bibinfo {title} {Single artificial atoms in silicon emitting at telecom wavelengths},\ }\href {https://doi.org/10.1038/s41928-020-00499-0} {\bibfield  {journal} {\bibinfo  {journal} {Nature Electronics}\ }\textbf {\bibinfo {volume} {3}},\ \bibinfo {pages} {738} (\bibinfo {year} {2020})}\BibitemShut {NoStop}%
\bibitem [{\citenamefont {Komza}\ \emph {et~al.}(2022)\citenamefont {Komza}, \citenamefont {Samutpraphoot}, \citenamefont {Odeh}, \citenamefont {Tang}, \citenamefont {Mathew}, \citenamefont {Chang}, \citenamefont {Song}, \citenamefont {Kim}, \citenamefont {Xiong}, \citenamefont {Hautier},\ and\ \citenamefont {Sipahigil}}]{komza_2022}%
  \BibitemOpen
  \bibfield  {author} {\bibinfo {author} {\bibfnamefont {L.}~\bibnamefont {Komza}}, \bibinfo {author} {\bibfnamefont {P.}~\bibnamefont {Samutpraphoot}}, \bibinfo {author} {\bibfnamefont {M.}~\bibnamefont {Odeh}}, \bibinfo {author} {\bibfnamefont {Y.-L.}\ \bibnamefont {Tang}}, \bibinfo {author} {\bibfnamefont {M.}~\bibnamefont {Mathew}}, \bibinfo {author} {\bibfnamefont {J.}~\bibnamefont {Chang}}, \bibinfo {author} {\bibfnamefont {H.}~\bibnamefont {Song}}, \bibinfo {author} {\bibfnamefont {M.-K.}\ \bibnamefont {Kim}}, \bibinfo {author} {\bibfnamefont {Y.}~\bibnamefont {Xiong}}, \bibinfo {author} {\bibfnamefont {G.}~\bibnamefont {Hautier}},\ and\ \bibinfo {author} {\bibfnamefont {A.}~\bibnamefont {Sipahigil}},\ }\href@noop {} {\bibinfo {title} {Indistinguishable photons from an artificial atom in silicon photonics}} (\bibinfo {year} {2022}),\ \Eprint {https://arxiv.org/abs/2211.09305} {arxiv:2211.09305} \BibitemShut {NoStop}%
\bibitem [{\citenamefont {Higginbottom}\ \emph {et~al.}(2022)\citenamefont {Higginbottom}, \citenamefont {Kurkjian}, \citenamefont {Chartrand}, \citenamefont {Kazemi}, \citenamefont {Brunelle}, \citenamefont {MacQuarrie}, \citenamefont {Klein}, \citenamefont {{Lee-Hone}}, \citenamefont {Stacho}, \citenamefont {Ruether}, \citenamefont {Bowness}, \citenamefont {Bergeron}, \citenamefont {DeAbreu}, \citenamefont {Harrigan}, \citenamefont {Kanaganayagam}, \citenamefont {Marsden}, \citenamefont {Richards}, \citenamefont {Stott}, \citenamefont {Roorda}, \citenamefont {Morse}, \citenamefont {Thewalt},\ and\ \citenamefont {Simmons}}]{higginbottom_2022a}%
  \BibitemOpen
  \bibfield  {author} {\bibinfo {author} {\bibfnamefont {D.~B.}\ \bibnamefont {Higginbottom}}, \bibinfo {author} {\bibfnamefont {A.~T.~K.}\ \bibnamefont {Kurkjian}}, \bibinfo {author} {\bibfnamefont {C.}~\bibnamefont {Chartrand}}, \bibinfo {author} {\bibfnamefont {M.}~\bibnamefont {Kazemi}}, \bibinfo {author} {\bibfnamefont {N.~A.}\ \bibnamefont {Brunelle}}, \bibinfo {author} {\bibfnamefont {E.~R.}\ \bibnamefont {MacQuarrie}}, \bibinfo {author} {\bibfnamefont {J.~R.}\ \bibnamefont {Klein}}, \bibinfo {author} {\bibfnamefont {N.~R.}\ \bibnamefont {{Lee-Hone}}}, \bibinfo {author} {\bibfnamefont {J.}~\bibnamefont {Stacho}}, \bibinfo {author} {\bibfnamefont {M.}~\bibnamefont {Ruether}}, \bibinfo {author} {\bibfnamefont {C.}~\bibnamefont {Bowness}}, \bibinfo {author} {\bibfnamefont {L.}~\bibnamefont {Bergeron}}, \bibinfo {author} {\bibfnamefont {A.}~\bibnamefont {DeAbreu}}, \bibinfo {author} {\bibfnamefont {S.~R.}\ \bibnamefont {Harrigan}}, \bibinfo {author} {\bibfnamefont {J.}~\bibnamefont {Kanaganayagam}},
  \bibinfo {author} {\bibfnamefont {D.~W.}\ \bibnamefont {Marsden}}, \bibinfo {author} {\bibfnamefont {T.~S.}\ \bibnamefont {Richards}}, \bibinfo {author} {\bibfnamefont {L.~A.}\ \bibnamefont {Stott}}, \bibinfo {author} {\bibfnamefont {S.}~\bibnamefont {Roorda}}, \bibinfo {author} {\bibfnamefont {K.~J.}\ \bibnamefont {Morse}}, \bibinfo {author} {\bibfnamefont {M.~L.~W.}\ \bibnamefont {Thewalt}},\ and\ \bibinfo {author} {\bibfnamefont {S.}~\bibnamefont {Simmons}},\ }\bibfield  {title} {\bibinfo {title} {Optical observation of single spins in silicon},\ }\href {https://doi.org/10.1038/s41586-022-04821-y} {\bibfield  {journal} {\bibinfo  {journal} {Nature}\ }\textbf {\bibinfo {volume} {607}},\ \bibinfo {pages} {266} (\bibinfo {year} {2022})}\BibitemShut {NoStop}%
\bibitem [{\citenamefont {Mi}\ \emph {et~al.}(2018)\citenamefont {Mi}, \citenamefont {Benito}, \citenamefont {Putz}, \citenamefont {Zajac}, \citenamefont {Taylor}, \citenamefont {Burkard},\ and\ \citenamefont {Petta}}]{mi_2018}%
  \BibitemOpen
  \bibfield  {author} {\bibinfo {author} {\bibfnamefont {X.}~\bibnamefont {Mi}}, \bibinfo {author} {\bibfnamefont {M.}~\bibnamefont {Benito}}, \bibinfo {author} {\bibfnamefont {S.}~\bibnamefont {Putz}}, \bibinfo {author} {\bibfnamefont {D.~M.}\ \bibnamefont {Zajac}}, \bibinfo {author} {\bibfnamefont {J.~M.}\ \bibnamefont {Taylor}}, \bibinfo {author} {\bibfnamefont {G.}~\bibnamefont {Burkard}},\ and\ \bibinfo {author} {\bibfnamefont {J.~R.}\ \bibnamefont {Petta}},\ }\bibfield  {title} {\bibinfo {title} {A coherent spin{\textendash}photon interface in silicon},\ }\href {https://doi.org/10.1038/nature25769} {\bibfield  {journal} {\bibinfo  {journal} {Nature}\ }\textbf {\bibinfo {volume} {555}},\ \bibinfo {pages} {599} (\bibinfo {year} {2018})}\BibitemShut {NoStop}%
\bibitem [{\citenamefont {Eichenfield}\ \emph {et~al.}(2009)\citenamefont {Eichenfield}, \citenamefont {Chan}, \citenamefont {Camacho}, \citenamefont {Vahala},\ and\ \citenamefont {Painter}}]{eichenfield_2009a}%
  \BibitemOpen
  \bibfield  {author} {\bibinfo {author} {\bibfnamefont {M.}~\bibnamefont {Eichenfield}}, \bibinfo {author} {\bibfnamefont {J.}~\bibnamefont {Chan}}, \bibinfo {author} {\bibfnamefont {R.~M.}\ \bibnamefont {Camacho}}, \bibinfo {author} {\bibfnamefont {K.~J.}\ \bibnamefont {Vahala}},\ and\ \bibinfo {author} {\bibfnamefont {O.}~\bibnamefont {Painter}},\ }\bibfield  {title} {\bibinfo {title} {Optomechanical crystals},\ }\href {https://doi.org/10.1038/nature08524} {\bibfield  {journal} {\bibinfo  {journal} {Nature}\ }\textbf {\bibinfo {volume} {462}},\ \bibinfo {pages} {78} (\bibinfo {year} {2009})}\BibitemShut {NoStop}%
\bibitem [{\citenamefont {Gambetta}\ \emph {et~al.}(2017)\citenamefont {Gambetta}, \citenamefont {Murray}, \citenamefont {Fung}, \citenamefont {McClure}, \citenamefont {Dial}, \citenamefont {Shanks}, \citenamefont {Sleight},\ and\ \citenamefont {Steffen}}]{gambetta_2017}%
  \BibitemOpen
  \bibfield  {author} {\bibinfo {author} {\bibfnamefont {J.~M.}\ \bibnamefont {Gambetta}}, \bibinfo {author} {\bibfnamefont {C.~E.}\ \bibnamefont {Murray}}, \bibinfo {author} {\bibfnamefont {Y.-K.-K.}\ \bibnamefont {Fung}}, \bibinfo {author} {\bibfnamefont {D.~T.}\ \bibnamefont {McClure}}, \bibinfo {author} {\bibfnamefont {O.}~\bibnamefont {Dial}}, \bibinfo {author} {\bibfnamefont {W.}~\bibnamefont {Shanks}}, \bibinfo {author} {\bibfnamefont {J.~W.}\ \bibnamefont {Sleight}},\ and\ \bibinfo {author} {\bibfnamefont {M.}~\bibnamefont {Steffen}},\ }\bibfield  {title} {\bibinfo {title} {Investigating {{Surface Loss Effects}} in {{Superconducting Transmon Qubits}}},\ }\href {https://doi.org/10.1109/TASC.2016.2629670} {\bibfield  {journal} {\bibinfo  {journal} {IEEE Transactions on Applied Superconductivity}\ }\textbf {\bibinfo {volume} {27}},\ \bibinfo {pages} {1} (\bibinfo {year} {2017})}\BibitemShut {NoStop}%
\bibitem [{\citenamefont {Woods}\ \emph {et~al.}(2019)\citenamefont {Woods}, \citenamefont {Calusine}, \citenamefont {Melville}, \citenamefont {Sevi}, \citenamefont {Golden}, \citenamefont {Kim}, \citenamefont {Rosenberg}, \citenamefont {Yoder},\ and\ \citenamefont {Oliver}}]{woods_2019}%
  \BibitemOpen
  \bibfield  {author} {\bibinfo {author} {\bibfnamefont {W.}~\bibnamefont {Woods}}, \bibinfo {author} {\bibfnamefont {G.}~\bibnamefont {Calusine}}, \bibinfo {author} {\bibfnamefont {A.}~\bibnamefont {Melville}}, \bibinfo {author} {\bibfnamefont {A.}~\bibnamefont {Sevi}}, \bibinfo {author} {\bibfnamefont {E.}~\bibnamefont {Golden}}, \bibinfo {author} {\bibfnamefont {D.}~\bibnamefont {Kim}}, \bibinfo {author} {\bibfnamefont {D.}~\bibnamefont {Rosenberg}}, \bibinfo {author} {\bibfnamefont {J.}~\bibnamefont {Yoder}},\ and\ \bibinfo {author} {\bibfnamefont {W.}~\bibnamefont {Oliver}},\ }\bibfield  {title} {\bibinfo {title} {Determining {{Interface Dielectric Losses}} in {{Superconducting Coplanar-Waveguide Resonators}}},\ }\href {https://doi.org/10.1103/PhysRevApplied.12.014012} {\bibfield  {journal} {\bibinfo  {journal} {Physical Review Applied}\ }\textbf {\bibinfo {volume} {12}},\ \bibinfo {pages} {014012} (\bibinfo {year} {2019})}\BibitemShut {NoStop}%
\bibitem [{\citenamefont {Melville}\ \emph {et~al.}(2020)\citenamefont {Melville}, \citenamefont {Calusine}, \citenamefont {Woods}, \citenamefont {Serniak}, \citenamefont {Golden}, \citenamefont {Niedzielski}, \citenamefont {Kim}, \citenamefont {Sevi}, \citenamefont {Yoder}, \citenamefont {Dauler},\ and\ \citenamefont {Oliver}}]{melville_2020}%
  \BibitemOpen
  \bibfield  {author} {\bibinfo {author} {\bibfnamefont {A.}~\bibnamefont {Melville}}, \bibinfo {author} {\bibfnamefont {G.}~\bibnamefont {Calusine}}, \bibinfo {author} {\bibfnamefont {W.}~\bibnamefont {Woods}}, \bibinfo {author} {\bibfnamefont {K.}~\bibnamefont {Serniak}}, \bibinfo {author} {\bibfnamefont {E.}~\bibnamefont {Golden}}, \bibinfo {author} {\bibfnamefont {B.~M.}\ \bibnamefont {Niedzielski}}, \bibinfo {author} {\bibfnamefont {D.~K.}\ \bibnamefont {Kim}}, \bibinfo {author} {\bibfnamefont {A.}~\bibnamefont {Sevi}}, \bibinfo {author} {\bibfnamefont {J.~L.}\ \bibnamefont {Yoder}}, \bibinfo {author} {\bibfnamefont {E.~A.}\ \bibnamefont {Dauler}},\ and\ \bibinfo {author} {\bibfnamefont {W.~D.}\ \bibnamefont {Oliver}},\ }\bibfield  {title} {\bibinfo {title} {Comparison of dielectric loss in titanium nitride and aluminum superconducting resonators},\ }\href {https://doi.org/10.1063/5.0021950} {\bibfield  {journal} {\bibinfo  {journal} {Applied Physics Letters}\ }\textbf {\bibinfo {volume} {117}},\
  \bibinfo {pages} {124004} (\bibinfo {year} {2020})}\BibitemShut {NoStop}%
\bibitem [{\citenamefont {J{\"a}ckle}(1972)}]{jackle_1972}%
  \BibitemOpen
  \bibfield  {author} {\bibinfo {author} {\bibfnamefont {J.}~\bibnamefont {J{\"a}ckle}},\ }\bibfield  {title} {\bibinfo {title} {On the ultrasonic attenuation in glasses at low temperatures},\ }\href {https://doi.org/10.1007/BF01401204} {\bibfield  {journal} {\bibinfo  {journal} {Zeitschrift f{\"u}r Physik A Hadrons and nuclei}\ }\textbf {\bibinfo {volume} {257}},\ \bibinfo {pages} {212} (\bibinfo {year} {1972})}\BibitemShut {NoStop}%
\bibitem [{\citenamefont {Sarabi}\ \emph {et~al.}(2016)\citenamefont {Sarabi}, \citenamefont {Ramanayaka}, \citenamefont {Burin}, \citenamefont {Wellstood},\ and\ \citenamefont {Osborn}}]{sarabi_2016b}%
  \BibitemOpen
  \bibfield  {author} {\bibinfo {author} {\bibfnamefont {B.}~\bibnamefont {Sarabi}}, \bibinfo {author} {\bibfnamefont {A.~N.}\ \bibnamefont {Ramanayaka}}, \bibinfo {author} {\bibfnamefont {A.~L.}\ \bibnamefont {Burin}}, \bibinfo {author} {\bibfnamefont {F.~C.}\ \bibnamefont {Wellstood}},\ and\ \bibinfo {author} {\bibfnamefont {K.~D.}\ \bibnamefont {Osborn}},\ }\bibfield  {title} {\bibinfo {title} {Projected {{Dipole Moments}} of {{Individual Two-Level Defects Extracted Using Circuit Quantum Electrodynamics}}},\ }\href {https://doi.org/10.1103/PhysRevLett.116.167002} {\bibfield  {journal} {\bibinfo  {journal} {Physical Review Letters}\ }\textbf {\bibinfo {volume} {116}},\ \bibinfo {pages} {167002} (\bibinfo {year} {2016})}\BibitemShut {NoStop}%
\bibitem [{\citenamefont {Phillips}(1987)}]{phillips_1987a}%
  \BibitemOpen
  \bibfield  {author} {\bibinfo {author} {\bibfnamefont {W.~A.}\ \bibnamefont {Phillips}},\ }\bibfield  {title} {\bibinfo {title} {Two-level states in glasses},\ }\href {https://doi.org/10.1088/0034-4885/50/12/003} {\bibfield  {journal} {\bibinfo  {journal} {Reports on Progress in Physics}\ }\textbf {\bibinfo {volume} {50}},\ \bibinfo {pages} {1657} (\bibinfo {year} {1987})}\BibitemShut {NoStop}%
\bibitem [{\citenamefont {Muhonen}\ \emph {et~al.}(2014)\citenamefont {Muhonen}, \citenamefont {Dehollain}, \citenamefont {Laucht}, \citenamefont {Hudson}, \citenamefont {Kalra}, \citenamefont {Sekiguchi}, \citenamefont {Itoh}, \citenamefont {Jamieson}, \citenamefont {McCallum}, \citenamefont {Dzurak},\ and\ \citenamefont {Morello}}]{muhonen_2014}%
  \BibitemOpen
  \bibfield  {author} {\bibinfo {author} {\bibfnamefont {J.~T.}\ \bibnamefont {Muhonen}}, \bibinfo {author} {\bibfnamefont {J.~P.}\ \bibnamefont {Dehollain}}, \bibinfo {author} {\bibfnamefont {A.}~\bibnamefont {Laucht}}, \bibinfo {author} {\bibfnamefont {F.~E.}\ \bibnamefont {Hudson}}, \bibinfo {author} {\bibfnamefont {R.}~\bibnamefont {Kalra}}, \bibinfo {author} {\bibfnamefont {T.}~\bibnamefont {Sekiguchi}}, \bibinfo {author} {\bibfnamefont {K.~M.}\ \bibnamefont {Itoh}}, \bibinfo {author} {\bibfnamefont {D.~N.}\ \bibnamefont {Jamieson}}, \bibinfo {author} {\bibfnamefont {J.~C.}\ \bibnamefont {McCallum}}, \bibinfo {author} {\bibfnamefont {A.~S.}\ \bibnamefont {Dzurak}},\ and\ \bibinfo {author} {\bibfnamefont {A.}~\bibnamefont {Morello}},\ }\bibfield  {title} {\bibinfo {title} {Storing quantum information for 30 seconds in a nanoelectronic device},\ }\href {https://doi.org/10.1038/nnano.2014.211} {\bibfield  {journal} {\bibinfo  {journal} {Nature Nanotechnology}\ }\textbf {\bibinfo {volume} {9}},\ \bibinfo
  {pages} {986} (\bibinfo {year} {2014})}\BibitemShut {NoStop}%
\bibitem [{\citenamefont {Wolfowicz}\ \emph {et~al.}(2013)\citenamefont {Wolfowicz}, \citenamefont {Tyryshkin}, \citenamefont {George}, \citenamefont {Riemann}, \citenamefont {Abrosimov}, \citenamefont {Becker}, \citenamefont {Pohl}, \citenamefont {Thewalt}, \citenamefont {Lyon},\ and\ \citenamefont {Morton}}]{wolfowicz_2013}%
  \BibitemOpen
  \bibfield  {author} {\bibinfo {author} {\bibfnamefont {G.}~\bibnamefont {Wolfowicz}}, \bibinfo {author} {\bibfnamefont {A.~M.}\ \bibnamefont {Tyryshkin}}, \bibinfo {author} {\bibfnamefont {R.~E.}\ \bibnamefont {George}}, \bibinfo {author} {\bibfnamefont {H.}~\bibnamefont {Riemann}}, \bibinfo {author} {\bibfnamefont {N.~V.}\ \bibnamefont {Abrosimov}}, \bibinfo {author} {\bibfnamefont {P.}~\bibnamefont {Becker}}, \bibinfo {author} {\bibfnamefont {H.-J.}\ \bibnamefont {Pohl}}, \bibinfo {author} {\bibfnamefont {M.~L.~W.}\ \bibnamefont {Thewalt}}, \bibinfo {author} {\bibfnamefont {S.~A.}\ \bibnamefont {Lyon}},\ and\ \bibinfo {author} {\bibfnamefont {J.~J.~L.}\ \bibnamefont {Morton}},\ }\bibfield  {title} {\bibinfo {title} {Atomic clock transitions in silicon-based spin qubits},\ }\href {https://doi.org/10.1038/nnano.2013.117} {\bibfield  {journal} {\bibinfo  {journal} {Nature Nanotechnology}\ }\textbf {\bibinfo {volume} {8}},\ \bibinfo {pages} {561} (\bibinfo {year} {2013})}\BibitemShut {NoStop}%
\bibitem [{\citenamefont {Kobayashi}\ \emph {et~al.}(2021)\citenamefont {Kobayashi}, \citenamefont {Salfi}, \citenamefont {Chua}, \citenamefont {Van Der~Heijden}, \citenamefont {House}, \citenamefont {Culcer}, \citenamefont {Hutchison}, \citenamefont {Johnson}, \citenamefont {McCallum}, \citenamefont {Riemann}, \citenamefont {Abrosimov}, \citenamefont {Becker}, \citenamefont {Pohl}, \citenamefont {Simmons},\ and\ \citenamefont {Rogge}}]{kobayashi_2021a}%
  \BibitemOpen
  \bibfield  {author} {\bibinfo {author} {\bibfnamefont {T.}~\bibnamefont {Kobayashi}}, \bibinfo {author} {\bibfnamefont {J.}~\bibnamefont {Salfi}}, \bibinfo {author} {\bibfnamefont {C.}~\bibnamefont {Chua}}, \bibinfo {author} {\bibfnamefont {J.}~\bibnamefont {Van Der~Heijden}}, \bibinfo {author} {\bibfnamefont {M.~G.}\ \bibnamefont {House}}, \bibinfo {author} {\bibfnamefont {D.}~\bibnamefont {Culcer}}, \bibinfo {author} {\bibfnamefont {W.~D.}\ \bibnamefont {Hutchison}}, \bibinfo {author} {\bibfnamefont {B.~C.}\ \bibnamefont {Johnson}}, \bibinfo {author} {\bibfnamefont {J.~C.}\ \bibnamefont {McCallum}}, \bibinfo {author} {\bibfnamefont {H.}~\bibnamefont {Riemann}}, \bibinfo {author} {\bibfnamefont {N.~V.}\ \bibnamefont {Abrosimov}}, \bibinfo {author} {\bibfnamefont {P.}~\bibnamefont {Becker}}, \bibinfo {author} {\bibfnamefont {H.-J.}\ \bibnamefont {Pohl}}, \bibinfo {author} {\bibfnamefont {M.~Y.}\ \bibnamefont {Simmons}},\ and\ \bibinfo {author} {\bibfnamefont {S.}~\bibnamefont {Rogge}},\ }\bibfield  {title}
  {\bibinfo {title} {Engineering long spin coherence times of spin{\textendash}orbit qubits in silicon},\ }\href {https://doi.org/10.1038/s41563-020-0743-3} {\bibfield  {journal} {\bibinfo  {journal} {Nature Materials}\ }\textbf {\bibinfo {volume} {20}},\ \bibinfo {pages} {38} (\bibinfo {year} {2021})}\BibitemShut {NoStop}%
\bibitem [{\citenamefont {Mansir}\ \emph {et~al.}(2018)\citenamefont {Mansir}, \citenamefont {Conti}, \citenamefont {Zeng}, \citenamefont {Pla}, \citenamefont {Bertet}, \citenamefont {Swift}, \citenamefont {Van De~Walle}, \citenamefont {Thewalt}, \citenamefont {Sklenard}, \citenamefont {Niquet},\ and\ \citenamefont {Morton}}]{mansir_2018}%
  \BibitemOpen
  \bibfield  {author} {\bibinfo {author} {\bibfnamefont {J.}~\bibnamefont {Mansir}}, \bibinfo {author} {\bibfnamefont {P.}~\bibnamefont {Conti}}, \bibinfo {author} {\bibfnamefont {Z.}~\bibnamefont {Zeng}}, \bibinfo {author} {\bibfnamefont {J.~J.}\ \bibnamefont {Pla}}, \bibinfo {author} {\bibfnamefont {P.}~\bibnamefont {Bertet}}, \bibinfo {author} {\bibfnamefont {M.~W.}\ \bibnamefont {Swift}}, \bibinfo {author} {\bibfnamefont {C.~G.}\ \bibnamefont {Van De~Walle}}, \bibinfo {author} {\bibfnamefont {M.~L.~W.}\ \bibnamefont {Thewalt}}, \bibinfo {author} {\bibfnamefont {B.}~\bibnamefont {Sklenard}}, \bibinfo {author} {\bibfnamefont {Y.~M.}\ \bibnamefont {Niquet}},\ and\ \bibinfo {author} {\bibfnamefont {J.~J.~L.}\ \bibnamefont {Morton}},\ }\bibfield  {title} {\bibinfo {title} {Linear {{Hyperfine Tuning}} of {{Donor Spins}} in {{Silicon Using Hydrostatic Strain}}},\ }\href {https://doi.org/10.1103/PhysRevLett.120.167701} {\bibfield  {journal} {\bibinfo  {journal} {Physical Review Letters}\ }\textbf {\bibinfo
  {volume} {120}},\ \bibinfo {pages} {167701} (\bibinfo {year} {2018})}\BibitemShut {NoStop}%
\bibitem [{\citenamefont {Song}\ and\ \citenamefont {Golding}(2011)}]{song_2011a}%
  \BibitemOpen
  \bibfield  {author} {\bibinfo {author} {\bibfnamefont {Y.~P.}\ \bibnamefont {Song}}\ and\ \bibinfo {author} {\bibfnamefont {B.}~\bibnamefont {Golding}},\ }\bibfield  {title} {\bibinfo {title} {Manipulation and decoherence of acceptor states in silicon},\ }\href {https://doi.org/10.1209/0295-5075/95/47004} {\bibfield  {journal} {\bibinfo  {journal} {Europhysics Letters}\ }\textbf {\bibinfo {volume} {95}},\ \bibinfo {pages} {47004} (\bibinfo {year} {2011})}\BibitemShut {NoStop}%
\bibitem [{\citenamefont {K{\"o}pf}\ and\ \citenamefont {Lassmann}(1992)}]{kopf_1992a}%
  \BibitemOpen
  \bibfield  {author} {\bibinfo {author} {\bibfnamefont {A.}~\bibnamefont {K{\"o}pf}}\ and\ \bibinfo {author} {\bibfnamefont {K.}~\bibnamefont {Lassmann}},\ }\bibfield  {title} {\bibinfo {title} {Linear {{Stark}} and nonlinear {{Zeeman}} coupling to the ground state of effective mass acceptors in silicon},\ }\href {https://doi.org/10.1103/PhysRevLett.69.1580} {\bibfield  {journal} {\bibinfo  {journal} {Physical Review Letters}\ }\textbf {\bibinfo {volume} {69}},\ \bibinfo {pages} {1580} (\bibinfo {year} {1992})}\BibitemShut {NoStop}%
\bibitem [{\citenamefont {Neubrand}(1978)}]{neubrand_1978}%
  \BibitemOpen
  \bibfield  {author} {\bibinfo {author} {\bibfnamefont {H.}~\bibnamefont {Neubrand}},\ }\bibfield  {title} {\bibinfo {title} {{{ESR From}} boron in silicon at zero and small external stress {{I}}. {{Line}} positions and line structure},\ }\href {https://doi.org/10.1002/pssb.2220860131} {\bibfield  {journal} {\bibinfo  {journal} {physica status solidi (b)}\ }\textbf {\bibinfo {volume} {86}},\ \bibinfo {pages} {269} (\bibinfo {year} {1978})}\BibitemShut {NoStop}%
\bibitem [{\citenamefont {Bir}\ \emph {et~al.}(1974)\citenamefont {Bir}, \citenamefont {Pikus}, \citenamefont {Hensel}, \citenamefont {Shelnitz},\ and\ \citenamefont {Louvish}}]{bir_pikus_1974}%
  \BibitemOpen
  \bibfield  {author} {\bibinfo {author} {\bibfnamefont {G.~L.}\ \bibnamefont {Bir}}, \bibinfo {author} {\bibfnamefont {G.~E.}\ \bibnamefont {Pikus}}, \bibinfo {author} {\bibfnamefont {J.~C.}\ \bibnamefont {Hensel}}, \bibinfo {author} {\bibfnamefont {P.}~\bibnamefont {Shelnitz}},\ and\ \bibinfo {author} {\bibfnamefont {D.}~\bibnamefont {Louvish}},\ }\href@noop {} {\emph {\bibinfo {title} {Symmetry and strain-induced effects in semiconductors}}}\ (\bibinfo  {publisher} {Wiley},\ \bibinfo {address} {New York},\ \bibinfo {year} {1974})\BibitemShut {NoStop}%
\bibitem [{Sup()}]{Supp}%
  \BibitemOpen
  \href@noop {} {\bibinfo {title} {{See Supplemental Material for supplementary methods, measurements, and theoretical analyses.}}}\BibitemShut {Stop}%
\bibitem [{\citenamefont {Pla}\ \emph {et~al.}(2018)\citenamefont {Pla}, \citenamefont {Bienfait}, \citenamefont {Pica}, \citenamefont {Mansir}, \citenamefont {Mohiyaddin}, \citenamefont {Zeng}, \citenamefont {Niquet}, \citenamefont {Morello}, \citenamefont {Schenkel}, \citenamefont {Morton},\ and\ \citenamefont {Bertet}}]{pla_2018}%
  \BibitemOpen
  \bibfield  {author} {\bibinfo {author} {\bibfnamefont {J.~J.}\ \bibnamefont {Pla}}, \bibinfo {author} {\bibfnamefont {A.}~\bibnamefont {Bienfait}}, \bibinfo {author} {\bibfnamefont {G.}~\bibnamefont {Pica}}, \bibinfo {author} {\bibfnamefont {J.}~\bibnamefont {Mansir}}, \bibinfo {author} {\bibfnamefont {F.~A.}\ \bibnamefont {Mohiyaddin}}, \bibinfo {author} {\bibfnamefont {Z.}~\bibnamefont {Zeng}}, \bibinfo {author} {\bibfnamefont {Y.~M.}\ \bibnamefont {Niquet}}, \bibinfo {author} {\bibfnamefont {A.}~\bibnamefont {Morello}}, \bibinfo {author} {\bibfnamefont {T.}~\bibnamefont {Schenkel}}, \bibinfo {author} {\bibfnamefont {J.~J.~L.}\ \bibnamefont {Morton}},\ and\ \bibinfo {author} {\bibfnamefont {P.}~\bibnamefont {Bertet}},\ }\bibfield  {title} {\bibinfo {title} {Strain-{{Induced Spin-Resonance Shifts}} in {{Silicon Devices}}},\ }\href {https://doi.org/10.1103/PhysRevApplied.9.044014} {\bibfield  {journal} {\bibinfo  {journal} {Physical Review Applied}\ }\textbf {\bibinfo {volume} {9}},\ \bibinfo {pages}
  {044014} (\bibinfo {year} {2018})}\BibitemShut {NoStop}%
\bibitem [{\citenamefont {Checchin}\ \emph {et~al.}(2022)\citenamefont {Checchin}, \citenamefont {Frolov}, \citenamefont {Lunin}, \citenamefont {Grassellino},\ and\ \citenamefont {Romanenko}}]{checchin_2022c}%
  \BibitemOpen
  \bibfield  {author} {\bibinfo {author} {\bibfnamefont {M.}~\bibnamefont {Checchin}}, \bibinfo {author} {\bibfnamefont {D.}~\bibnamefont {Frolov}}, \bibinfo {author} {\bibfnamefont {A.}~\bibnamefont {Lunin}}, \bibinfo {author} {\bibfnamefont {A.}~\bibnamefont {Grassellino}},\ and\ \bibinfo {author} {\bibfnamefont {A.}~\bibnamefont {Romanenko}},\ }\bibfield  {title} {\bibinfo {title} {Measurement of the {{Low-Temperature Loss Tangent}} of {{High-Resistivity Silicon Using}} a {{High- Q Superconducting Resonator}}},\ }\href {https://doi.org/10.1103/PhysRevApplied.18.034013} {\bibfield  {journal} {\bibinfo  {journal} {Physical Review Applied}\ }\textbf {\bibinfo {volume} {18}},\ \bibinfo {pages} {034013} (\bibinfo {year} {2022})}\BibitemShut {NoStop}%
\bibitem [{\citenamefont {Aggarwal}(1964)}]{AGGARWAL1964163}%
  \BibitemOpen
  \bibfield  {author} {\bibinfo {author} {\bibfnamefont {R.}~\bibnamefont {Aggarwal}},\ }\bibfield  {title} {\bibinfo {title} {Optical determination of the valley-orbit splitting of the ground state of donors in silicon},\ }\href {https://doi.org/10.1016/0038-1098(64)90105-X} {\bibfield  {journal} {\bibinfo  {journal} {Solid State Communications}\ }\textbf {\bibinfo {volume} {2}},\ \bibinfo {pages} {163} (\bibinfo {year} {1964})}\BibitemShut {NoStop}%
\bibitem [{\citenamefont {Stan}\ \emph {et~al.}(2004)\citenamefont {Stan}, \citenamefont {Field},\ and\ \citenamefont {Martinis}}]{stan_2004}%
  \BibitemOpen
  \bibfield  {author} {\bibinfo {author} {\bibfnamefont {G.}~\bibnamefont {Stan}}, \bibinfo {author} {\bibfnamefont {S.~B.}\ \bibnamefont {Field}},\ and\ \bibinfo {author} {\bibfnamefont {J.~M.}\ \bibnamefont {Martinis}},\ }\bibfield  {title} {\bibinfo {title} {Critical {{Field}} for {{Complete Vortex Expulsion}} from {{Narrow Superconducting Strips}}},\ }\href {https://doi.org/10.1103/PhysRevLett.92.097003} {\bibfield  {journal} {\bibinfo  {journal} {Physical Review Letters}\ }\textbf {\bibinfo {volume} {92}},\ \bibinfo {pages} {097003} (\bibinfo {year} {2004})}\BibitemShut {NoStop}%
\bibitem [{\citenamefont {Bir}\ \emph {et~al.}(1963{\natexlab{a}})\citenamefont {Bir}, \citenamefont {Butikov},\ and\ \citenamefont {Pikus}}]{birSpinCombinedResonance1963b}%
  \BibitemOpen
  \bibfield  {author} {\bibinfo {author} {\bibfnamefont {G.}~\bibnamefont {Bir}}, \bibinfo {author} {\bibfnamefont {E.}~\bibnamefont {Butikov}},\ and\ \bibinfo {author} {\bibfnamefont {G.}~\bibnamefont {Pikus}},\ }\bibfield  {title} {\bibinfo {title} {Spin and combined resonance on acceptor centres in {{Ge}} and {{Si}} type crystals---{{I}}},\ }\href {https://doi.org/10.1016/0022-3697(63)90086-6} {\bibfield  {journal} {\bibinfo  {journal} {Journal of Physics and Chemistry of Solids}\ }\textbf {\bibinfo {volume} {24}},\ \bibinfo {pages} {1467} (\bibinfo {year} {1963}{\natexlab{a}})}\BibitemShut {NoStop}%
\bibitem [{\citenamefont {Bir}\ \emph {et~al.}(1963{\natexlab{b}})\citenamefont {Bir}, \citenamefont {Butikov},\ and\ \citenamefont {Pikus}}]{birSpinCombinedResonance1963c}%
  \BibitemOpen
  \bibfield  {author} {\bibinfo {author} {\bibfnamefont {G.}~\bibnamefont {Bir}}, \bibinfo {author} {\bibfnamefont {E.}~\bibnamefont {Butikov}},\ and\ \bibinfo {author} {\bibfnamefont {G.}~\bibnamefont {Pikus}},\ }\bibfield  {title} {\bibinfo {title} {Spin and combined resonance on acceptor centres in {{Ge}} and {{Si}} type crystals---{{II}}},\ }\href {https://doi.org/10.1016/0022-3697(63)90087-8} {\bibfield  {journal} {\bibinfo  {journal} {Journal of Physics and Chemistry of Solids}\ }\textbf {\bibinfo {volume} {24}},\ \bibinfo {pages} {1475} (\bibinfo {year} {1963}{\natexlab{b}})}\BibitemShut {NoStop}%
\bibitem [{\citenamefont {Zemlicka}\ \emph {et~al.}(2023)\citenamefont {Zemlicka}, \citenamefont {Redchenko}, \citenamefont {Peruzzo}, \citenamefont {Hassani}, \citenamefont {Trioni}, \citenamefont {Barzanjeh},\ and\ \citenamefont {Fink}}]{zemlicka_2023}%
  \BibitemOpen
  \bibfield  {author} {\bibinfo {author} {\bibfnamefont {M.}~\bibnamefont {Zemlicka}}, \bibinfo {author} {\bibfnamefont {E.}~\bibnamefont {Redchenko}}, \bibinfo {author} {\bibfnamefont {M.}~\bibnamefont {Peruzzo}}, \bibinfo {author} {\bibfnamefont {F.}~\bibnamefont {Hassani}}, \bibinfo {author} {\bibfnamefont {A.}~\bibnamefont {Trioni}}, \bibinfo {author} {\bibfnamefont {S.}~\bibnamefont {Barzanjeh}},\ and\ \bibinfo {author} {\bibfnamefont {J.}~\bibnamefont {Fink}},\ }\bibfield  {title} {\bibinfo {title} {Compact vacuum-gap transmon qubits: {{Selective}} and sensitive probes for superconductor surface losses},\ }\href {https://doi.org/10.1103/PhysRevApplied.20.044054} {\bibfield  {journal} {\bibinfo  {journal} {Physical Review Applied}\ }\textbf {\bibinfo {volume} {20}},\ \bibinfo {pages} {044054} (\bibinfo {year} {2023})}\BibitemShut {NoStop}%
\bibitem [{\citenamefont {Chen}\ \emph {et~al.}(2023)\citenamefont {Chen}, \citenamefont {Owens}, \citenamefont {Putterman}, \citenamefont {Sch{\"a}fer},\ and\ \citenamefont {Painter}}]{chen_2023a}%
  \BibitemOpen
  \bibfield  {author} {\bibinfo {author} {\bibfnamefont {M.}~\bibnamefont {Chen}}, \bibinfo {author} {\bibfnamefont {J.~C.}\ \bibnamefont {Owens}}, \bibinfo {author} {\bibfnamefont {H.}~\bibnamefont {Putterman}}, \bibinfo {author} {\bibfnamefont {M.}~\bibnamefont {Sch{\"a}fer}},\ and\ \bibinfo {author} {\bibfnamefont {O.}~\bibnamefont {Painter}},\ }\href@noop {} {\bibinfo {title} {Phonon engineering of atomic-scale defects in superconducting quantum circuits}} (\bibinfo {year} {2023}),\ \Eprint {https://arxiv.org/abs/2310.03929} {arxiv:2310.03929} \BibitemShut {NoStop}%
\bibitem [{\citenamefont {Odeh}\ \emph {et~al.}(2023)\citenamefont {Odeh}, \citenamefont {Godeneli}, \citenamefont {Li}, \citenamefont {Tangirala}, \citenamefont {Zhou}, \citenamefont {Zhang}, \citenamefont {Zhang},\ and\ \citenamefont {Sipahigil}}]{odeh_2023}%
  \BibitemOpen
  \bibfield  {author} {\bibinfo {author} {\bibfnamefont {M.}~\bibnamefont {Odeh}}, \bibinfo {author} {\bibfnamefont {K.}~\bibnamefont {Godeneli}}, \bibinfo {author} {\bibfnamefont {E.}~\bibnamefont {Li}}, \bibinfo {author} {\bibfnamefont {R.}~\bibnamefont {Tangirala}}, \bibinfo {author} {\bibfnamefont {H.}~\bibnamefont {Zhou}}, \bibinfo {author} {\bibfnamefont {X.}~\bibnamefont {Zhang}}, \bibinfo {author} {\bibfnamefont {Z.-H.}\ \bibnamefont {Zhang}},\ and\ \bibinfo {author} {\bibfnamefont {A.}~\bibnamefont {Sipahigil}},\ }\href@noop {} {\bibinfo {title} {Non-{{Markovian}} dynamics of a superconducting qubit in a phononic bandgap}} (\bibinfo {year} {2023}),\ \Eprint {https://arxiv.org/abs/2312.01031} {arxiv:2312.01031} \BibitemShut {NoStop}%
\bibitem [{\citenamefont {Ruskov}\ and\ \citenamefont {Tahan}(2013)}]{ruskov_2013a}%
  \BibitemOpen
  \bibfield  {author} {\bibinfo {author} {\bibfnamefont {R.}~\bibnamefont {Ruskov}}\ and\ \bibinfo {author} {\bibfnamefont {C.}~\bibnamefont {Tahan}},\ }\bibfield  {title} {\bibinfo {title} {On-chip cavity quantum phonodynamics with an acceptor qubit in silicon},\ }\href {https://doi.org/10.1103/PhysRevB.88.064308} {\bibfield  {journal} {\bibinfo  {journal} {Physical Review B}\ }\textbf {\bibinfo {volume} {88}},\ \bibinfo {pages} {064308} (\bibinfo {year} {2013})}\BibitemShut {NoStop}%
\bibitem [{\citenamefont {Van Der~Heijden}\ \emph {et~al.}(2014)\citenamefont {Van Der~Heijden}, \citenamefont {Salfi}, \citenamefont {Mol}, \citenamefont {Verduijn}, \citenamefont {Tettamanzi}, \citenamefont {Hamilton}, \citenamefont {Collaert},\ and\ \citenamefont {Rogge}}]{vanderheijden_2014}%
  \BibitemOpen
  \bibfield  {author} {\bibinfo {author} {\bibfnamefont {J.}~\bibnamefont {Van Der~Heijden}}, \bibinfo {author} {\bibfnamefont {J.}~\bibnamefont {Salfi}}, \bibinfo {author} {\bibfnamefont {J.~A.}\ \bibnamefont {Mol}}, \bibinfo {author} {\bibfnamefont {J.}~\bibnamefont {Verduijn}}, \bibinfo {author} {\bibfnamefont {G.~C.}\ \bibnamefont {Tettamanzi}}, \bibinfo {author} {\bibfnamefont {A.~R.}\ \bibnamefont {Hamilton}}, \bibinfo {author} {\bibfnamefont {N.}~\bibnamefont {Collaert}},\ and\ \bibinfo {author} {\bibfnamefont {S.}~\bibnamefont {Rogge}},\ }\bibfield  {title} {\bibinfo {title} {Probing the {{Spin States}} of a {{Single Acceptor Atom}}},\ }\href {https://doi.org/10.1021/nl4047015} {\bibfield  {journal} {\bibinfo  {journal} {Nano Letters}\ }\textbf {\bibinfo {volume} {14}},\ \bibinfo {pages} {1492} (\bibinfo {year} {2014})}\BibitemShut {NoStop}%
\bibitem [{\citenamefont {Salfi}\ \emph {et~al.}(2016{\natexlab{a}})\citenamefont {Salfi}, \citenamefont {Mol}, \citenamefont {Culcer},\ and\ \citenamefont {Rogge}}]{salfi_2016}%
  \BibitemOpen
  \bibfield  {author} {\bibinfo {author} {\bibfnamefont {J.}~\bibnamefont {Salfi}}, \bibinfo {author} {\bibfnamefont {J.~A.}\ \bibnamefont {Mol}}, \bibinfo {author} {\bibfnamefont {D.}~\bibnamefont {Culcer}},\ and\ \bibinfo {author} {\bibfnamefont {S.}~\bibnamefont {Rogge}},\ }\bibfield  {title} {\bibinfo {title} {Charge-{{Insensitive Single-Atom Spin-Orbit Qubit}} in {{Silicon}}},\ }\href {https://doi.org/10.1103/PhysRevLett.116.246801} {\bibfield  {journal} {\bibinfo  {journal} {Physical Review Letters}\ }\textbf {\bibinfo {volume} {116}},\ \bibinfo {pages} {246801} (\bibinfo {year} {2016}{\natexlab{a}})}\BibitemShut {NoStop}%
\bibitem [{\citenamefont {Salfi}\ \emph {et~al.}(2016{\natexlab{b}})\citenamefont {Salfi}, \citenamefont {Tong}, \citenamefont {Rogge},\ and\ \citenamefont {Culcer}}]{salfi_2016b}%
  \BibitemOpen
  \bibfield  {author} {\bibinfo {author} {\bibfnamefont {J.}~\bibnamefont {Salfi}}, \bibinfo {author} {\bibfnamefont {M.}~\bibnamefont {Tong}}, \bibinfo {author} {\bibfnamefont {S.}~\bibnamefont {Rogge}},\ and\ \bibinfo {author} {\bibfnamefont {D.}~\bibnamefont {Culcer}},\ }\bibfield  {title} {\bibinfo {title} {Quantum computing with acceptor spins in silicon},\ }\href {https://doi.org/10.1088/0957-4484/27/24/244001} {\bibfield  {journal} {\bibinfo  {journal} {Nanotechnology}\ }\textbf {\bibinfo {volume} {27}},\ \bibinfo {pages} {244001} (\bibinfo {year} {2016}{\natexlab{b}})}\BibitemShut {NoStop}%
\bibitem [{\citenamefont {Zhang}\ \emph {et~al.}(2023)\citenamefont {Zhang}, \citenamefont {He},\ and\ \citenamefont {Huang}}]{zhang_2023a}%
  \BibitemOpen
  \bibfield  {author} {\bibinfo {author} {\bibfnamefont {S.}~\bibnamefont {Zhang}}, \bibinfo {author} {\bibfnamefont {Y.}~\bibnamefont {He}},\ and\ \bibinfo {author} {\bibfnamefont {P.}~\bibnamefont {Huang}},\ }\bibfield  {title} {\bibinfo {title} {Acceptor-based qubit in silicon with tunable strain},\ }\href {https://doi.org/10.1103/PhysRevB.107.155301} {\bibfield  {journal} {\bibinfo  {journal} {Physical Review B}\ }\textbf {\bibinfo {volume} {107}},\ \bibinfo {pages} {155301} (\bibinfo {year} {2023})}\BibitemShut {NoStop}%
\bibitem [{\citenamefont {Samkharadze}\ \emph {et~al.}(2016)\citenamefont {Samkharadze}, \citenamefont {Bruno}, \citenamefont {Scarlino}, \citenamefont {Zheng}, \citenamefont {DiVincenzo}, \citenamefont {DiCarlo},\ and\ \citenamefont {Vandersypen}}]{samkharadze_2016b}%
  \BibitemOpen
  \bibfield  {author} {\bibinfo {author} {\bibfnamefont {N.}~\bibnamefont {Samkharadze}}, \bibinfo {author} {\bibfnamefont {A.}~\bibnamefont {Bruno}}, \bibinfo {author} {\bibfnamefont {P.}~\bibnamefont {Scarlino}}, \bibinfo {author} {\bibfnamefont {G.}~\bibnamefont {Zheng}}, \bibinfo {author} {\bibfnamefont {D.~P.}\ \bibnamefont {DiVincenzo}}, \bibinfo {author} {\bibfnamefont {L.}~\bibnamefont {DiCarlo}},\ and\ \bibinfo {author} {\bibfnamefont {L.~M.~K.}\ \bibnamefont {Vandersypen}},\ }\bibfield  {title} {\bibinfo {title} {High-{{Kinetic-Inductance Superconducting Nanowire Resonators}} for {{Circuit QED}} in a {{Magnetic Field}}},\ }\href {https://doi.org/10.1103/PhysRevApplied.5.044004} {\bibfield  {journal} {\bibinfo  {journal} {Physical Review Applied}\ }\textbf {\bibinfo {volume} {5}},\ \bibinfo {pages} {044004} (\bibinfo {year} {2016})}\BibitemShut {NoStop}%
\end{thebibliography}%


\providecommand{\noopsort}[1]{}\providecommand{\singleletter}[1]{#1}%
\begin{thebibliography}{17}%
\makeatletter
\providecommand \@ifxundefined [1]{%
 \@ifx{#1\undefined}
}%
\providecommand \@ifnum [1]{%
 \ifnum #1\expandafter \@firstoftwo
 \else \expandafter \@secondoftwo
 \fi
}%
\providecommand \@ifx [1]{%
 \ifx #1\expandafter \@firstoftwo
 \else \expandafter \@secondoftwo
 \fi
}%
\providecommand \natexlab [1]{#1}%
\providecommand \enquote  [1]{``#1''}%
\providecommand \bibnamefont  [1]{#1}%
\providecommand \bibfnamefont [1]{#1}%
\providecommand \citenamefont [1]{#1}%
\providecommand \href@noop [0]{\@secondoftwo}%
\providecommand \href [0]{\begingroup \@sanitize@url \@href}%
\providecommand \@href[1]{\@@startlink{#1}\@@href}%
\providecommand \@@href[1]{\endgroup#1\@@endlink}%
\providecommand \@sanitize@url [0]{\catcode `\\12\catcode `\$12\catcode `\&12\catcode `\#12\catcode `\^12\catcode `\_12\catcode `\%12\relax}%
\providecommand \@@startlink[1]{}%
\providecommand \@@endlink[0]{}%
\providecommand \url  [0]{\begingroup\@sanitize@url \@url }%
\providecommand \@url [1]{\endgroup\@href {#1}{\urlprefix }}%
\providecommand \urlprefix  [0]{URL }%
\providecommand \Eprint [0]{\href }%
\providecommand \doibase [0]{https://doi.org/}%
\providecommand \selectlanguage [0]{\@gobble}%
\providecommand \bibinfo  [0]{\@secondoftwo}%
\providecommand \bibfield  [0]{\@secondoftwo}%
\providecommand \translation [1]{[#1]}%
\providecommand \BibitemOpen [0]{}%
\providecommand \bibitemStop [0]{}%
\providecommand \bibitemNoStop [0]{.\EOS\space}%
\providecommand \EOS [0]{\spacefactor3000\relax}%
\providecommand \BibitemShut  [1]{\csname bibitem#1\endcsname}%
\let\auto@bib@innerbib\@empty
\bibitem [{\citenamefont {Murray}\ \emph {et~al.}(2018)\citenamefont {Murray}, \citenamefont {Gambetta}, \citenamefont {McClure},\ and\ \citenamefont {Steffen}}]{murray_2018a}%
  \BibitemOpen
  \bibfield  {author} {\bibinfo {author} {\bibfnamefont {C.~E.}\ \bibnamefont {Murray}}, \bibinfo {author} {\bibfnamefont {J.~M.}\ \bibnamefont {Gambetta}}, \bibinfo {author} {\bibfnamefont {D.~T.}\ \bibnamefont {McClure}},\ and\ \bibinfo {author} {\bibfnamefont {M.}~\bibnamefont {Steffen}},\ }\bibfield  {title} {\bibinfo {title} {Analytical {{Determination}} of {{Participation}} in {{Superconducting Coplanar Architectures}}},\ }\href {https://doi.org/10.1109/TMTT.2018.2841829} {\bibfield  {journal} {\bibinfo  {journal} {IEEE Transactions on Microwave Theory and Techniques}\ }\textbf {\bibinfo {volume} {66}},\ \bibinfo {pages} {3724} (\bibinfo {year} {2018})}\BibitemShut {NoStop}%
\bibitem [{\citenamefont {Aspelmeyer}\ \emph {et~al.}(2014)\citenamefont {Aspelmeyer}, \citenamefont {Kippenberg},\ and\ \citenamefont {Marquardt}}]{aspelmeyer_2014a}%
  \BibitemOpen
  \bibfield  {author} {\bibinfo {author} {\bibfnamefont {M.}~\bibnamefont {Aspelmeyer}}, \bibinfo {author} {\bibfnamefont {T.~J.}\ \bibnamefont {Kippenberg}},\ and\ \bibinfo {author} {\bibfnamefont {F.}~\bibnamefont {Marquardt}},\ }\bibfield  {title} {\bibinfo {title} {Cavity optomechanics},\ }\href {https://doi.org/10.1103/RevModPhys.86.1391} {\bibfield  {journal} {\bibinfo  {journal} {Reviews of Modern Physics}\ }\textbf {\bibinfo {volume} {86}},\ \bibinfo {pages} {1391} (\bibinfo {year} {2014})}\BibitemShut {NoStop}%
\bibitem [{\citenamefont {Bruno}\ \emph {et~al.}(2015)\citenamefont {Bruno}, \citenamefont {De~Lange}, \citenamefont {Asaad}, \citenamefont {Van Der~Enden}, \citenamefont {Langford},\ and\ \citenamefont {DiCarlo}}]{bruno_2015}%
  \BibitemOpen
  \bibfield  {author} {\bibinfo {author} {\bibfnamefont {A.}~\bibnamefont {Bruno}}, \bibinfo {author} {\bibfnamefont {G.}~\bibnamefont {De~Lange}}, \bibinfo {author} {\bibfnamefont {S.}~\bibnamefont {Asaad}}, \bibinfo {author} {\bibfnamefont {K.~L.}\ \bibnamefont {Van Der~Enden}}, \bibinfo {author} {\bibfnamefont {N.~K.}\ \bibnamefont {Langford}},\ and\ \bibinfo {author} {\bibfnamefont {L.}~\bibnamefont {DiCarlo}},\ }\bibfield  {title} {\bibinfo {title} {Reducing intrinsic loss in superconducting resonators by surface treatment and deep etching of silicon substrates},\ }\href {https://doi.org/10.1063/1.4919761} {\bibfield  {journal} {\bibinfo  {journal} {Applied Physics Letters}\ }\textbf {\bibinfo {volume} {106}},\ \bibinfo {pages} {182601} (\bibinfo {year} {2015})}\BibitemShut {NoStop}%
\bibitem [{\citenamefont {Bir}\ \emph {et~al.}(1974)\citenamefont {Bir}, \citenamefont {Pikus}, \citenamefont {Hensel}, \citenamefont {Shelnitz},\ and\ \citenamefont {Louvish}}]{bir_pikus_1974}%
  \BibitemOpen
  \bibfield  {author} {\bibinfo {author} {\bibfnamefont {G.~L.}\ \bibnamefont {Bir}}, \bibinfo {author} {\bibfnamefont {G.~E.}\ \bibnamefont {Pikus}}, \bibinfo {author} {\bibfnamefont {J.~C.}\ \bibnamefont {Hensel}}, \bibinfo {author} {\bibfnamefont {P.}~\bibnamefont {Shelnitz}},\ and\ \bibinfo {author} {\bibfnamefont {D.}~\bibnamefont {Louvish}},\ }\href@noop {} {\emph {\bibinfo {title} {Symmetry and strain-induced effects in semiconductors}}}\ (\bibinfo  {publisher} {Wiley},\ \bibinfo {address} {New York},\ \bibinfo {year} {1974})\BibitemShut {NoStop}%
\bibitem [{\citenamefont {Kobayashi}\ \emph {et~al.}(2021)\citenamefont {Kobayashi}, \citenamefont {Salfi}, \citenamefont {Chua}, \citenamefont {Van Der~Heijden}, \citenamefont {House}, \citenamefont {Culcer}, \citenamefont {Hutchison}, \citenamefont {Johnson}, \citenamefont {McCallum}, \citenamefont {Riemann}, \citenamefont {Abrosimov}, \citenamefont {Becker}, \citenamefont {Pohl}, \citenamefont {Simmons},\ and\ \citenamefont {Rogge}}]{kobayashi_2021a}%
  \BibitemOpen
  \bibfield  {author} {\bibinfo {author} {\bibfnamefont {T.}~\bibnamefont {Kobayashi}}, \bibinfo {author} {\bibfnamefont {J.}~\bibnamefont {Salfi}}, \bibinfo {author} {\bibfnamefont {C.}~\bibnamefont {Chua}}, \bibinfo {author} {\bibfnamefont {J.}~\bibnamefont {Van Der~Heijden}}, \bibinfo {author} {\bibfnamefont {M.~G.}\ \bibnamefont {House}}, \bibinfo {author} {\bibfnamefont {D.}~\bibnamefont {Culcer}}, \bibinfo {author} {\bibfnamefont {W.~D.}\ \bibnamefont {Hutchison}}, \bibinfo {author} {\bibfnamefont {B.~C.}\ \bibnamefont {Johnson}}, \bibinfo {author} {\bibfnamefont {J.~C.}\ \bibnamefont {McCallum}}, \bibinfo {author} {\bibfnamefont {H.}~\bibnamefont {Riemann}}, \bibinfo {author} {\bibfnamefont {N.~V.}\ \bibnamefont {Abrosimov}}, \bibinfo {author} {\bibfnamefont {P.}~\bibnamefont {Becker}}, \bibinfo {author} {\bibfnamefont {H.-J.}\ \bibnamefont {Pohl}}, \bibinfo {author} {\bibfnamefont {M.~Y.}\ \bibnamefont {Simmons}},\ and\ \bibinfo {author} {\bibfnamefont {S.}~\bibnamefont {Rogge}},\ }\bibfield  {title}
  {\bibinfo {title} {Engineering long spin coherence times of spin{\textendash}orbit qubits in silicon},\ }\href {https://doi.org/10.1038/s41563-020-0743-3} {\bibfield  {journal} {\bibinfo  {journal} {Nature Materials}\ }\textbf {\bibinfo {volume} {20}},\ \bibinfo {pages} {38} (\bibinfo {year} {2021})}\BibitemShut {NoStop}%
\bibitem [{\citenamefont {Aggarwal}(1964)}]{AGGARWAL1964163}%
  \BibitemOpen
  \bibfield  {author} {\bibinfo {author} {\bibfnamefont {R.}~\bibnamefont {Aggarwal}},\ }\bibfield  {title} {\bibinfo {title} {Optical determination of the valley-orbit splitting of the ground state of donors in silicon},\ }\href {https://doi.org/10.1016/0038-1098(64)90105-X} {\bibfield  {journal} {\bibinfo  {journal} {Solid State Communications}\ }\textbf {\bibinfo {volume} {2}},\ \bibinfo {pages} {163} (\bibinfo {year} {1964})}\BibitemShut {NoStop}%
\bibitem [{\citenamefont {Muhonen}\ \emph {et~al.}(2014)\citenamefont {Muhonen}, \citenamefont {Dehollain}, \citenamefont {Laucht}, \citenamefont {Hudson}, \citenamefont {Kalra}, \citenamefont {Sekiguchi}, \citenamefont {Itoh}, \citenamefont {Jamieson}, \citenamefont {McCallum}, \citenamefont {Dzurak},\ and\ \citenamefont {Morello}}]{muhonen_2014}%
  \BibitemOpen
  \bibfield  {author} {\bibinfo {author} {\bibfnamefont {J.~T.}\ \bibnamefont {Muhonen}}, \bibinfo {author} {\bibfnamefont {J.~P.}\ \bibnamefont {Dehollain}}, \bibinfo {author} {\bibfnamefont {A.}~\bibnamefont {Laucht}}, \bibinfo {author} {\bibfnamefont {F.~E.}\ \bibnamefont {Hudson}}, \bibinfo {author} {\bibfnamefont {R.}~\bibnamefont {Kalra}}, \bibinfo {author} {\bibfnamefont {T.}~\bibnamefont {Sekiguchi}}, \bibinfo {author} {\bibfnamefont {K.~M.}\ \bibnamefont {Itoh}}, \bibinfo {author} {\bibfnamefont {D.~N.}\ \bibnamefont {Jamieson}}, \bibinfo {author} {\bibfnamefont {J.~C.}\ \bibnamefont {McCallum}}, \bibinfo {author} {\bibfnamefont {A.~S.}\ \bibnamefont {Dzurak}},\ and\ \bibinfo {author} {\bibfnamefont {A.}~\bibnamefont {Morello}},\ }\bibfield  {title} {\bibinfo {title} {Storing quantum information for 30 seconds in a nanoelectronic device},\ }\href {https://doi.org/10.1038/nnano.2014.211} {\bibfield  {journal} {\bibinfo  {journal} {Nature Nanotechnology}\ }\textbf {\bibinfo {volume} {9}},\ \bibinfo
  {pages} {986} (\bibinfo {year} {2014})}\BibitemShut {NoStop}%
\bibitem [{\citenamefont {Wolfowicz}\ \emph {et~al.}(2013)\citenamefont {Wolfowicz}, \citenamefont {Tyryshkin}, \citenamefont {George}, \citenamefont {Riemann}, \citenamefont {Abrosimov}, \citenamefont {Becker}, \citenamefont {Pohl}, \citenamefont {Thewalt}, \citenamefont {Lyon},\ and\ \citenamefont {Morton}}]{wolfowicz_2013}%
  \BibitemOpen
  \bibfield  {author} {\bibinfo {author} {\bibfnamefont {G.}~\bibnamefont {Wolfowicz}}, \bibinfo {author} {\bibfnamefont {A.~M.}\ \bibnamefont {Tyryshkin}}, \bibinfo {author} {\bibfnamefont {R.~E.}\ \bibnamefont {George}}, \bibinfo {author} {\bibfnamefont {H.}~\bibnamefont {Riemann}}, \bibinfo {author} {\bibfnamefont {N.~V.}\ \bibnamefont {Abrosimov}}, \bibinfo {author} {\bibfnamefont {P.}~\bibnamefont {Becker}}, \bibinfo {author} {\bibfnamefont {H.-J.}\ \bibnamefont {Pohl}}, \bibinfo {author} {\bibfnamefont {M.~L.~W.}\ \bibnamefont {Thewalt}}, \bibinfo {author} {\bibfnamefont {S.~A.}\ \bibnamefont {Lyon}},\ and\ \bibinfo {author} {\bibfnamefont {J.~J.~L.}\ \bibnamefont {Morton}},\ }\bibfield  {title} {\bibinfo {title} {Atomic clock transitions in silicon-based spin qubits},\ }\href {https://doi.org/10.1038/nnano.2013.117} {\bibfield  {journal} {\bibinfo  {journal} {Nature Nanotechnology}\ }\textbf {\bibinfo {volume} {8}},\ \bibinfo {pages} {561} (\bibinfo {year} {2013})}\BibitemShut {NoStop}%
\bibitem [{\citenamefont {Mansir}\ \emph {et~al.}(2018)\citenamefont {Mansir}, \citenamefont {Conti}, \citenamefont {Zeng}, \citenamefont {Pla}, \citenamefont {Bertet}, \citenamefont {Swift}, \citenamefont {Van De~Walle}, \citenamefont {Thewalt}, \citenamefont {Sklenard}, \citenamefont {Niquet},\ and\ \citenamefont {Morton}}]{mansir_2018}%
  \BibitemOpen
  \bibfield  {author} {\bibinfo {author} {\bibfnamefont {J.}~\bibnamefont {Mansir}}, \bibinfo {author} {\bibfnamefont {P.}~\bibnamefont {Conti}}, \bibinfo {author} {\bibfnamefont {Z.}~\bibnamefont {Zeng}}, \bibinfo {author} {\bibfnamefont {J.~J.}\ \bibnamefont {Pla}}, \bibinfo {author} {\bibfnamefont {P.}~\bibnamefont {Bertet}}, \bibinfo {author} {\bibfnamefont {M.~W.}\ \bibnamefont {Swift}}, \bibinfo {author} {\bibfnamefont {C.~G.}\ \bibnamefont {Van De~Walle}}, \bibinfo {author} {\bibfnamefont {M.~L.~W.}\ \bibnamefont {Thewalt}}, \bibinfo {author} {\bibfnamefont {B.}~\bibnamefont {Sklenard}}, \bibinfo {author} {\bibfnamefont {Y.~M.}\ \bibnamefont {Niquet}},\ and\ \bibinfo {author} {\bibfnamefont {J.~J.~L.}\ \bibnamefont {Morton}},\ }\bibfield  {title} {\bibinfo {title} {Linear {{Hyperfine Tuning}} of {{Donor Spins}} in {{Silicon Using Hydrostatic Strain}}},\ }\href {https://doi.org/10.1103/PhysRevLett.120.167701} {\bibfield  {journal} {\bibinfo  {journal} {Physical Review Letters}\ }\textbf {\bibinfo
  {volume} {120}},\ \bibinfo {pages} {167701} (\bibinfo {year} {2018})}\BibitemShut {NoStop}%
\bibitem [{\citenamefont {Pla}\ \emph {et~al.}(2018)\citenamefont {Pla}, \citenamefont {Bienfait}, \citenamefont {Pica}, \citenamefont {Mansir}, \citenamefont {Mohiyaddin}, \citenamefont {Zeng}, \citenamefont {Niquet}, \citenamefont {Morello}, \citenamefont {Schenkel}, \citenamefont {Morton},\ and\ \citenamefont {Bertet}}]{pla_2018}%
  \BibitemOpen
  \bibfield  {author} {\bibinfo {author} {\bibfnamefont {J.~J.}\ \bibnamefont {Pla}}, \bibinfo {author} {\bibfnamefont {A.}~\bibnamefont {Bienfait}}, \bibinfo {author} {\bibfnamefont {G.}~\bibnamefont {Pica}}, \bibinfo {author} {\bibfnamefont {J.}~\bibnamefont {Mansir}}, \bibinfo {author} {\bibfnamefont {F.~A.}\ \bibnamefont {Mohiyaddin}}, \bibinfo {author} {\bibfnamefont {Z.}~\bibnamefont {Zeng}}, \bibinfo {author} {\bibfnamefont {Y.~M.}\ \bibnamefont {Niquet}}, \bibinfo {author} {\bibfnamefont {A.}~\bibnamefont {Morello}}, \bibinfo {author} {\bibfnamefont {T.}~\bibnamefont {Schenkel}}, \bibinfo {author} {\bibfnamefont {J.~J.~L.}\ \bibnamefont {Morton}},\ and\ \bibinfo {author} {\bibfnamefont {P.}~\bibnamefont {Bertet}},\ }\bibfield  {title} {\bibinfo {title} {Strain-{{Induced Spin-Resonance Shifts}} in {{Silicon Devices}}},\ }\href {https://doi.org/10.1103/PhysRevApplied.9.044014} {\bibfield  {journal} {\bibinfo  {journal} {Physical Review Applied}\ }\textbf {\bibinfo {volume} {9}},\ \bibinfo {pages}
  {044014} (\bibinfo {year} {2018})}\BibitemShut {NoStop}%
\bibitem [{\citenamefont {White}(1962)}]{white_1962}%
  \BibitemOpen
  \bibfield  {author} {\bibinfo {author} {\bibfnamefont {G.}~\bibnamefont {White}},\ }\bibfield  {title} {\bibinfo {title} {Thermal expansion of vanadium, niobium, and tantalum at low temperatures},\ }\href {https://doi.org/10.1016/0011-2275(62)90014-0} {\bibfield  {journal} {\bibinfo  {journal} {Cryogenics}\ }\textbf {\bibinfo {volume} {2}},\ \bibinfo {pages} {292} (\bibinfo {year} {1962})}\BibitemShut {NoStop}%
\bibitem [{\citenamefont {Swenson}(1983)}]{swenson_1983}%
  \BibitemOpen
  \bibfield  {author} {\bibinfo {author} {\bibfnamefont {C.~A.}\ \bibnamefont {Swenson}},\ }\bibfield  {title} {\bibinfo {title} {Recommended {{Values}} for the {{Thermal Expansivity}} of {{Silicon}} from 0 to 1000 {{K}}},\ }\href {https://doi.org/10.1063/1.555681} {\bibfield  {journal} {\bibinfo  {journal} {Journal of Physical and Chemical Reference Data}\ }\textbf {\bibinfo {volume} {12}},\ \bibinfo {pages} {179} (\bibinfo {year} {1983})}\BibitemShut {NoStop}%
\bibitem [{\citenamefont {Barends}\ \emph {et~al.}(2013)\citenamefont {Barends}, \citenamefont {Kelly}, \citenamefont {Megrant}, \citenamefont {Sank}, \citenamefont {Jeffrey}, \citenamefont {Chen}, \citenamefont {Yin}, \citenamefont {Chiaro}, \citenamefont {Mutus}, \citenamefont {Neill}, \citenamefont {O'Malley}, \citenamefont {Roushan}, \citenamefont {Wenner}, \citenamefont {White}, \citenamefont {Cleland},\ and\ \citenamefont {Martinis}}]{barends_2013a}%
  \BibitemOpen
  \bibfield  {author} {\bibinfo {author} {\bibfnamefont {R.}~\bibnamefont {Barends}}, \bibinfo {author} {\bibfnamefont {J.}~\bibnamefont {Kelly}}, \bibinfo {author} {\bibfnamefont {A.}~\bibnamefont {Megrant}}, \bibinfo {author} {\bibfnamefont {D.}~\bibnamefont {Sank}}, \bibinfo {author} {\bibfnamefont {E.}~\bibnamefont {Jeffrey}}, \bibinfo {author} {\bibfnamefont {Y.}~\bibnamefont {Chen}}, \bibinfo {author} {\bibfnamefont {Y.}~\bibnamefont {Yin}}, \bibinfo {author} {\bibfnamefont {B.}~\bibnamefont {Chiaro}}, \bibinfo {author} {\bibfnamefont {J.}~\bibnamefont {Mutus}}, \bibinfo {author} {\bibfnamefont {C.}~\bibnamefont {Neill}}, \bibinfo {author} {\bibfnamefont {P.}~\bibnamefont {O'Malley}}, \bibinfo {author} {\bibfnamefont {P.}~\bibnamefont {Roushan}}, \bibinfo {author} {\bibfnamefont {J.}~\bibnamefont {Wenner}}, \bibinfo {author} {\bibfnamefont {T.~C.}\ \bibnamefont {White}}, \bibinfo {author} {\bibfnamefont {A.~N.}\ \bibnamefont {Cleland}},\ and\ \bibinfo {author} {\bibfnamefont {J.~M.}\ \bibnamefont
  {Martinis}},\ }\bibfield  {title} {\bibinfo {title} {Coherent {{Josephson Qubit Suitable}} for {{Scalable Quantum Integrated Circuits}}},\ }\href {https://doi.org/10.1103/PhysRevLett.111.080502} {\bibfield  {journal} {\bibinfo  {journal} {Physical Review Letters}\ }\textbf {\bibinfo {volume} {111}},\ \bibinfo {pages} {080502} (\bibinfo {year} {2013})}\BibitemShut {NoStop}%
\bibitem [{\citenamefont {Yariv}(1989)}]{yariv_1989}%
  \BibitemOpen
  \bibfield  {author} {\bibinfo {author} {\bibfnamefont {A.}~\bibnamefont {Yariv}},\ }\href@noop {} {\emph {\bibinfo {title} {Quantum electronics, 3rd Edition}}}\ (\bibinfo  {publisher} {Wiley},\ \bibinfo {address} {New York},\ \bibinfo {year} {1989})\BibitemShut {NoStop}%
\bibitem [{\citenamefont {Zhang}\ \emph {et~al.}(2023)\citenamefont {Zhang}, \citenamefont {He},\ and\ \citenamefont {Huang}}]{zhang_2023a}%
  \BibitemOpen
  \bibfield  {author} {\bibinfo {author} {\bibfnamefont {S.}~\bibnamefont {Zhang}}, \bibinfo {author} {\bibfnamefont {Y.}~\bibnamefont {He}},\ and\ \bibinfo {author} {\bibfnamefont {P.}~\bibnamefont {Huang}},\ }\bibfield  {title} {\bibinfo {title} {Acceptor-based qubit in silicon with tunable strain},\ }\href {https://doi.org/10.1103/PhysRevB.107.155301} {\bibfield  {journal} {\bibinfo  {journal} {Physical Review B}\ }\textbf {\bibinfo {volume} {107}},\ \bibinfo {pages} {155301} (\bibinfo {year} {2023})}\BibitemShut {NoStop}%
\bibitem [{\citenamefont {K{\"o}pf}\ and\ \citenamefont {Lassmann}(1992)}]{kopf_1992a}%
  \BibitemOpen
  \bibfield  {author} {\bibinfo {author} {\bibfnamefont {A.}~\bibnamefont {K{\"o}pf}}\ and\ \bibinfo {author} {\bibfnamefont {K.}~\bibnamefont {Lassmann}},\ }\bibfield  {title} {\bibinfo {title} {Linear {{Stark}} and nonlinear {{Zeeman}} coupling to the ground state of effective mass acceptors in silicon},\ }\href {https://doi.org/10.1103/PhysRevLett.69.1580} {\bibfield  {journal} {\bibinfo  {journal} {Physical Review Letters}\ }\textbf {\bibinfo {volume} {69}},\ \bibinfo {pages} {1580} (\bibinfo {year} {1992})}\BibitemShut {NoStop}%
\bibitem [{\citenamefont {Odeh}\ \emph {et~al.}(2023)\citenamefont {Odeh}, \citenamefont {Godeneli}, \citenamefont {Li}, \citenamefont {Tangirala}, \citenamefont {Zhou}, \citenamefont {Zhang}, \citenamefont {Zhang},\ and\ \citenamefont {Sipahigil}}]{odeh_2023}%
  \BibitemOpen
  \bibfield  {author} {\bibinfo {author} {\bibfnamefont {M.}~\bibnamefont {Odeh}}, \bibinfo {author} {\bibfnamefont {K.}~\bibnamefont {Godeneli}}, \bibinfo {author} {\bibfnamefont {E.}~\bibnamefont {Li}}, \bibinfo {author} {\bibfnamefont {R.}~\bibnamefont {Tangirala}}, \bibinfo {author} {\bibfnamefont {H.}~\bibnamefont {Zhou}}, \bibinfo {author} {\bibfnamefont {X.}~\bibnamefont {Zhang}}, \bibinfo {author} {\bibfnamefont {Z.-H.}\ \bibnamefont {Zhang}},\ and\ \bibinfo {author} {\bibfnamefont {A.}~\bibnamefont {Sipahigil}},\ }\href@noop {} {\bibinfo {title} {Non-{{Markovian}} dynamics of a superconducting qubit in a phononic bandgap}} (\bibinfo {year} {2023}),\ \Eprint {https://arxiv.org/abs/2312.01031} {arxiv:2312.01031} \BibitemShut {NoStop}%
\end{thebibliography}%

\end{document}


\title{Supplemental Material for \\``Acceptor-induced bulk dielectric loss in superconducting circuits on silicon''}

\author{Zi-Huai Zhang}
\thanks{Z.Z. and K.G. contributed equally.}
\affiliation{
Department of Electrical Engineering and Computer Sciences, University of California,  Berkeley, Berkeley, California 94720, USA
}
\affiliation{
 Materials Sciences Division, Lawrence Berkeley National Laboratory, Berkeley, California 94720, USA
}
\affiliation{
Department of Physics, University of California, Berkeley, Berkeley, California 94720, USA
}%

\author{Kadircan Godeneli}
\thanks{Z.Z. and K.G. contributed equally.}
\affiliation{
Department of Electrical Engineering and Computer Sciences, University of California,  Berkeley, Berkeley, California 94720, USA
}

\affiliation{
 Materials Sciences Division, Lawrence Berkeley National Laboratory, Berkeley, California 94720, USA
}%

\author{Justin He}
\affiliation{
Department of Electrical Engineering and Computer Sciences, University of California,  Berkeley, Berkeley, California 94720, USA
}

\author{Mutasem~Odeh}
\affiliation{
Department of Electrical Engineering and Computer Sciences, University of California,  Berkeley, Berkeley, California 94720, USA
}

\affiliation{
 Materials Sciences Division, Lawrence Berkeley National Laboratory, Berkeley, California 94720, USA
}%

\author{Haoxin Zhou}
\affiliation{
Department of Electrical Engineering and Computer Sciences, University of California,  Berkeley, Berkeley, California 94720, USA
}
\affiliation{
 Materials Sciences Division, Lawrence Berkeley National Laboratory, Berkeley, California 94720, USA
}
\affiliation{
Department of Physics, University of California, Berkeley, Berkeley, California 94720, USA
}%

\author{Srujan Meesala}
\affiliation{Institute for Quantum Information and Matter, California Institute of Technology, Pasadena, California 91125, USA.}
\author{Alp Sipahigil}
\email{Corresponding author: alp@berkeley.edu}

\affiliation{
Department of Electrical Engineering and Computer Sciences, University of California,  Berkeley, Berkeley, California 94720, USA
}
\affiliation{
 Materials Sciences Division, Lawrence Berkeley National Laboratory, Berkeley, California 94720, USA
}
\affiliation{
Department of Physics, University of California, Berkeley, Berkeley, California 94720, USA
}%
\date{\today}
\maketitle
\label{Sec:SI}

\tableofcontents
\setcounter{figure}{0}
\setcounter{section}{0}
\renewcommand{\thefigure}{S\arabic{figure}}
\renewcommand{\thetable}{\Roman{table}}
\renewcommand{\thesection}{\Roman{section}}

\section{\label{SI_Setup} SUPPLEMENTARY EXPERIMENTAL METHODS}
\subsection{Experimental setup}
The devices were measured using a vector network analyzer (Copper Mountain C1209) in a dilution refrigerator (Bluefors, BF-LD250) with a base temperature of $\sim 9$~mK at the mixing chamber. The output level of the network analyzer was sometimes attenuated (Vaunix LDA-908V or Mini-Circuits fixed attenuators) or amplified (Mini-Circuits ZX60-83LN-S+) to increase the dynamic range of the excitation power. The microwave excitation on the input line of the dilution refrigerator was attenuated at different stages (20~dB at 4~K, 20~dB at 100~mK, 20~dB at mixing chamber) using cryogenic attenuators (XMA Corporation, 2082-6418-20-CRYO Bulk) with a total attenuation of 60~dB. After attenuation, the excitation was filtered using an Eccosorb IR filter (QMC-CRYOIRF-002MF-S). We connected the input line and the output line to a pair of RF switches (Radiall R583423141) for asynchronous device multiplexing. For microwave detection, the signal was isolated with two isolators (Low Noise Factory LNF-CIC4$\_$8A) and filtered with a bandpass filter (Keenlion KBF-4/8-2S). After isolation and filtering, the signal was amplified with a HEMT amplifier (Low Noise Factory LNF-$\text{LNC4}\_\text{8C}$) at 4~K followed by a low noise amplifier at room temperature (Mini-Circuits ZX60-123LN-S+). 

The devices were packaged in a copper box and mounted vertically inside two cylindrical magnetic shields (Cryo-NETIC) for most measurements. For magnetic field dependence measurements, the devices were mounted outside the magnetic shields. A superconducting coil (field factor $\approx$300 Gauss/Amp for the experimental sample-coil distance) made from NbTi wires (Supercon SC-54S43-0.152mm) was mounted below the device to provide an in-plane magnetic field.

\subsection{\label{SI_fab}Device fabrication}
Prior to metallization, we cleaned the 6'' silicon wafers in piranha solution (5~min) and dilute (25:1) HF. The cleaned wafers were loaded into a DC sputtering system (MRC~943). The wafers were in-situ sputter-etch cleaned and sputtered with $\sim200$ nm of niobium. Coplanar-waveguide resonators were patterned with optical lithography using a maskless aligner (Heidelberg MLA150), and subsequently dry etched (Lam Research) with Cl$_2$ chemistry, followed by deionized water passivation and photoresist stripping (1165 Remover). The wafers were then diced (Disco DAD3240) into 1~cm $\times$ 1~cm chips with a protective photoresist layer. Upon photoresist stripping, the chips were dipped in buffered oxide etch (5:1) for $\sim 35$ min, wirebonded onto a PCB, packaged in a copper box, and cooled down in a dilution refrigerator with a base temperature of 8 mK.

\subsection{Silicon substrate for bulk loss characterization}
To study the excess dielectric loss induced from bulk crystalline defects, we fabricated superconducting resonators following procedures in Sec.~\ref{SI_fab} on prime-grade silicon wafers with different doping conditions. Each wafer was specified with a wafer resistivity. Secondary ion mass spectrometry (EAG Eurofins) was used to accurately determine the boron concentration in each wafer (detection limit [B] = $8\times 10^{11} \,\text{cm}^{-3}$). The specifications of different wafers are summarized here: 
\begin{table}[h]
\begin{center}
    \caption{Specifications of silicon wafers.}
    \begin{tabular}{l|l|l|l|l}
     Wafer Type & Vendor &Resistivity ($\Omega \cdot \text{cm}$) & Growth method & Boron concentration (cm$^{-3}$)\\
     \hline
      Undoped & NOVA Electronic Materials &$> 10000$ &  Float zone &  $<8\times 10^{11}$ \\
      Phosphorus doped & University Wafer &50 - 70  &  Float zone & $<8\times 10^{11}$\\
      Boron doped $\#$1& University Wafer & $> 2500$  &  Float zone & 4.6 $ \times \ 10^{12}$\\
      Boron doped $\#$2& University Wafer & $> 50$  &  Float zone & 6.7 $\times \ 10^{13}$\\
      Boron doped $\#$3& University Wafer & $8-13$  &  Float zone & 7.4 $\times \ 10^{14}$\\
      Boron doped $\#$4 & Waferpro & $1-10$  &  Czochralski & 2.5 $\times  \ 10^{15}$\\    
    \end{tabular}
\end{center} 
\end{table}

\subsection{Resonator design}
\begin{figure}[ht]
    \centering
    \includegraphics[width=172mm]{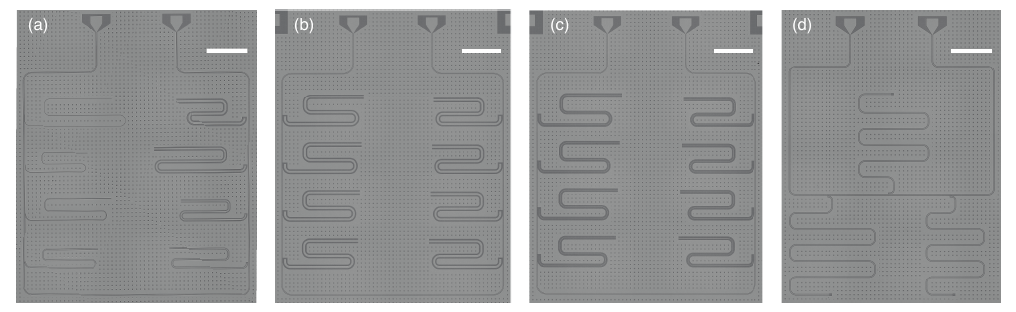}
    \caption{Optical images of samples with different resonator designs. (a) Resonator design with different surface participation ratio. (b) Resonator design with low surface participation ratio. (c) Resonator design with narrow trace and wide gap for magnetic field study. (d) Resonator design for magnetic field study of surface loss. Scale bar represents 1~mm.}
        \label{figDevice}
\end{figure}

Coplanar-waveguide resonators of different geometries were fabricated for different studies throughout the work. The optical images of samples with different resonator designs are shown in Fig.~\ref{figDevice}.

\textbf{Surface loss study}: For the loss study with different surface participation, we swept the trace and gap widths of quarter-wave resonators by more than an order of magnitude while maintaining a 50~$\Omega$ impedance (Fig.~\ref{figDevice}(a)). The resonators were designed to have a coupling quality factor of $\approx 300\times 10^3$. Their lengths were swept for frequency multiplexing from 4 to 7.5~GHz. We estimated the total surface participation ratio for each resonator geometry using the analytical formalism developed in Ref.~\cite{murray_2018a}, assuming a 3~nm surface contamination layer with a dielectric constant of 10. We also calculated the energy participation ratio in bulk silicon using electrostatic simulation with COMSOL Multiphysics. The parameters of the resonators are summarized in Table.~\ref{SPRTable}.
\begin{table}
\begin{center}
   \caption{Resonator parameters for geometric sweeping.}
    \begin{tabular}{l|l|l|l|l}
    \label{SPRTable}
     Resonator $\#$ & Trace width ($\mu$m) & Gap width ($\mu$m) & Surface participation ratio & Bulk participation ratio\\
     \hline
      1 & 2 &  1.42 & $7.75\times 10^{-3}$ & $9.04 \times 10^{-1}$\\
      2 & 4  &  2.78 &$4.27\times 10^{-3}$ & $9.12 \times 10^{-1}$\\
      3 & 8  &  5.4 &$2.35\times 10^{-3}$ & $9.16 \times 10^{-1}$\\
      4 & 13  &  8.8 &$1.52\times 10^{-3}$ & $9.18 \times 10^{-1}$\\
      5 & 20 &  13.5 &$1.03\times 10^{-3}$  & $9.19 \times 10^{-1}$\\
      6 & 30  &  20.25 &$7.2\times 10^{-4}$ & $9.20 \times 10^{-1}$\\  
      7 & 40  &  27 &$5.5\times 10^{-4}$ & $9.20 \times 10^{-1}$\\   
      8 & 50  &  33.75 &$4.5\times 10^{-4}$ & $9.20 \times 10^{-1}$\\   
    \end{tabular}
\end{center} 
\end{table}

\textbf{Bulk loss study}: To characterize the bulk dielectric loss, we fabricated quarter-wave resonators with trace width of 50~$\mu$m and gap width of 33.75~$\mu$m~(Fig.~\ref{figDevice}(b)). Each chip contains eight resonators, and the resonator lengths were swept for frequency multiplexing in a 1~GHz band centered around 6~GHz. The coupling quality factor was reduced to around $40 \times 10^3$ to improve the sensitivity to detect low-Q resonances on boron doped substrates.

\textbf{Magnetic field study (bulk loss)}: To probe the magnetic field response of loss on boron doped substrates, we fabricated quarter-wave resonators with a trace width of 10~$\mu$m and a gap width of 40~$\mu$m~(Fig.~\ref{figDevice}(c)). Such dimensions were chosen to mitigate vortex losses due to magnetic field while maintaining a relatively low surface participation. Each chip contains eight resonators, and the resonator lengths were swept for frequency multiplexing in a 1~GHz band centered around 6~GHz. The coupling quality factor was designed to be around $50 \times 10^3$.

\textbf{Magnetic field study (surface loss)}: To probe the magnetic field response of loss induced by surface TLSs on an undoped substrate, we fabricated half-wave resonators with a trace width of 16.3~$\mu$m and a gap width of 10~$\mu$m~(Fig.~\ref{figDevice}(d)). After fabrication, we intentionally introduce excessive amorphous TLS on the surface by drop casting Hydrogen Silsesquioxane. The chip contains three resonators with frequencies at 4.8, 5.7, and 6.6 GHz. The coupling quality factor was designed to be around $300 \times 10^3$. 

The different chip designs together with the measurements performed on each of them are summarized here:
\begin{table}[h]
\begin{center}
   \caption{Summary of parameters and measurements for different resonator designs.}
    \begin{tabular}{l|l|l|l|l}
     Chip design & Trace ($\mu$m) & Gap ($\mu$m) & Surface participation ratio & Figures\\
     \hline
      Surface loss study & $2-50$ &  $1.42 - 33.75$ & $7.75\times 10^{-3} - 4.5\times 10^{-4}$ & 2(b), S2\\
      Bulk loss study & 50  &  33.75 & $4.5\times 10^{-4}$& 2(c), 3, S3\\
      Magnetic field study (bulk loss) & 10  &  40 &$9.8\times 10^{-4}$ & 4(c,d,f), S4, S5, S10\\
      Magnetic field study (surface loss) & 16.3  &  10 & $1.3\times 10^{-3}$ & 4(e)\\
    \end{tabular}
\end{center}
\end{table}

\subsection{Resonator measurement and data analysis}
We probe the resonator loss by measuring the transmission through the microwave feedline on each chip. With the side-coupled configuration, the resonance of each resonator shows up as a narrow dip in $\lvert \text{S}_{21}\rvert$. With the large variation of resonator loss on different substrates, the measurement bandwidth for each resonator were adjusted with a total number of $500 - 1000$ sampling points using an IF bandwidth of 10~Hz. S$_{21}$ traces were averaged for noise reduction for low power measurements. 

The complex S$_{21}$ traces were fitted using: 
\begin{equation}
\text{S}_{21}(f) = a\times e^{i[\phi + 2\pi(f-f_{\text{start}})\tau]}\frac{1-Q/Q_{c}\times (1+2i\times df/f_0)}{1+2i\times Q(f-f_0)/f_0}
\end{equation}
where $a$ represents the baseline transmission away from the resonance, $\phi$ represents the phase offset, $f_{\text{start}}$ is the starting frequency of the measurement, $\tau$ represents the group delay, $Q$ represents the total quality factor, $Q_c$ represents the coupling quality factor, $f_0$ is the center frequency of the resonator, and $df$ is the asymmetric factor to account for impedance mismatch. 

The average photon number ($\langle n \rangle$) inside the resonator with input power $P_{\text{in}}$ can be estimated from the input-output theory~\cite{aspelmeyer_2014a}. Here we describe the formalism for a generic coupling scheme where a cavity couples to two waveguide ports. The dynamics of the cavity field $\hat{a}$ is:
\begin{equation}
\frac{d\hat{a}}{dt} = -i(\omega_0 - \omega)\hat{a} - \frac{\kappa_i + \kappa_{e1} + \kappa_{e2}}{2}\hat{a} + \sqrt{\kappa_{e1}}s_{1+} + \sqrt{\kappa_{e2}}s_{2+}
\end{equation}
where $\omega_0$ ($\omega$) denotes the cavity (drive) frequency, $\kappa_{e1}$ and $\kappa_{e2}$ denote the coupling rate from the cavity to the two waveguide ports, $s_{1+}$ ($s_{2+}$) denotes the input field from waveguide port 1 (port 2), and $\kappa_i$ denotes the intrinsic decay rate of the cavity, respectively. In the steady-state ($d\hat{a}/dt = 0$), the cavity field is:
\begin{equation}
\hat{a} =  \frac{\sqrt{\kappa_{e1}}s_{1+} + \sqrt{\kappa_{e2}}s_{2+}}{i(\omega_0 - \omega) + (\kappa_i + \kappa_{e1} + \kappa_{e2})/2}
\end{equation}

For our resonators with a side-coupled configuration and single-sided excitation, we have $\kappa_{e1} = \kappa_{e2} = \kappa_e /2$ and $s_{2+} = 0$. On resonance ($\omega_0 = \omega$), the steady state cavity field is:
\begin{equation}
\langle \hat{a} \rangle =  \frac{\sqrt{\kappa_{e}/2}s_{1+}}{(\kappa_i + \kappa_{e})/2}
\end{equation}
and we can convert input power ($P_{\text{in}} = \hbar \omega_0\lvert s_{1+} \rvert^2$) to average photon number ($\langle n \rangle$) following~\cite{bruno_2015}: 
\begin{equation}
\langle n \rangle = \lvert \langle \hat{a}\rangle \rvert^2 = \frac{2\kappa_e}{(\kappa_e+\kappa_i)^2} \frac{P_{\text{in}}}{\hbar\omega_0}=
\frac{2}{\hbar \omega_0^2} \frac{Q^2}{Q_e}P_{\text{in}}
\end{equation}
where $Q$ ($Q_e$) denotes the loaded (coupling) quality factor, and $P_{\text{in}}$ is the input power at the resonator. By measuring the loss through a closed loop of cables when the dilution fridge is at its base temperature, we estimated a total attenuation of 85~dB in the input microwave chain due to cryogenic attenuators, as well as attenuation from coaxial cables, and insertion loss from filters, RF switch, and PCB.

\section{\label{SI_Setup} SUPPLEMENTARY DATA AND MEASUREMENTS}
\subsection{Surface participation dependence of $Q_i$}
\begin{figure}[ht]
    \centering
    \includegraphics[width=129mm]{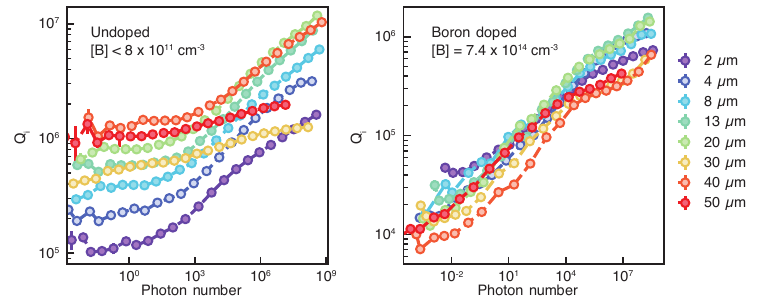}
    \caption{Power-dependent measurement on individual resonators on undoped silicon (left) and boron doped silicon (right) with varying surface participation. The legend denotes the signal trace width for each resonator.}
        \label{figSPR}
\end{figure}

In Fig.~2(b) of the main text, we presented the low-power $Q_i$ of resonators with different surface participation. The power dependent $Q_i$ measurements on each resonator are shown in Fig.~\ref{figSPR}.

\subsection{Doping dependence of $Q_i$}
\begin{figure}[ht]
    \centering
    \includegraphics[width=129mm]{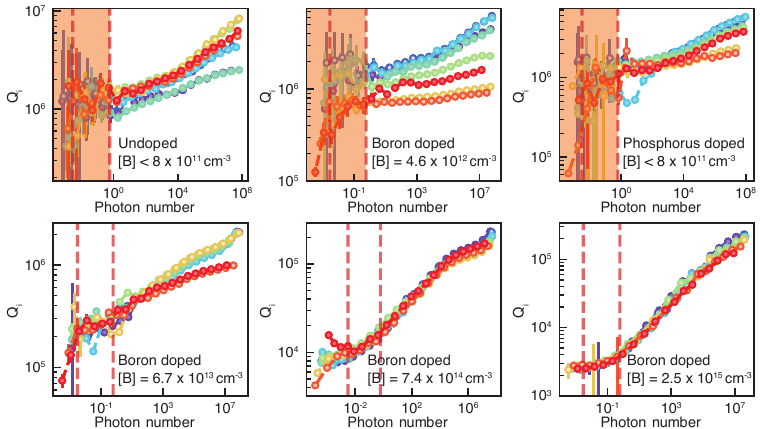}
    \caption{Power dependence of $Q_i$ for individual resonators on different silicon substrates. For devices with low loss, accurately extracting high $Q_i$ at low powers is challenging due to the extremely overcoupled resonator design ($Q_e \ll Q_i$). The orange shaded regions indicate such low power regions with fitting results dominated by noise. The dashed vertical lines indicate $\langle n \rangle \approx 0.003$ and $\langle n \rangle \approx 0.6$, respectively.}
        \label{figDoping}
\end{figure}

In Fig.~3 of the main text, we presented the average low-power $Q_i$ of resonators fabricated on different silicon substrates. The power dependent $Q_i$ measurements on each resonator are shown in Fig.~\ref{figDoping}. 

\subsection{Magnetic field dependence of acceptor-induced dielectric loss}
\begin{figure}[ht]
    \centering
    \includegraphics[width=129mm]{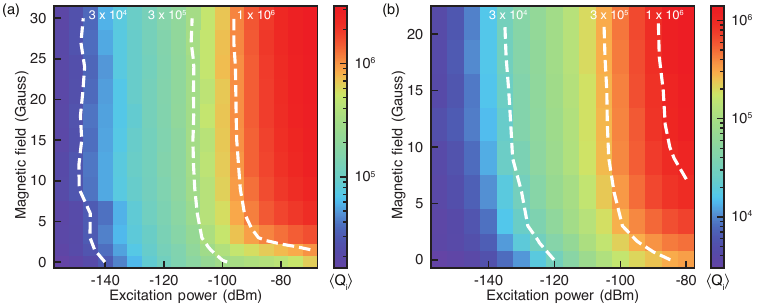}
    \caption{Magnetic field dependence of $Q_i$ on resonators fabricated on (a) a float zone grown substrate with [B] = $7.4\times 10^{14}\,\text{cm}^{-3}$, and (b) a Czochralski grown substrate with [B] = $2.5\times 10^{15}\, \text{cm}^{-3}$. The dashed lines are interpolated contour lines for $Q_i = 3\times 10^4$, $Q_i = 3\times 10^5$, and $Q_i = 1\times 10^6$ as guides for eye.}
        \label{figMag}
\end{figure}

We performed magnetic field dependence study with resonators on two types of boron doped wafers. The results presented in Fig.~4 of the main text were measured on boron doped wafer $\#$3 (Float zone, [B] = $7.4\times 10^{14}\,\text{cm}^{-3}$). Additional measurements were performed on resonators fabricated on boron doped wafer $\#$4 (Czochralski, [B] = $2.5\times 10^{15}\,\text{cm}^{-3}$). The magnetic field dependence of $Q_i$ is summarized in Fig.~\ref{figMag}. The boron-induced dielectric loss is reduced upon applying a small magnetic field for both types of substrates.

\begin{figure}[ht]
    \centering
    \includegraphics[width=172mm]{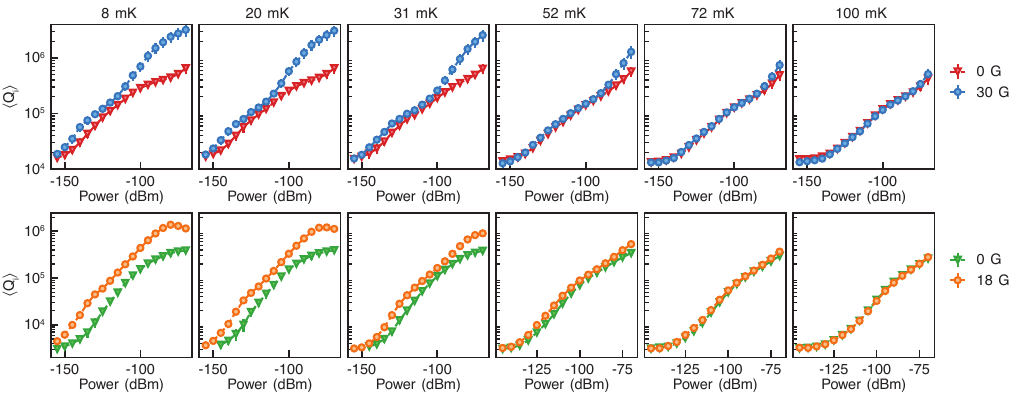}
    \caption{Temperature dependence of the magnetic field response on resonators fabricated on a float zone grown substrate with [B] = $7.4\times 10^{14}\, \text{cm}^{-3}$ (top row), and a Czochralski grown substrate with [B] = $2.5\times 10^{15}\, \text{cm}^{-3}$ (bottom row). The temperatures are specified at the mixing chamber plate.}
        \label{figMT}
\end{figure}

The magnetic field induced loss reduction on boron doped silicon shows a strong temperature dependence. At higher temperatures, the magnetic field response is suppressed. The temperature-dependent magnetic field response on resonators fabricated on two types of boron doped substrates is summarized in Fig.~\ref{figMT}.

\section{SUPPLEMENTARY ANALYSIS}
\subsection{Electronic structure of acceptors and donors}
Donor and acceptor defects in silicon have been widely studied for quantum applications. Here, we compare the electronic structure of boron acceptors and phosphorus donors, and discuss their impact on bulk dielectric loss in silicon.

\begin{figure}[ht]
    \centering
    \includegraphics[width=172mm]{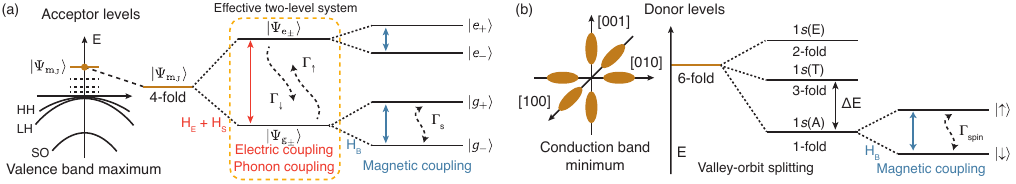}
    \caption{Electronic structures of (a) acceptor and (b) donor defects.}
        \label{figDA}
\end{figure}

The electronic structure of acceptor defects inherits the structure of the valence band maximum of silicon with a two-fold orbital degeneracy and strong spin-orbit coupling~(Fig.~\ref{figDA}(a))~\cite{bir_pikus_1974}. The acceptor based spin-orbit system couples strongly to environmental perturbations from strain ($H_S$) and electric ($H_E$) fields. Therefore, the spin-orbit system based on acceptors can be described as an effective two-level system causing dielectric loss. In the context of spin qubits, these strong environmental sensitivities negatively influence the lifetime and coherence of acceptor spins~\cite{kobayashi_2021a}. Therefore, acceptor defects normally cannot constitute a good qubit system. The extreme sensitivity of acceptor defects to their local environment also results in a broad inhomogeneous distribution of the orbital splittings. This distribution can be particularly large in typical nanofabricated devices due to additional lattice strain and electric fields. 

On the other hand, the electronic structure of donor spins inherits the structure of conduction band minimum of silicon with a six-fold valley degeneracy~(Fig.~\ref{figDA}(b)). With valley-orbit splitting, the $1s$ orbital is split into $1s(\text{A})$, $1s(\text{T})$, and $1s(\text{E})$ orbitals. The lowest orbital, $1s(\text{A})$, for phosphorus defects is 11.8~meV ($\approx$3 THz, or an equivalent thermal energy of $\approx$150~K) below the higher $1s(\text{T})$ orbital~\cite{AGGARWAL1964163}. Therefore, at cryogenic temperatures, phosphorus defects effectively provide an isolated spin degree of freedom in a singlet orbital. The lack of strong spin-orbit coupling makes phosphorus spins a good qubit system with long lifetimes and coherence times~\cite{muhonen_2014,wolfowicz_2013,mansir_2018}. Regarding dielectric loss from phosphorus, no loss is expected from orbital transitions due to the large energy mismatch between the orbital splitting ($\approx$3 THz) and microwave frequencies of interest for typical superconducting devices ($<$10~GHz). We note that donor-induced loss can occur at the spin resonance frequency due to magnetic coupling between donor spins and microwave resonators. However, such effects are only significant at a large magnetic field for phosphorus ($g=2$) spins ($\approx$2000 Gauss for a 5.6~GHz spin resonance). Additionally, such donor spin-induced loss will be a narrow-band effect due to the narrow inhomogeneous distribution of the spin resonance frequencies. 

\subsection{Strain at the metal-silicon interface}

In our experiment, we observe a broadband ($> 1$~GHz) boron-induced dielectric loss at all samples. This indicates that the orbital transition frequency of boron acceptors has a large inhomogeneous distribution, even though no intentional strain or electric field biases were applied. Such a large inhomogeneous distribution can originate from strain at the metal-silicon interface due to differential thermal contraction of the thin metal film and silicon substrate upon cooling the sample from the film deposition temperature to cryogenic temperatures~\cite{pla_2018}. 

At room temperature, we characterize the niobium film stress to be a compressive stress of 150~MPa on a prime-grade silicon wafer. Using the film stress at room temperature as the initial condition, we perform finite-element simulation of strain induced by differential thermal contraction using COMSOL Multiphysics. For niobium, we used a Young's modulus of 105~GPa, a Poisson ratio of 0.4, and a temperature dependent coefficient of thermal expansion~\cite{white_1962}. For silicon, we used an anisotropic model for its elasticity matrix ($D_{11} =$~166~GPa, $D_{12} = $~64~GPa, $D_{44}=$~80~GPa), and a temperature dependent coefficient of thermal expansion~\cite{swenson_1983}.

\begin{figure}[ht]
    \centering
    \includegraphics[width=172mm]{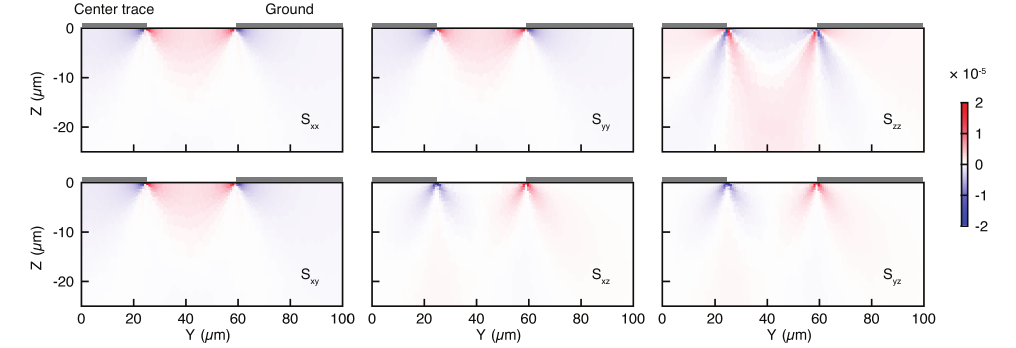}
    \caption{Strain distribution at the metal-silicon interface from differential thermal contraction. Half of the cross-section is shown due to the symmetry of the geometry. The strain tensor components are given in the cubic [100] basis (xyz). The Z direction is aligned with the silicon [001] direction while the Y direction is aligned with silicon [110] direction, based on the silicon wafer orientation and the sample cut directions.}
        \label{figStrain}
\end{figure}

In Fig.~\ref{figStrain}, we show the simulated strain distribution for the low-surface participation ratio resonator (trace width of 50~$\mu$m, gap width of 33.75~$\mu$m) at 10~mK. Near the edges of the metal film, the strain can be as high as $2\times 10^{-5}$, and can extend deep into the bulk. Therefore, this thermal strain is a possible origin of the broad inhomogeneous distribution of boron orbital splittings. We note that our simulation uses linear elastic approximation for thin-film niobium, and does not consider its yield behavior (transition from elastic behavior to plastic behavior). Therefore, the simulated thermal strain only provides a simple, order-of-magnitude estimation for a rather complicated strain environment.

\subsection{First-principles estimation of acceptor-induced loss}
Before providing a first-principles estimation of acceptor-induced loss, we first qualitatively estimate the loss from amorphous TLSs and from boron defects based on their effective density and the respective energy participation ratio for our low surface participation ratio resonator. For TLSs, we assume a density of $10^{14}\ \text{cm}^{-3}\text{GHz}^{-1}$~\cite{barends_2013a}. For boron doped silicon, the effective density of boron defects depends on the inhomogeneous distribution of boron resonances which strongly depends on static stray strain and electric field in the silicon substrate. For a doping concentration of $1.0\times 10^{15} \text{cm}^{-3}$ ($15~\Omega\cdot \text{cm}$) and a conservative inhomogeneous distribution of 100~GHz, the effective density is $10^{13}\ \text{cm}^{-3}\text{GHz}^{-1}$. The energy participation ratio for surface TLSs in the low surface participation ratio design is $4.5\times 10^{-4}$ while the energy participation ratio for boron in bulk defects silicon is $\epsilon_r/(1+\epsilon_r) \approx 0.92$, where $\epsilon_r$ is the dielectric constant of silicon. With these parameters and assuming similar electric dipole moments for TLSs and boron defects, the acceptor-induced loss will be 200 times more compared to TLS-induced loss. 

\begin{figure}[ht]
    \centering
    \includegraphics[width=172mm]{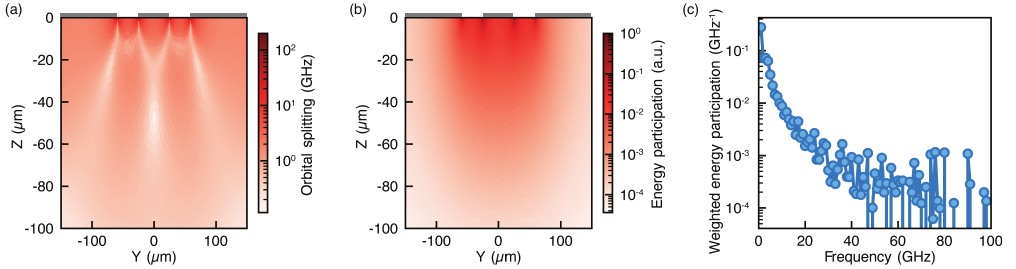}
    \caption{(a) Spatial distribution of boron orbital splittings due to thermal strain at metal-silicon interface. (b) Spatial distribution of electrical energy in silicon. (c) Weighted energy participation ($P(f_0)$) of boron acceptors as a function of boron orbital splittings.}
        \label{figLinewidth}
\end{figure}

From the above discussion, we see that acceptor-induced loss can far exceed TLS-induced loss at high doping concentrations. The loss estimation depends strongly on the inhomogeneous distribution of boron resonances. We calculate the spatial distribution of boron orbital splittings for the low surface participation ratio resonator using the thermal strain and boron Hamiltonian (Eq.~(2-4) in the main text, Fig.~\ref{figLinewidth}(a)). The spatial distribution shows that the orbital splitting can be as high as 100~GHz near the surface, and strain-induced splittings in the boron ensemble can extend deep in the bulk. The energy participation in the bulk is also spatially dependent (Fig.~\ref{figLinewidth}(b)). Therefore, it is critical to include spatial dependences of both energy participation and boron resonances to estimate the boron-induced dielectric loss.

Defect-induced dielectric loss can be estimated from the first principles using the imaginary part of atomic susceptibility: $\chi^{\prime \prime}(f,f_0) = \mu^2Ng(f,f_0)/(2\epsilon_0\hbar)$, where $\mu$ is the electric transition dipole moment, $\epsilon_0$ is the vacuum permittivity, $N$ represents the volumetric dopant density, and $g(f,f_0)$ represents the normalized Lorentzian lineshape of the atomic transition with a resonance frequency $f_0$~\cite{yariv_1989}.

For a generic device geometry, the observed loss tangent at frequency $f$ is a participation-ratio averaged loss tangent: 
\begin{equation}
\label{tanloss1}
\overline{\text{tan}\delta_0}(f) = \int_V p(r)\text{tan}\delta_0(f,r)dV
\end{equation}
where $p(r)$ represents the volumetric density of bulk energy participation ratio ($\int_V p(r)dV \approx 0.92$) and $\text{tan}\delta_0(f,r)$ denotes a spatially varying loss tangent as a function of spatial location $r$. 

The spatial dependence in loss tangent arises from the spectral inhomogeneity of boron resonances due to the spatially varying strain. The defect-induced loss tangent can be calculated from the imaginary part of atomic susceptibility: 
\begin{equation}
\label{tanloss2}
\text{tan}\delta_0(f,r) = \int_{f_0}\frac{q(f_0,r)\chi^{\prime \prime}(f,f_0)}{\epsilon_r}df_0
\end{equation}
where $q(f_0,r)$ denotes the spectral distribution of boron resonances at location $r$ ($\int q(f_0,r)df_0 = 1$), and $\epsilon_r$ represents the dielectric constant of silicon. Combining Eq.~\ref{tanloss1} and Eq.~\ref{tanloss2}, the observed loss tangent is represented by a double integral over space and frequency: 
\begin{equation}
\overline{\text{tan}\delta_0}(f) = \int_V\int_{f_0}\frac{p(r)q(f_0,r)\chi^{\prime \prime}(f,f_0)}{\epsilon_r}df_0 dV
\end{equation}
We can evaluate the volume integral of $p(r)q(f_0,r)$ numerically using a uniform sampling of the spatial distribution of boron orbital splittings (Fig.~\ref{figLinewidth}(a)) and energy participation (Fig.~\ref{figLinewidth}(b)): 
\begin{equation}
P(f_0) = \int_V p(r)q(f_0,r)dV = \sum_{V}p(r)\Delta V \times q(f_0,r)
\end{equation}
where $p(r)\Delta V$ represents the energy participation ratio in volume $\Delta V$. For each small volume $\Delta V$, we assume all the boron acceptors inside share the same orbital splitting. Therefore, we can numerically evaluate the weighted energy participation $P(f_0)$ in the unit of $\text{GHz}^{-1}$ as a function of frequency $f_0$ (Fig.~\ref{figLinewidth}(c)). Near our device frequencies ($\approx 6$~GHz), $P(f_0 = \text{6 GHz}) \approx 0.03/\text{GHz}$. Physically, $P(f_0 = \text{6 GHz}) \approx 0.03/\text{GHz}$ means 3\% of the total electrical energy is experiencing dielectric loss from boron defects distributed in a 1~GHz band around 6~GHz.

In the low power limit free from atomic saturation effects, the dielectric loss tangent is:
\begin{equation}\overline{\text{tan}\delta_0}(f) = \int \chi^{\prime \prime}(f,f_0)P(f_0)df_0/\epsilon_r
\end{equation}
With a smooth, slowly varying $P(f_0)$, we have:
\begin{equation}
\overline{\text{tan}\delta_0}(f) = \int \chi^{\prime\prime}(f,f_0)P(f_0)df_0 /\epsilon_r = P(f)\int \chi^{\prime \prime}(f,f_0)df_0 /\epsilon_r = \frac{\mu^2 P(f)}{2\epsilon_0\epsilon_r\hbar}N
\end{equation}
For our silicon wafer with the highest boron concentration ($N = 2.5 \times  10^{15} $~cm$^{-3}$), using $P(\nu = \text{6 GHz}) = 0.03/\text{GHz}$ and $\mu = \sqrt{1/3}\times 0.26$~D, we estimate the boron-limited quality factor ($1/\overline{\text{tan}\delta_0}(f = \text{6 GHz})$) to be approximately $1\times 10^6$. The factor of $\sqrt{1/3}$ in $\mu$ accounts for the random alignment of the electric dipole moment with respect to the direction of the electric field from the resonator. Interestingly, this first-principles estimation of quality factor is about 400 times higher than our experimental observation. Further investigation is needed to resolve the discrepancy. The difference may be accounted by a combination of (1) a different inhomogeneous distribution from unaccounted strain contributions beyond the simplified thermal strain, (2) the modification of electric dipole moment with the static stray strain~\cite{zhang_2023a}, (3) the electric dipole moment being much higher than the literature values in Ref.~\cite{kopf_1992a}. We note that the electric dipole moment was estimated from dielectric loss measurement on unstrained silicon in Ref.~\cite{kopf_1992a}. In future work, it would be interesting to perform a Stark shift measurement on single boron acceptors to extract the electric field coupling coefficients more accurately.

\subsection{Theoretical analysis of saturation behavior for a four-level system}\label{twolevelME} 
\begin{figure}[ht]
    \centering
    \includegraphics[width=86mm]{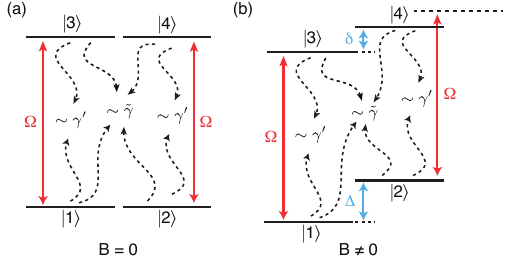}
    \caption{Level diagram with a resonant single-tone drive for a four-level system (a) at zero magnetic field and (b) at a non-zero magnetic field. $\Omega$ denotes the driving strength. $\delta$ and $\Delta$ denote the Zeeman splittings within the two orbital branches. $\gamma^\prime$ and $\tilde{\gamma}$ denote the relevant spontaneous decay rates. The actual orbital decay rates also include thermally-assisted processes.}
        \label{leveldiagram}
\end{figure}

Defect-induced dielectric loss can be saturated with microwave excitation and/or thermal excitation. The saturation behavior depends strongly on the detailed electronic structure of the defect. For the case of boron acceptors, the electronic structure under static crystal strain is a four-level system with a two-fold degeneracy in the two orbital branches~(Fig.~\ref{leveldiagram}(a)). Under a small magnetic field, the level degeneracy is lifted, and the electronic structure contains four distinct levels with generalized spin sublevels~(Fig.~\ref{leveldiagram}(b)). In this case, the generalized spin states can have long lifetimes~\cite{kobayashi_2021a}. We observe that applying a small magnetic field can lead to strong reduction of boron-induced loss (Fig.~4 in the main text and Fig.~\ref{figMag}). At the same time, the magnetic response disappears at elevated temperatures (Fig.~\ref{figMT}). In this section, we perform theoretical analysis of saturation power of a four-level system under different conditions. 

The electric and strain couplings of boron levels depend strongly on the details of static crystal strain. To reduce the complexity of the problem, our analysis is performed assuming static tensile strain with a single tone drive. Under tensile strain, only $\ket{1} \leftrightarrow \ket{3}$ and $\ket{2} \leftrightarrow \ket{4}$ transitions are electric dipole allowed. At the same time, $\ket{1} \leftrightarrow \ket{3}$ and $\ket{2} \leftrightarrow \ket{4}$ ($\ket{1} \leftrightarrow \ket{4}$ and $\ket{2} \leftrightarrow \ket{3}$) transitions share the same spontaneous phonon emission rate $\gamma^\prime$ ($\tilde{\gamma}$).

\textbf{Four-level system at zero magnetic field}: The saturation behavior of defect-induced dielectric loss can be related to the population difference of the ground and excited states in steady state~\cite{yariv_1989}. The Hamiltonian of a four-level system (Fig.~\ref{leveldiagram}(a)) in the rotating frame of a resonant drive is: 
\begin{equation}
    H/\hbar = \Omega^{*} (\ket{1} \bra{3} + \ket{2} \bra{4}) + \Omega (\ket{3} \bra{1} + \ket{4} \bra{2})
\end{equation}
The system evolves following the Lindblad master equation:
\begin{equation}\label{mastereqn}
    \dot{\rho} = -\frac{i}{\hbar}[H,\rho]+\sum_{i}{\mathcal{D}[C_i]\rho}
\end{equation}
with the superoperator defined as $\mathcal{D}[C] \rho= C \rho C^\dagger - \frac{1}{2}(C^\dagger C \rho + \rho C^\dagger C)$. There are eight collapse operators denoting the phonon assisted relaxation paths: $C_1 = \sqrt{\bar{n}\gamma^\prime} \ket{3}\bra{1}$, $C_2 = \sqrt{(\bar{n}+1)\gamma^\prime} \ket{1}\bra{3}$, $C_3 = \sqrt{\bar{n}\tilde{\gamma}} \ket{3}\bra{2}$, $C_4 = \sqrt{(\bar{n}+1)\tilde{\gamma}} \ket{2}\bra{3}$, $C_5 = \sqrt{\bar{n}\tilde{\gamma}} \ket{4}\bra{1}$, $C_6 = \sqrt{(\bar{n}+1)\tilde{\gamma}} \ket{1}\bra{4}$, $C_7 = \sqrt{\bar{n}\gamma^\prime} \ket{4}\bra{2}$ and $C_8 = \sqrt{(\bar{n}+1)\gamma^\prime} \ket{2}\bra{4}$. $\bar{n}$ denotes the thermal phonon occupancy, and $\gamma^\prime$ and $\tilde{\gamma}$ are the spontaneous phonon emission rates from $\ket{3} \rightarrow \ket{1}$ ($\ket{4} \rightarrow \ket{2}$) and $\ket{3} \rightarrow \ket{2}$ ($\ket{4} \rightarrow \ket{1}$), respectively. Based on the master equation, the equations of motion describing the evolution of population and coherences can be written as

\begin{equation}
\dot{\rho}_{11} = -(\gamma^\prime + \tilde{\gamma})\bar{n}\rho_{11} + (1+\bar{n})(\gamma^\prime \rho_{33} + \tilde{\gamma}\rho_{44})+ i\Omega \rho_{13}- i\Omega^*\rho_{31}
\end{equation}
\begin{equation}
\dot{\rho}_{22} = -(\gamma^\prime + \tilde{\gamma})\bar{n}\rho_{22} + (1+\bar{n})(\tilde{\gamma} \rho_{33} + \gamma^\prime \rho_{44}) + i\Omega\rho_{24}-i\Omega^*\rho_{42}
\end{equation}
\begin{equation}
\dot{\rho}_{33} = -(\gamma^\prime + \tilde{\gamma})(1+\bar{n})\rho_{33} + \bar{n}(\tilde{\gamma} \rho_{22} + \gamma^\prime \rho_{11}) + i\Omega^*\rho_{31}-i\Omega\rho_{13}
\end{equation}
\begin{equation}
\dot{\rho}_{13} = -(\gamma^\prime + \tilde{\gamma})(1/2+\bar{n})\rho_{13} + i\Omega^*(\rho_{11}-\rho_{33})
\end{equation}
\begin{equation}
\dot{\rho}_{24} = -(\gamma^\prime + \tilde{\gamma})(1/2+\bar{n})\rho_{24} + i\Omega^*(\rho_{22}-\rho_{44})
\end{equation}
\begin{equation}
\dot{\rho}_{12} = -(\gamma^\prime + \tilde{\gamma})\bar{n}\rho_{12} + i(\Omega\rho_{14}-\Omega^*\rho_{32})
\end{equation}
\begin{equation}
\dot{\rho}_{14} = -(\gamma^\prime + \tilde{\gamma})(1/2+\bar{n})\rho_{14} + i\Omega^*(\rho_{12}-\rho_{34})
\end{equation}
\begin{equation}
\dot{\rho}_{23} = -(\gamma^\prime + \tilde{\gamma})(1/2+\bar{n})\rho_{23} + i\Omega^*(\rho_{21}-\rho_{43})
\end{equation}
\begin{equation}
\dot{\rho}_{34} = -(\gamma^\prime + \tilde{\gamma})(1+\bar{n})\rho_{34} + i\Omega^*\rho_{32}- i\Omega\rho_{14}
\end{equation}

We note that $\rho_{ij} = \rho_{ji}^*$ and $\rho_{11}+\rho_{22}+\rho_{33}+\rho_{44}=1$. Population in the steady state ($\dot{\rho}_{ij}=0$ and $\dot{\rho}_{ii}=0$) can be solved as
\begin{equation}
\rho_{11} = \rho_{22} = \frac{1}{2}\times \frac{(1+\bar{n})(\gamma^\prime + \tilde{\gamma})+R}{(1+2\bar{n})(\gamma^\prime + \tilde{\gamma}) + 2R}
\end{equation}
\begin{equation}
\rho_{33} = \rho_{44} = \frac{1}{2}\times \frac{\bar{n}(\gamma^\prime + \tilde{\gamma}) + R}{(1+2\bar{n})(\gamma^\prime + \tilde{\gamma})+ 2R}
\end{equation}
where $R = 4|\Omega|^2/[(1+2\bar{n})(\gamma^\prime + \tilde{\gamma})]$ is a pumping rate. Note that if $R \gg (\gamma^\prime + \tilde{\gamma})(\bar{n}+\frac{1}{2})$, we achieve equal population in the ground and excited states ($\rho_{11} = \rho_{22} = \rho_{33} = \rho_{44} = 1/4$).

In order to quantify the critical Rabi frequency, we define the saturation condition as $\rho_{11}-\rho_{33} = (\rho_{11}-\rho_{33})_0/\sqrt{2}$, where $(\rho_{11}-\rho_{33})_0 = 1/(4\bar{n}+2)$ is the initial population difference at thermal equilibrium under no drive. The critical Rabi frequency that achieves this condition is
\begin{equation}\label{twolevelsat}
    |\Omega_c(B=0)|^2 = \frac{(\gamma^\prime + \tilde{\gamma})^2}{8}(2\bar{n}+1)^2(\sqrt{2}-1)
\end{equation}
We note that Eq.~\ref{twolevelsat} also holds for a pure two-level system with a spontaneous decay rate of $\gamma^\prime + \tilde{\gamma}$. This validates the two-level system approximation when no magnetic field is applied.

\textbf{Four-level system in a magnetic field}: Under a magnetic field, the electronic structure of boron is a four level system with different Zeeman splittings ($\Delta$ and $\delta$) in the two orbital branches. The Hamiltonian of such four-level system~(Fig.~\ref{leveldiagram}(b)) in the rotating frame of the drive can be written as:
\begin{equation}
    H/\hbar = \Delta \ket{2}\bra{2} + \delta \ket{4}\bra{4}+ \Omega (\ket{3} \bra{1} + \ket{4} \bra{2}) + \Omega^{*} (\ket{1} \bra{3} + \ket{2} \bra{4})
\end{equation}
Similar to the zero magnetic field case, there are eight collapse operators denoting the phonon assisted decay paths: $C_1 = \sqrt{\bar{n}\gamma^\prime} \ket{3}\bra{1}$, $C_2 = \sqrt{(\bar{n}+1)\gamma^\prime} \ket{1}\bra{3}$, $C_3 = \sqrt{\bar{n}\tilde{\gamma}} \ket{3}\bra{2}$, $C_4 = \sqrt{(\bar{n}+1)\tilde{\gamma}} \ket{2}\bra{3}$, $C_5 = \sqrt{\bar{n}\tilde{\gamma}} \ket{4}\bra{1}$, $C_6 = \sqrt{(\bar{n}+1)\tilde{\gamma}} \ket{1}\bra{4}$, $C_7 = \sqrt{\bar{n}\gamma^\prime} \ket{4}\bra{2}$ and $C_8 = \sqrt{(\bar{n}+1)\gamma^\prime} \ket{2}\bra{4}$. $\bar{n}$ denotes the thermal phonon occupancy, and $\gamma^\prime$ and $\tilde{\gamma}$ are the spontaneous emission rates from $\ket{3} \rightarrow \ket{1}$ ($\ket{4} \rightarrow \ket{2}$) and $\ket{3} \rightarrow \ket{2}$ ($\ket{4} \rightarrow \ket{1}$), respectively. The spin lifetime within each orbital branch is much longer than orbital lifetimes~\cite{kobayashi_2021a}. Therefore, we neglect the spin relaxation process between $\ket{2} \leftrightarrow \ket{1}$.

Assuming radiative decoherence only, we obtain the equations of motion for the populations and coherences using the Lindblad Master Equation (Eq.~\ref{mastereqn}):
\begin{equation}
\dot{\rho}_{11} = -(\gamma^\prime + \tilde{\gamma})\bar{n}\rho_{11} + (1+\bar{n})(\gamma^\prime \rho_{33} + \tilde{\gamma}\rho_{44})+ i\Omega\rho_{13}-i\Omega^*\rho_{31}
\end{equation}
\begin{equation}
\dot{\rho}_{22} = -(\gamma^\prime + \tilde{\gamma})\bar{n}\rho_{22} + (1+\bar{n})(\tilde{\gamma} \rho_{33} + \gamma^\prime \rho_{44}) + i\Omega\rho_{24}-i\Omega^*\rho_{42}
\end{equation}
\begin{equation}
\dot{\rho}_{33} = -(\gamma^\prime + \tilde{\gamma})(1+\bar{n})\rho_{33} + \bar{n}(\tilde{\gamma} \rho_{22} + \gamma^\prime \rho_{11}) - i\Omega\rho_{13} + i\Omega^*\rho_{31}
\end{equation}
\begin{equation}
\dot{\rho}_{13} = -(\gamma^\prime + \tilde{\gamma})(1/2+\bar{n})\rho_{13} + i\Omega^*(\rho_{11}-\rho_{33})
\end{equation}
\begin{equation}
\dot{\rho}_{24} = -(\gamma^\prime + \tilde{\gamma})(1/2+\bar{n})\rho_{24} + i(\delta-\Delta)\rho_{24}+ i\Omega^*(\rho_{22}-\rho_{44})
\end{equation}
\begin{equation}
\dot{\rho}_{12} = -(\gamma^\prime + \tilde{\gamma})\bar{n}\rho_{12} + i\Delta\rho_{12}+ i\Omega\rho_{14}-i\Omega^*\rho_{32}
\end{equation}
\begin{equation}
\dot{\rho}_{14} = -(\gamma^\prime + \tilde{\gamma})(1/2+\bar{n})\rho_{14} + i\delta\rho_{14}+ i\Omega^*(\rho_{12}-\rho_{34})
\end{equation}
\begin{equation}
\dot{\rho}_{23} = -(\gamma^\prime + \tilde{\gamma})(1/2+\bar{n})\rho_{23} - i\Delta\rho_{23}+ i\Omega^*(\rho_{21}-\rho_{43})
\end{equation}
\begin{equation}
\dot{\rho}_{34} = -(\gamma^\prime + \tilde{\gamma})(1+\bar{n})\rho_{34} + i\delta\rho_{34}+ i\Omega^*\rho_{32}-i\Omega\rho_{14}
\end{equation}

For large detuning ($\lvert \Delta - \delta \rvert \gg \lvert \Omega \rvert$), we can ignore the coherences between transitions detuned from the drive ($\rho_{24}\approx 0$, $\rho_{23}\approx 0$, $\rho_{14}\approx 0$, $\rho_{12}\approx 0$). The steady state populations and coherences under drive can then be evaluated with $\dot{\rho}_{11} = \dot{\rho}_{22}= \dot{\rho}_{33}= \dot{\rho}_{13} = 0$ (steady state condition) and $\rho_{11} + \rho_{22}+\rho_{33}+\rho_{44} = 1$ (conservation of population):
\begin{equation}
\rho_{11} = \frac{\bar{n}(1+\bar{n})[2(1+\bar{n})\gamma^\prime(\gamma^\prime + \tilde{\gamma})+R(2\gamma^\prime + \tilde{\gamma})]}{R[1+8\bar{n}(1+\bar{n})]\gamma^\prime+R(1+2\bar{n})^2\tilde{\gamma}+4\bar{n}(1+\bar{n})(1+2\bar{n})\gamma^\prime(\gamma^\prime+\tilde{\gamma})}
\end{equation}

\begin{equation}
\rho_{22} = \frac{(1+\bar{n})[2\bar{n}(1+\bar{n})\gamma^\prime(\gamma^\prime+\tilde{\gamma})+R(\gamma^\prime+2\bar{n}\gamma^\prime +\tilde{\gamma}+\bar{n}\tilde{\gamma})]}{R[1+8\bar{n}(1+\bar{n})]\gamma^\prime+R(1+2\bar{n})^2\tilde{\gamma}+4\bar{n}(1+\bar{n})(1+2\bar{n})\gamma^\prime(\gamma^\prime+\tilde{\gamma})}
\end{equation}

\begin{equation}
\rho_{33} = \frac{\bar{n}(1+\bar{n})[2\bar{n}\gamma^\prime(\gamma^\prime+\tilde{\gamma})+R(2\gamma^\prime + \tilde{\gamma})]}{R[1+8\bar{n}(1+\bar{n})]\gamma^\prime+R(1+2\bar{n})^2\tilde{\gamma}+4\bar{n}(1+\bar{n})(1+2\bar{n})\gamma^\prime(\gamma^\prime+\tilde{\gamma})}
\end{equation}

\begin{equation}
\rho_{44} = \frac{\bar{n}[2\bar{n}(1+\bar{n})\gamma^\prime(\gamma^\prime + \tilde{\gamma})+R(\gamma^\prime + 2\bar{n}\gamma^\prime + \bar{n}\tilde{\gamma})]}{R[1+8\bar{n}(1+\bar{n})]\gamma^\prime+R(1+2\bar{n})^2\tilde{\gamma}+4\bar{n}(1+\bar{n})(1+2\bar{n})\gamma^\prime(\gamma^\prime+\tilde{\gamma})}
\end{equation}
where $R = 4|\Omega|^2/[(1+2\bar{n})(\gamma^\prime + \tilde{\gamma})]$ is a pumping rate. We observe that $\rho_{22} > \rho_{11}$ holds with $R>0$. This shows the generation of ground state population imbalance with a selective drive. The optical pumping process is competing with the phonon-assisted relaxation process that depletes $\rho_{22}$ at a rate of $\bar{n}(\gamma^\prime + \tilde{\gamma})$. When $R \gg \bar{n}(\gamma^\prime + \tilde{\gamma})$ and $\bar{n} \ll 1$, the ground state is fully polarized ($\rho_{22} \approx 1$).

In order to quantitatively calculate the critical Rabi frequency, we define the saturation condition as $\rho_{11} - \rho_{33} = (\rho_{11} - \rho_{33})_0/\sqrt{2}$ where $(\rho_{11} - \rho_{33})_0 = 1/(4\bar{n}+2)$ is the initial population difference at thermal equilibrium under no drive. The critical Rabi frequency that achieves this condition is:
\begin{equation}
\label{foursat}
|\Omega_c (B>0)|^2 = \frac{\bar{n}(1+\bar{n})(1+2\bar{n})^2 \gamma^\prime (\gamma^\prime + \tilde{\gamma})^2(\sqrt{2}-1)}{\gamma^{\prime}+8\bar{n}(1+\bar{n})\gamma^{\prime}+\tilde{\gamma}+4\bar{n}(1+\bar{n})\tilde{\gamma}}
\end{equation}

Comparing Eq.~\ref{twolevelsat} and Eq.~\ref{foursat}, we can calculate the ratio of saturation power for a four-level system with and without magnetic field:
\begin{equation}\label{p_ratio}
\frac{P_c(B=0)}{P_c(B>0)} = \frac{n_{c}(B=0)}{n_{c}(B>0)} = \left\lvert \frac{\Omega_c(B=0)}{\Omega_c(B>0)}\right\rvert^2 = 1 + \frac{1}{8\bar{n}(\bar{n}+1)} + \frac{\tilde{\gamma}}{8\bar{n}(\bar{n}+1)\gamma^\prime} + \frac{\tilde{\gamma}}{2\gamma^\prime}
\end{equation}

Two remarks can be made here: (1) the ratio of saturation power with and without a magnetic field is greater than 1. This is consistent with the intuition that the ability to pump into a dark state reduces saturation power, (2) the magnetic field induced contrast has a $1/\bar{n}$ dependence for $n \ll 1$, and is therefore strongly temperature dependent. 

\begin{figure}[ht]
    \centering
    \includegraphics[width=172mm]{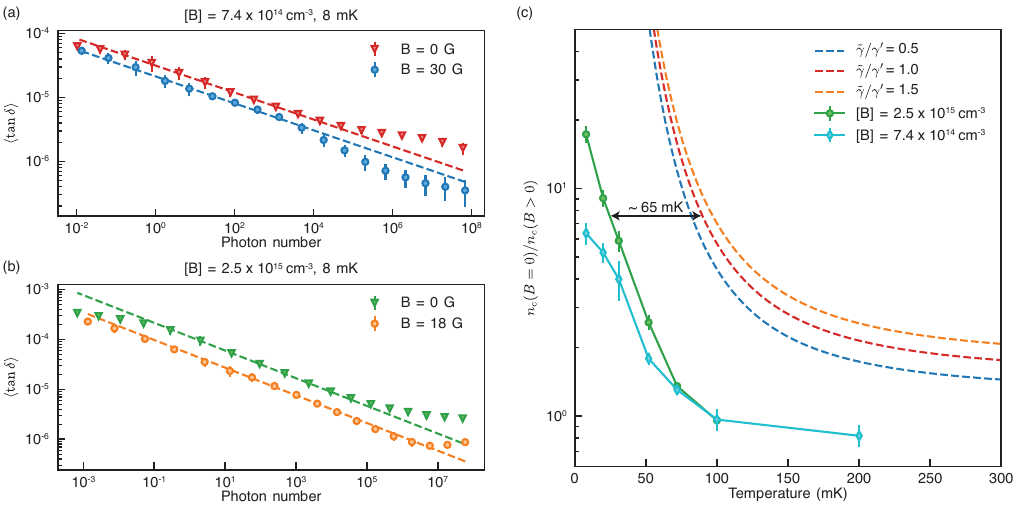}
    \caption{Power dependence of average loss tangent of eight resonators at $\text{T}_{\text{MXC}}=8$~mK with/without magnetic field in (a) a float-zone grown substrate with [B] = $7.4\times 10^{14} \,\text{cm}^{-3}$ and (b) a Czochralski grown substrate with [B] = $2.5\times 10^{15} \,\text{cm}^{-3}$. The dashed lines are linear fits between $\text{log}(\tan\delta)$ and $\text{log}(n)$ based on Eq.~\ref{fit}. (c) Theoretical and experimental critical photon number ratio with and without magnetic field as a function of temperature for two boron doping concentrations and three branching ratios ($\tilde{\gamma}/\gamma^\prime$). In theoretical calculations, we use the average frequency ($\approx 6.1$~GHz) of our resonators to convert thermal occupancy to temperature.}
        \label{nc_ratio}
\end{figure}

We fit the temperature dependence of the saturation power change induced by a magnetic field on two samples with high boron doping. The generic form of power dependent loss tangent for superconducting resonators is expressed as:
\begin{equation}\label{fit1}
     \tan\delta = \tan\delta_0\frac{A(T)}{\sqrt{1+\left(\frac{n}{n_c}\right)^\beta}}
\end{equation}
where $\tan\delta_0$ indicates the loss tangent with no drive at zero temperature, and $n_c$ denotes the critical photon number that suppresses the loss by $\sqrt{2}$, $\beta$ is a geometry-dependent fit parameter, and $A(T) = \text{tanh}(\hbar\omega/(2k_B T))$ represents thermal saturation. The fitting parameter $\beta$, which is not captured in theory, is phenomenologically added to account for electrical field distribution. We perform our analysis in a regime where $n/n_c \gg 1$, such that:
\begin{equation}\label{fit}
     \text{log}(\tan\delta) = \text{log}\left(A(T)\tan{\delta_0}\right) -\frac{1}{2}\text{log}\left(1+\left(\frac{n}{n_c}\right)^\beta\right)\approx \text{log}\left(A(T)\tan{\delta_0}\right)-\frac{\beta}{2}\text{log}(n) + \frac{\beta}{2}\text{log}(n_c)
\end{equation}
We fit the data in Fig.~\ref{figMT} to $\text{log}(\tan\delta) = -a~\text{log}(n) + b$, where $a=\beta/2$ and $b=\text{log}(A(T)\tan{\delta_0})+\frac{\beta}{2}\text{log}(n_c)$ are fit parameters, as seen in Fig.~\ref{nc_ratio}(a,b). The fit parameter $a$ (resonator geometry dependent) is shared at a given temperature for both $B=0$ and $B>0$ cases. Note that $\Delta b = b(B=0) - b(B>0) = a~\text{log}\left(n_{c}(B=0)/n_{c}(B>0)\right)$ when we assume the zero-power loss tangent is magnetic field independent ($A(T)\tan{\delta_0}(B=0) = A(T)\tan{\delta_0}(B>0)$). Therefore, the ratio between critical photon numbers can be calculated as
\begin{equation}\label{p_ratio_exp}
    \frac{n_{c}(B=0)}{n_{c}(B>0)} = 10^{\Delta b / a}
\end{equation}
In Fig.~\ref{nc_ratio}(c), we plot the experimentally extracted ratios of critical photon numbers as a function of temperature for two boron-doped samples and compare them to the theoretical prediction based on Eq.~\ref{p_ratio}. For both samples, we observe that the critical photon number is reduced by approximately an order of magnitude with an applied magnetic field at our base temperature (8~mK). At the same time, the magnetic field response disappears at slightly elevated temperatures around 100~mK. The experimental results qualitatively share the similar temperature-dependent behavior  as the theoretical prediction. However, the experimental temperature is about 65~mK lower than the theoretical prediction to achieve the same ratio of critical photon numbers. This discrepancy can be accounted for if the actual sample temperature is higher than the temperature reading of the mixing chamber of the fridge. Indeed, we observed a non-negligible thermal excited state population (2.8\%) for a superconducting transmon qubit at 6.3~GHz measured in the same setup~\cite{odeh_2023}. This excited state population translates to a chip temperature of 84~mK when the mixing chamber temperature reads 10~mK. Therefore, we expect a higher contrast in magnetic field response when the sample is better thermalized.

Even though Eq.~\ref{p_ratio} explains the qualitative trend of the saturation photon number ratio, it does not capture all the experimental observations. We note that the high temperature critical photon number ratio is estimated higher in theory. The discrepancy between our experiment and the theory may originate from: (1) our master equation model only takes into account the boron defects on resonance with the resonator. The large bath of off-resonant boron defects can also lead to loss and are not accounted, (2) the fitting parameter $\beta$ is not predicted by theory but rather a phenomenologically included parameter to account for electric field distribution in a real device.

\clearpage
\bibliography{main}